\documentclass[aps,pra,twocolumn,groupedaddress]{revtex4-2}
\usepackage{physics}
\usepackage{graphicx}
\usepackage{amsfonts} 
\usepackage{hyperref}
\usepackage{bbold}
\usepackage{xcolor}

\newcommand{\epss}{\epsilon_\mathrm{s}}

\newcommand{\Es}{E} 

\newcommand{\cinv}{\vb{c}^{\mathrm{in}}} 
\newcommand{\cin}{c^{\mathrm{in}}} 
\newcommand{\coutv}{\vb{c}^{\mathrm{out}}} 
\newcommand{\cout}{c^{\mathrm{out}}} 

\newcommand{\cinout}{c^{\mathrm{in \, (out)}}}
\newcommand{\cinoutv}{\vb{c}^{\mathrm{in \, (out)}}}

\newcommand{\gin}{g^{\mathrm{in}}}
\newcommand{\gout}{g^{\mathrm{out}}}
\newcommand{\ginout}{g^{\mathrm{in \, (out)}}}

\newcommand{\ctildein}{\Tilde{c}^{\mathrm{in}}}
\newcommand{\ctildeout}{\Tilde{c}^{\mathrm{out}}}
\newcommand{\ctildeinout}{\Tilde{c}^{\mathrm{in/out}}}

\newcommand{\lead}{\sigma}
\newcommand{\indexLR}{{\lead,n}}
\newcommand{\indexL}{{\mathrm{l},n}}
\newcommand{\indexR}{{\mathrm{r},n}}

\newcommand{\pt}{\partial_t}
\newcommand{\px}{\partial_x}
\newcommand{\pa}{\partial_\alpha}

\newcommand{\ptA}{\dot{A}}

\newcommand{\SF}{S_\mathrm{F}} 

\newcommand{\QF}[1]{Q_{\mathrm{F}\!,#1}} 
\newcommand{\evF}[1]{\theta_{\mathrm{F}\!,#1}}

\DeclareMathOperator{\sign}{sign}

\newcommand{\cor}[1]{#1}
\newcommand{\corr}[1]{#1} 

\begin{document}

\title{\cor{Pseudounitary} Floquet Scattering Matrix for Wavefront-Shaping in Time-Periodic \cor{Photonic} Media}
\author{David Globosits}
\author{Jakob Hüpfl}
\author{Stefan Rotter}
\affiliation{Institute for Theoretical Physics, Vienna University of Technology (TU Wien), 1040 Vienna, Austria}

\begin{abstract}
The physics of waves in time-varying media provides numerous opportunities for wave control that are unattainable with static media. In particular, Floquet systems with a periodic time modulation are currently of considerable interest. Here, we demonstrate how the scattering properties of a finite Floquet medium can be correctly described by a static Floquet scattering matrix, which satisfies a pseudounitary relation. This algebraic property is a consequence of the conservation of wave action for which we formulate here a continuity equation. \cor{Using this Floquet scattering matrix, we further demonstrate how it can be used to transfer concepts for wavefront-shaping based on the Wigner-Smith operator from static to Floquet systems.} The eigenstates of the corresponding Floquet Wigner-Smith matrix are shown to be light pulses that are optimally shaped in both their spatial and temporal degrees of freedom for the optical micromanipulation of time-varying media.
\end{abstract}

\maketitle

\section{Introduction}
The scattering matrix not only is a valuable tool for storing information about reflection and transmission of waves, but also marks the central building block for several wave engineering protocols \corr{\cite{Mosk,Kim,Rotter,Yoon,Gigan,Bertolotti,Cao22,Gigan22,Patel,Cao23,Asadova}}. Recent technological progress in measuring the scattering matrix for various systems has accelerated both theoretical and experimental investigations to exploit its information content \cite{Popoff,Shi,Yu,Chaigne,Andreoli,Dremeau,Mounaix,Horodynski,Boniface,Li21,Horodynski22,Choi}. \corr{All these developments are based on the understanding of the key characteristics of the scattering matrix, such as its unitarity. This property is fulfilled for time-independent scattering landscapes without gain and loss and reflects the conservation of energy during the scattering process due to time-translation invariance.}

A newly emerging field of research that is currently attracting great attention is that of waves in time-varying media. The special properties of waves inside a material that changes its properties with time open novel opportunities in the field of wave control \cite{Galiffi22,Pacheco,Won}. Several theoretical studies on the temporal scattering properties of waves inside a time-varying medium \cite{Biancalana,Xiao14,Hayrapetyan,Koutserimpas18} have recently been complemented by innovative experiments with water waves \cite{Bacot,Apffel} and with electromagnetic radiation, where effects like temporal reflection \cite{Moussa} and refraction \cite{Lustig23} or a temporal double slit \cite{Tirole} have been observed.

These exciting developments can be expected to lead to the realization of ever more sophisticated systems that will require a comprehensive description of wave scattering in such time-varying media. The question we want to address here is thus whether and how a scattering matrix can also be properly defined for photonic time-varying media, in particular for those with a periodic time dependence like photonic time crystals \cite{Lustig232,Wang23}. The first important developments in setting up a scattering matrix formalism for periodically modulated Floquet systems were presented for the case of electron scattering \cite{Reichl99,Reichl01,Reichl02}. While the conservation of probability in the scattering of electrons leads to a unitary Floquet scattering matrix, the situation is much more complex for electromagnetic waves: In bulk \cor{media} with a periodically time-varying impedance, the temporal reflection and refraction change the frequency of radiation propagating in these media. These frequency shifts are not energy preserving and thus render the associated scattering matrix non-unitary. \cor{In} the special case that a spatial translation invariance of the system leads to the conservation of the wave momentum \cite{Chegnizadeh}, characteristic algebraic properties of the temporal scattering matrix can be derived \cite{Li,Yin}. However, for the general case of a finite time-varying medium, neither the wave energy (due to the \cor{temporal inhomogeneity caused by the} time modulation) nor the wave momentum (due to the \cor{spatial inhomogeneity caused by the} finite extent \cor{of the slab}) is conserved \cor{(irrespective of the specific definition used for the wave momentum inside the slab \cite{PhysRevLett.104.070401})}. \cor{As a result, the Floquet scattering matrix does not obey a unitarity relation for these systems \cite{Zurita,Martinez16,Huanan,Martinez18,Pantazopoulos,Stefanou,Stefanou23,Ptitcyn} and entries are even found to be diverging at resonance \cite{Zurita10}, when expressed in an energy-flux normalized basis}. These shortcomings not only point to a missing element in the fundamental understanding of scattering at periodic media, but also hamper the transfer of well-established wave control techniques from time-invariant systems to time-varying ones.

\cor{Our central goal is to overcome these limitations by proving that the multispectral Floquet scattering matrix for electromagnetic waves obeys a so-called pseudounitary relation. On a fundamental level, this relation encapsulates that even for time-varying media a conserved quantity exists, which we identify as the wave action \cite{Brizard,Bellotti} notably \cor{incorporating} elastic as well as inelastic scattering process. Crucially, the pseudounitarity of the Floquet scattering matrix emerges through the coupling of positive- and negative-frequency channels, a characteristic of systems with fast and strong modulation. Previously, the Floquet scattering matrix has only been considered in regimes where these transitions can be neglected, in which case its \corr{pseudounitarity} reduces to the conventional unitarity relation \cite{Buddhiraju,Fan22}.} 

\section{The Floquet Scattering Matrix \label{sec:FSM}}
\begin{figure}
    \includegraphics[width=\columnwidth]{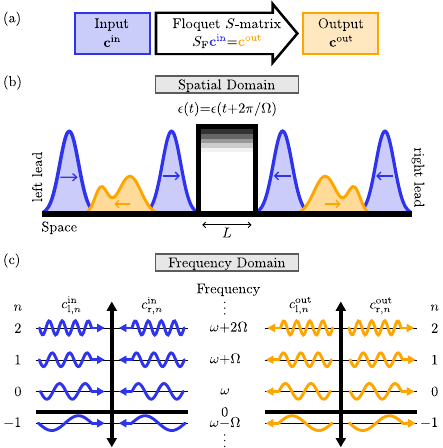}
    \caption{\label{fig:Fig1} Schematic representation of a Floquet scattering process. \corr{(a)} The Floquet scattering matrix $\SF$ transforms the vector of incoming channels $\cinv$ into a vector of outgoing channels $\coutv$. (b) In the spatial domain, an input pulse is scattered off a periodically time-varying medium \corr{of length $L$} described by the periodic permittivity $\epsilon(t)=\epsilon(t+2\pi/\Omega)$. (c) Due to the periodic driving, an input wave of quasifrequency $\omega$ can only change its frequency by a multiple of the driving frequency $\Omega$. Thus, in the frequency domain the scattering process can be described by a discrete set of channels $n$.}
\end{figure}
\begin{figure*}
    \includegraphics[width=\textwidth]{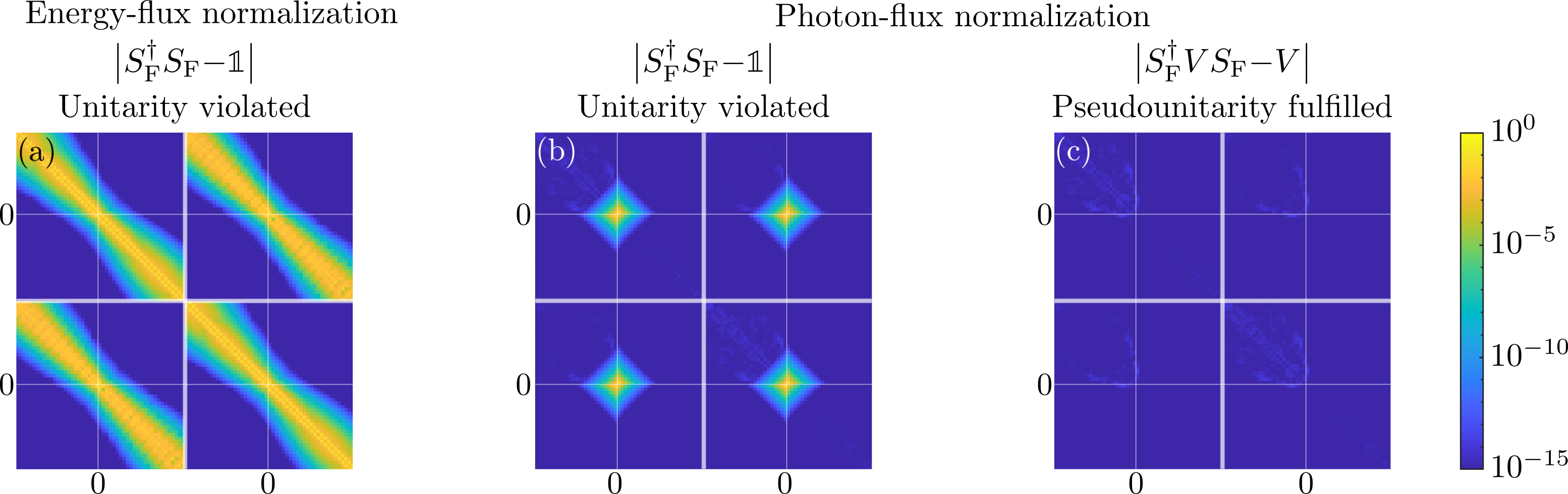}
    \caption{\label{fig:bstH_Unitarity} Unitarity vs pseudounitarity of the Floquet scattering matrix (logarithmic scale). (a) Absolute value of the deviation from unitarity of the Floquet scattering matrix for energy-flux normalized asymptotic states. Since energy is not conserved during the scattering process, large deviations from unitarity can be observed. (b) Absolute value of the deviation from unitarity for photon-flux normalized states. Large deviations from unitarity appear around $n=m=0$, where frequency channels of different signs couple to each other, rendering the Floquet scattering matrix nonunitary. (c) The Floquet scattering matrix is pseudounitary (errors are below $10^{-13}$ and thus below numerical precision), with the coupling between positive and negative frequencies now being taken into account properly. In all panels we assume a harmonic modulation of the form $\epsilon(t) = \corr{1}+\epss+\delta\epsilon \cos(\Omega t)$, \cor{with $\epss=1$ and $\delta\epsilon=0.3$.}}
\end{figure*}

The central object that we are after is the photonic Floquet scattering matrix, which is a generalized scattering matrix applicable to time-periodic scattering processes in electromagnetism. To discuss the unconventional properties of the Floquet scattering matrix, we reduce the complexity of the scattering system as much as possible while still maintaining the characteristic features of wave propagation in time-varying media. For this purpose, we consider a one-dimensional scattering problem and assume an instantaneous response of the material to the modulation \cite{Zurita} such that the system is described by the scalar wave equation of the form
\begin{equation}
    \partial_x^2 \Es(x,t)-\partial_t^2 \left[ \epsilon(x,t) \Es(x,t) \right]=0. \label{eq:scalar_waveeq_E}
\end{equation}
Here and in the following, we set the vacuum permittivity and permeability to unity, i.e., $\varepsilon_0=\mu_0=1$, such that the speed of light becomes $c=(\varepsilon_0 \mu_0)^{-1/2}=1$. Specifically, we describe a scattering landscape with a real time-periodic dielectric function $\epsilon(x,t)=\epsilon(x,t+T)$, where $T$ is the oscillation period and $\Omega=2\pi/T$ is the associated angular frequency. Furthermore, we assume that this oscillating object has a finite extent $L$ and is centered around $x_0$ such that its dielectric function may be written as
\begin{equation}
    \epsilon(x,t)=1+[\epss+\delta \epsilon \, g(t)]\Theta(L/2-\abs{x-x_0}), \label{eq:epsilon_general}
\end{equation}
where $\Theta$ is the Heaviside step function (outside the oscillating dielectric we assume $\epsilon=1$). The static permittivity of the scatterer is $1+\epss$ and the time-periodic modulations occur on top of this static background, where $\delta\epsilon \geq 0$ is the modulation strength and $g(t)=g(t+T)$ is the time-periodic modulation itself. In a periodically time-dependent scattering landscape energy is not conserved due to the external driving. However, in Floquet systems an incoming wave cannot exchange energy arbitrarily with the system, but only by a multiple of the driving frequency $n\Omega$ with $n\in \mathbb{Z}$ \cite{Joachain}. This gives rise to a quantity called quasifrequency $\omega$ (corresponding to the quasimomentum in spatially periodic systems). This characteristic frequency is restricted to an interval $l\Omega \leq \omega < (l+1) \Omega$ with arbitrary integer~$l$ (corresponding to the Brillouin zone). We choose $l=0$ here and throughout the paper. Such a particular choice of the quasifrequency allows us to label the available frequencies of the scattering process as $\omega_n=\omega+n\Omega$ (see Fig.~\ref{fig:Fig1}). These frequencies span a discrete set of scattering channels in the frequency domain with a corresponding time-dependence $e^{-i \omega_n t}$. We are thus able to describe the scattering behavior of several input channels, corresponding to plane waves with frequencies $\omega_n$, at once. Hence, the wave field in the left lead ($\sigma=\mathrm{l}$) and in the right lead ($\sigma=\mathrm{r}$) may be expressed as \cite{Zurita}
\begin{equation}
    E_\lead(x,t) = \corr{\sum_{n=-\infty}^{\infty}} \mathcal{N}_\indexLR \left( \cin_\indexLR e^{i k_\indexLR x} + \cout_\indexLR e^{-i k_\indexLR x} \right) e^{-i\omega_n t}.
    \label{eq:asym_Efield}
\end{equation}
We have a linear dispersion relation $k_\indexL = -k_\indexR = \omega_n$ in the leads and $\mathcal{N}_\indexLR$ are normalization constants, which we will discuss in the following. Details on the derivation of the wave field can be found in Appendix~\ref{sec:app_wavefield}. We introduce the amplitudes $\cin_\indexLR$ and $\cout_\indexLR$ corresponding to plane waves that are incoming and outgoing on both sides $\lead \in\{\mathrm{l},\mathrm{r}\}$ at the discrete frequencies $\omega_n$ (see Fig.~\ref{fig:Fig1}). Writing these coefficients as vectors
\begin{subequations}
    \begin{align}
        \cinv &= \mqty(\cinv_\mathrm{l} \\ \cinv_\mathrm{r}), \\
        \coutv&= \mqty(\coutv_\mathrm{l} \\ \coutv_\mathrm{r}),
    \end{align}
\end{subequations}
where \corr{$\cinoutv_\sigma = (\dots,\cinout_{\sigma,-1},\cinout_{\sigma,0},\cinout_{\sigma,1},\dots )^\mathrm{T}$}, the Floquet scattering matrix can now be defined via the relation
\begin{equation}
    \coutv=\SF \cinv.
\end{equation}
This multispectral scattering matrix not only incorporates information about the elastic scattering of waves of identical frequency, but additionally holds information about inelastic scattering processes to channels of different frequencies. Consequently, this matrix contains all asymptotically available information on the scattering of any multispectral wave field associated with a quasifrequency $\omega$ into an output wave field through a periodically time-varying medium.

To assess the general properties of the Floquet scattering matrix, we recall, that static scattering matrices $S$ exhibit specific algebraic properties, which are based on symmetries of the scattering process \cite{Rotter}. \cor{Crucially,} the static scattering matrix for time-independent systems without gain and loss is unitary, which embodies the conservation of the total energy in a scattering process. However, as in a Floquet system of finite size neither energy nor the linear momentum constitutes a conserved quantity, also the associated Floquet scattering matrix is not unitary \cite{Yin}. Nevertheless, as we will show below, a generalized unitary condition for the Floquet scattering matrix can be established, which is based on the conservation of the wave action.

For time-invariant energy-conserving systems, the unitarity of the scattering matrix follows from the source-free Poynting theorem that constitutes the fundamental continuity equation governing energy conservation for electromagnetic fields. The appearing energy-flux density (the Poynting vector) induces an appropriate normalization for the asymptotic electric fields, in which the scattering matrix is expressed. With this energy-flux normalization, the static scattering matrix is unitary since the incoming energy flux can only be distributed into the reflected and transmitted wave fields without changing the total energy of the field. Uncritically transferring the energy-flux normalization to a Floquet scattering process would result in the normalization constant~$\mathcal{N}_\indexLR = \sqrt{\abs{\omega_n/k_\indexLR}}=1$ in Eq.~\eqref{eq:asym_Efield}. Since, however, in Floquet scattering systems energy is not conserved, an energy-flux normalization of the asymptotic fields quite naturally yields a nonunitary Floquet scattering matrix. This is demonstrated in Fig.~\ref{fig:bstH_Unitarity}(a) and has already been observed by several authors before \cite{Zurita,Zurita10,Martinez16,Martinez18,Pantazopoulos,Stefanou,Stefanou23}.

To fix this problem, an alternative source-free continuity equation is thus required to hold also in the time-varying case. Indeed, there is a conserved quantity in Hermitian time-varying scattering systems: the action of the wave field \cite{Brizard,Bellotti}. Similar to the energy density and the energy-flux density in the Poynting theorem, we introduce here a wave-action density $u(x,t)$ and a wave-action-flux density $j(x,t)$, which obey the continuity equation
\begin{equation}
    \partial_t u(x,t) + \partial_x j(x,t) = 0. \label{eq:continuity}
\end{equation}
Specifically, we use the vector potential $A(x,t)$ to define them as
\begin{align}
    u(x,t) &= \epsilon(x,t)\left(A^*\partial_t A - A\partial_t A^*\right), \\
    j(x,t) &= A\partial_x A^*-A^*\partial_x A,
\end{align}
where the asterisk represents complex conjugation and we omitted the space and time dependence of the $A$ field for brevity. Details on the derivation of Eq.~\eqref{eq:continuity} are presented in Appendix~\ref{sec:app_continuity}. The electric field is recovered from the vector potential as $E(x,t)=-\partial_t A(x,t)$ in the Coulomb gauge \cite{Leonhardt}. As we show in the following, this continuity equation induces an appropriate normalization for the asymptotic wave fields in Floquet scattering problems, since the incoming action-flux density equals the outgoing one (``outgoing'' refers here to the sum of the reflected and transmitted quantities). Furthermore, expressing the Floquet scattering matrix in the basis of action-flux normalized input states, we derive in the following a generalized unitary condition for the Floquet scattering matrix.

We start by integrating Eq.~\eqref{eq:continuity} over one temporal period $0 \leq t \leq T$ and over the spatial extension of the finite scattering region $-L/2+x_0 \leq x \leq L/2+x_0$. By using the temporal periodicity of $u(x,t)$ and inserting the Fourier representation of the asymptotic fields [Eqs.~\eqref{eq:asym_Efield}], we arrive at (for details see Appendix~\ref{sec:app_continuity})
\begin{equation}
\begin{split}
    0 &= \int_{-T/2}^{T/2} \frac{\dd{t}}{T} \left[ j(L/2+x_0,t) - j(-L/2+x_0,t) \right] \\
     &= 2i \sum_\indexLR \frac{\abs{\mathcal{N}_\indexLR}^2}{\omega_n }\left( \abs{\cin_\indexLR}^2 - \abs{\cout_\indexLR}^2 \right). \label{eq:continuity_integral}
    \end{split}
\end{equation}
This relation holds two crucial insights. First, it suggests to choose for the normalization $\mathcal{N}_\indexLR=\sqrt{\abs{\omega_n}}$, which reduces the factor $\abs{\mathcal{N}_\indexLR}^2/\omega_n$ in Eq.~\eqref{eq:continuity_integral} to $\sign(\omega_n)$. The resulting action-flux normalization, which is also termed photon-flux normalization in the literature, is typically imposed on quantized electric fields \cite{Birrell}, but is applicable for classical Maxwell fields as well \cite{Cohen, New}. With the photon being considered as a quantized entity of the wave action, the conservation of wave action in the classical domain is equivalent to the conservation of \corr{pseudophotons} contained in the field \cite{Bellotti,Xiao11,Pendry23}.

Second, by applying the photon-flux normalization to Eq.~\eqref{eq:continuity_integral} and rewriting it in a matrix form, we arrive at the desired generalization of the unitary condition of the photonic Floquet scattering matrix
\begin{equation}
    \SF^{\dagger}V\SF=V. \label{eq:pseudounitarity}
\end{equation}
A matrix $\SF$ obeying this type of relation is called a $V$-pseudounitary matrix (the standard unitary relation is recovered if $V$ is the identity matrix) \cite{Mostafazadeh}. In our case, the matrix $V$ is a diagonal matrix, which assigns an additional minus sign to negative-frequency channels
\begin{equation}
    V = \mqty(-\mathbb{1}  & \mathbb{0} & \mathbb{0} & \mathbb{0} \\
          \mathbb{0} & \mathbb{1} & \mathbb{0} & \mathbb{0} \\
          \mathbb{0} & \mathbb{0} & -\mathbb{1} & \mathbb{0} \\
          \mathbb{0} & \mathbb{0} & \mathbb{0} & \mathbb{1} ),
\end{equation}
where the top left (bottom right) matrix quadrant is assigned to waves incoming from the left (right). \corr{Notably, when deriving Eq.~\eqref{eq:pseudounitarity} the only assumption imposed on the scattering landscape is that it is described by a real and time-periodic permittivity $\epsilon(t)$. This means that the Floquet scattering matrix is pseudounitary for arbitrary modulation schemes [corresponding to an arbitrary real time-periodic function $g(t)$ in Eq.~\eqref{eq:epsilon_general}].}

To test this pseudounitarity relation of the Floquet scattering matrix explicitly, we present numerical data in Fig.~\ref{fig:bstH_Unitarity}. \cor{Specifically, we assume a harmonic modulation, $g(t)=\cos(\Omega t)$, but we checked numerically (not shown) that these results \corr{indeed} hold for arbitrary real and time-periodic modulation protocols as well, even for those that involve several harmonics and for potentially nonreciprocal systems that break spatial and/or temporal symmetries.} In Fig.~\ref{fig:bstH_Unitarity}(a) the Floquet scattering matrix is expressed in an energy-flux normalized basis and, as expected, notable deviations from unitarity are present. Furthermore, even when a photon-flux normalization is used, the Floquet scattering matrix is not a unitary matrix [see Fig.~\ref{fig:bstH_Unitarity}(b)]. In particular, in the photon-flux basis the deviations from unitarity are most pronounced in the vicinity of the zero channel ($n=0$). Since this channel marks the border between positive and negative frequencies, processes that change the sign of the frequency are most likely to appear here. This strongly indicates, that the coupling between positive and negative channels is crucial in a Maxwell-based Floquet scattering process \cite{Serra} and has to be taken into account properly. Correspondingly, these deviations completely vanish in the case of pseudounitarity in a photon-flux normalized basis [see Fig.~\ref{fig:bstH_Unitarity}(c)]. The additional minus signs, which are necessary to describe the mixing of negative and positive frequencies correctly, enter here through the matrix $V$. \cor{Furthermore, if only positive frequencies are involved and no conversion between positive and negative frequencies takes place during the scattering process, e.g. due to a weak or slow modulation, only the positive-frequency part of the matrix $V$ is relevant. This part is in fact the identity matrix; thus the pseudounitarity condition reduces to the standard unitary condition \cite{Buddhiraju,Fan22} (the same argument holds true if only negative frequencies are considered).} \cor{Moreover,} in the static limit, i.e., for $\delta \epsilon=0$, only elastic scattering processes are present. Correspondingly, also the Floquet scattering matrix obeys both the standard unitary relation as well as the pseudounitary relation in the static limit. For details on the truncation of the Floquet scattering matrix and the choice of parameters we refer the reader to Appendix~\ref{sec:app_truncation}.

\cor{Next} we point out interesting differences between the photonic Floquet scattering matrix discussed here and the electronic Floquet scattering matrix introduced earlier for the scattering of matter waves off periodically modulated potentials \cite{Reichl99,Reichl01,Reichl02}. Since the total probability of finding a particle described by the Schrödinger equation can neither increase nor decrease even in modulated potentials, the Schrödinger-based Floquet scattering matrix is unitary in a basis of asymptotic scattering states when these are normalized using the probability flux. Moreover, we need to stress that Schrödinger waves with negative energies correspond to evanescent waves, which carry no probability flux. As a result, the electronic Floquet scattering matrix is restricted to positive-frequency channels for which a standard unitary relation holds. Our findings on the pseudounitarity of the Floquet scattering matrix are thus specific to photonic systems with a linear dispersion relation. In this case, negative-frequency channels are still propagating waves carrying both energy flux and photon flux to the asymptotic regions. Hence, negative-frequency channels have to be considered within the photonic Floquet scattering matrix. \cor{For waves with potentially different dispersion relations, a careful analysis along the lines above is recommended.}

\cor{We note that there exists another important property of scattering matrices, both in the static and in the time-variant case. If the scattering system is reciprocal, the static scattering matrix can be chosen to be transposition symmetric \cite{Rotter}. Conversely, asymmetric scattering matrices can be used to detect nonreciprocity in static systems \cite{Jalas}. Similarly, asymmetric Floquet scattering matrices have been used to identify nonreciprocity in periodically time-varying scattering systems \cite{Sounas,Koutserimpas182,Wang20,Wang21}. We remark that our findings on the pseudounitarity on the Floquet scattering matrix are however independent of reciprocity and thus hold for both reciprocal and nonreciprocal Floquet scattering systems, which are described by a real time-periodic permittivity function.}

We \cor{further} emphasize that the findings presented here for a one-dimensional problem also apply to higher-dimensional setups involving more than one spatial dimension and an arbitrary number of asymptotic scattering channels. For all of these cases, our formalism can be used to map well-established results from static scattering theory, which rely on the unitarity of the static scattering matrix, to Floquet systems.

\section{The Floquet Wigner-Smith Matrix}
In this section we aim to demonstrate that the photonic Floquet scattering matrix is not just an abstract theoretical concept but is indeed very useful for the implementation of various wave-front-shaping protocols. We recall that in the time-independent case the static monochromatic scattering matrix is the key quantity to determine those incoming wave fronts that are optimally suited to transmit through a given medium, to focus behind or inside it or to perform certain tasks like micromanipulating a specific object \corr{\cite{Mosk,Abboud,MaCheng,ZhouEdward,Kim,Rotter,Yoon,Gigan,Bertolotti,Cao22,Gigan22,Patel,Cao23,Sol}}.

A versatile tool in this context is the generalized Wigner-Smith (GWS) matrix
\begin{equation}
    Q_{\alpha}=-iS^\dagger \partial_\alpha S.
\end{equation}
\cor{The definition of this matrix only involves the static scattering matrix $S$ and its derivative with respect to a \corr{parameter $\alpha$ on which the scattering matrix depends} \cite{Brouwer97,Brouwer99,Ambichl,Horodynski}. We note, that originally this matrix was introduced by Wigner \cite{Wigner} and Smith \cite{Smith} with the particular choice of $\alpha$ being the frequency of the incident field.} The static scattering matrix is expressed here in an energy-flux normalized basis and is thus unitary. Correspondingly, the monochromatic $Q_\alpha$ matrix is Hermitian and evaluated at a single frequency only. The GWS matrix has successfully been applied in various settings ranging from the design of tractor beams \cite{Horodynski23PRA} and maximally stiff particle traps \cite{Butaite} to optimally deal with perturbations in multimode fibers \cite{Matthes} and cool levitated particles \cite{HupflPRL,HupflPRA} or for the inverse design of antireflection coatings for perfect transmission through complex media \cite{Horodynski22}. Here we make use of the feature that the GWS matrix yields optimal light fields for micromanipulation and optimal focusing \cite{Ambichl, Horodynski,Orazbayev}. \corr{We note that for the case of optimal focusing on time-independent targets, also several other scattering matrix-based schemes exist that can identify the corresponding optimal wave fields \cite{Abboud,MaCheng,ZhouEdward,Sol}. The GWS matrix comes with the advantage however that it lets us identify not only these optimal focusing states but also those} that apply, for example, a well-defined force or torque to a specific target scatterer even if this target is located inside a disordered medium. This is due to a close connection between the eigenvalues $\theta_{\alpha}$ of the GWS matrix and the near field of the target, which reads, for the time-invariant one-dimensional case,
\begin{equation}
    \theta_{\alpha}=\frac{\omega}{2}\int \dd{x} \abs{E(x)}^2 \partial_\alpha \epsilon(x,\alpha). \label{eq:static_GWS_eigval}
\end{equation}
Here $E(x)$ is the complex amplitude of the time-harmonic electric field with angular frequency $\omega$ and the permittivity $\epsilon(x,\alpha)$, which is a function of space but not of time and which depends parametrically on $\alpha$. The eigenvalue $\theta_\alpha$ of the GWS matrix $Q_\alpha$ on the left-hand side of this analytic relation is associated with a corresponding eigenvector that represents the far-field input light fields that give rise to the near fields appearing on the right-hand side of the equation. In general, this near field in the vicinity of the target scatterer gets weighted with the derivative of the permittivity with respect to the parameter $\alpha$. In this way \cite{Horodynski}, Eq.~\eqref{eq:static_GWS_eigval} links a specific manipulation task in the near field of an object with the far-field input wave field that performs it. Moreover, the GWS matrix $Q_\alpha$ connects each target parameter $\alpha$ with the corresponding conjugate quantity: When the parameter $\alpha$ corresponds to a target's position, the eigenvalue $\theta_\alpha$ quantifies the momentum transferred onto it by the corresponding eigenstate, and when $\alpha$ corresponds to the target's angular rotation, $\theta_\alpha$ measures the angular momentum transfer, etc. \cor{We note that the eigenstates of the GWS matrix are equivalent to the so-called optical eigenmodes \cite{Mazilu}. A detailed discussion about this connection can be found in \cite{Horodynski}.}

With a unitary static scattering matrix giving rise to a Hermitian GWS matrix already by definition, its extremal eigenstates that deliver the strongest momentum or angular momentum transfer for a fixed input power, are simply given by those eigenvectors of the GWS matrix associated with the highest or lowest (real) eigenvalue $\theta_\alpha$. Here we show how to generalize the GWS matrix concept to periodically time-varying media via the Floquet scattering matrix. In particular, we demonstrate that we are able to identify states that are optimally shaped not only in their spatial but also in their temporal degrees of freedom to execute micromanipulation tasks at the optimal level of efficiency. For this purpose we first define the Floquet Wigner-Smith (FWS) matrix with a structure that is very similar to the static GWS matrix:
\begin{equation}
    \QF{\alpha}= U^\dagger \left( -i \SF^\dagger V \partial_{\alpha} \SF \right) U. \label{eq:QF_definition}
\end{equation}
This FWS matrix consists of the Floquet scattering matrix $\SF$ and its derivative with respect to a parameter $\alpha$ of the scattering system. To account for the pseudounitarity of $\SF$ the matrix $V$ is included in Eq.~\eqref{eq:QF_definition}. Furthermore, we introduce the diagonal matrix 
\begin{equation}
    \begin{split}
        U = \mathrm{diag} \big( \dots& \, ,\sqrt{\abs{\omega_{-1}}}, \sqrt{\abs{\omega_{0}}}, \sqrt{\abs{\omega_{1}}}, \, \dots \, , \\
        \dots& \, , \sqrt{\abs{\omega_{-1}}}, \sqrt{\abs{\omega_{0}}}, \sqrt{\abs{\omega_{1}}}, \, \dots \big).
    \end{split}
\end{equation}
The matrix $U$ transforms photon-flux normalized incoming coefficients into energy-flux normalized ones, which has the advantage that input states have a fixed energy content (a fixed input power) instead of a fixed (pseudo)photon content. Importantly, due to the pseudounitarity of the Floquet scattering matrix [Eq.~\eqref{eq:pseudounitarity}], the FWS matrix is a Hermitian matrix by definition. Moreover, as its eigenstates are wave fields containing multiple frequencies, these far field input states are electromagnetic pulses. As we show below, the eigenvectors corresponding to the extremal eigenvalues constitute pulses that perform certain tasks at the optimal level in the near field. The specific spatiotemporal properties of the eigenstates of the FWS matrix $\QF{\alpha}$ can be understood when considering the connection of the corresponding eigenvalues $\evF{\alpha}$ to the near field of a time-modulated target scatterer (see Appendix~\ref{sec:app_continuity} for a derivation):
\begin{equation}
    \evF{\alpha} = \frac{1}{2} \int_{-T/2}^{T/2} \frac{\dd{t}}{T} \int \dd{x} \abs{\Es(x,t)}^2 \partial_{\alpha} \epsilon(x,t,\alpha). \label{eq:FWS_relation_general}
\end{equation}
We highlight that the permittivity depends on space and time as well as on $\alpha$ parametrically. From a conceptual point of view, Eq.~\eqref{eq:FWS_relation_general} is the Floquet generalization of Eq.~\eqref{eq:static_GWS_eigval} including an additional temporal integral over one period. The eigenvalues of the FWS matrix are related to the time average of the electric field weighted with the derivative of the permittivity with respect to $\alpha$. Correspondingly, the eigenvectors associated with the eigenvalues $\evF{\alpha}$ of $\QF{\alpha}$ represent far-field input light fields, which result in near fields of the scatterer in Eq.~\eqref{eq:FWS_relation_general}. Since the weighting function in this equation is time dependent, the eigenstates of the FWS matrix $\QF{\alpha}$ in general give rise to pulsed wave fields that interact with the target scatterer with an interaction strength that is weighted by the periodic Floquet modulation. In other words, for maximally positive eigenvalues $\evF{\alpha}$ the input pulses will produce a maximal near-field intensity when the Floquet modulation reaches its peak value. For maximally negative eigenvalues the near field will be maximal when the Floquet modulation reaches its minimal values. In this way, our approach constitutes an optimal micromanipulation scheme for spatiotemporal wave control of Floquet media. In Appendix~\ref{sec:app_RealFields} we show how to apply the relation~\eqref{eq:FWS_relation_general} when, instead of complex electric fields, the associated physical (real) fields are considered. In the following we employ this FWS matrix approach to two different modulation protocols and explicitly demonstrate its ability to provide optimal spatio-temporal light fields for focusing and micromanipulation.

\subsection{Harmonic Modulation \label{sec:harmonic_mod}}
\begin{figure*}
    \includegraphics[width=\textwidth]{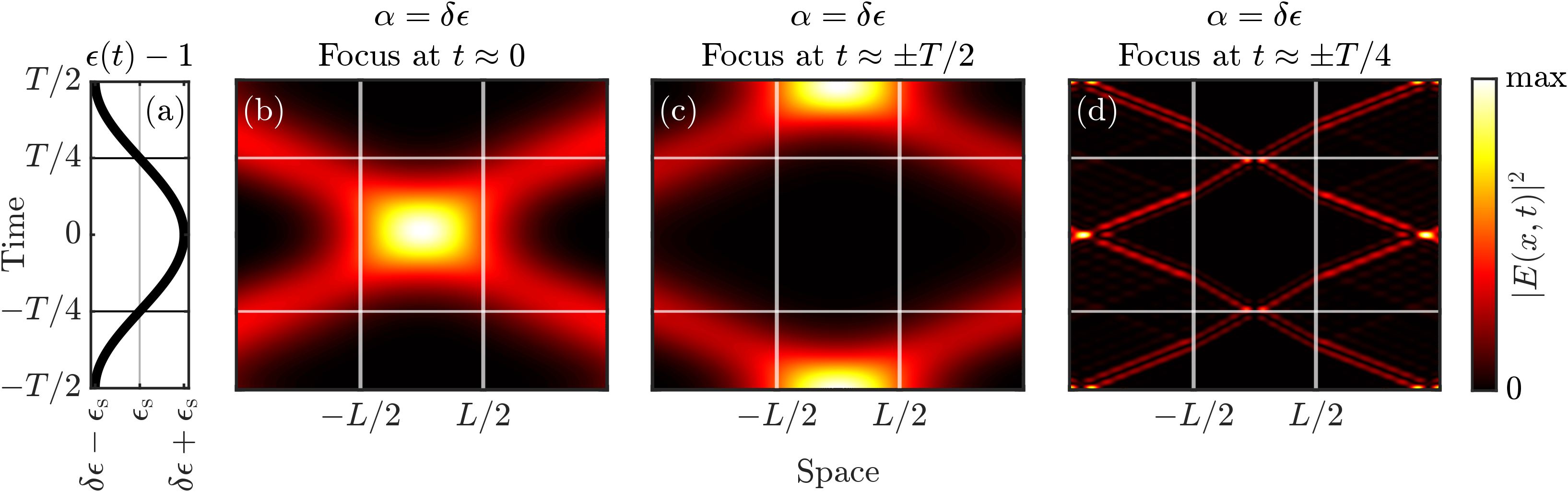}
    \caption{\label{fig:bstH_cos_deltaeps} Optimal spatiotemporal focusing of light. (a) One period of the harmonic modulation protocol $\epsilon(t)=1+\epss+\delta\epsilon \cos(\Omega t)$ of the target scatterer centered at the origin. (b) Intensity of the eigenstate corresponding to the largest eigenvalue of $\QF{\delta\epsilon}$. This optimal spatiotemporal state exhibits a focus inside the target during times when the permittivity $\epsilon(t)$ is high. (c) Intensity of the eigenstate corresponding to the most negative eigenvalue. This wave field builds up a focus when the permittivity is low. (d) The intensity of an eigenstate associated with an eigenvalue close to zero focuses at times when the modulation of the permittivity is zero.}
\end{figure*}
To start with, we consider a harmonic modulation of a target scatterer. This is the modulation protocol studied, e.g., in \cite{Zurita, Zurita10} as a model for a photonic time crystal, which we use here to demonstrate the capabilities of the FWS matrix. Specifically, we model the modulation as $g(t)=\cos(\Omega t)$ [see Fig.~\ref{fig:bstH_cos_deltaeps}(a)] such that Eq.~\eqref{eq:epsilon_general} takes the form
\begin{equation}
    \epsilon(x,t)=1+[\epss+\delta \epsilon \cos(\Omega t)]\Theta(L/2-\abs{x-x_0}).
\end{equation}
In the following we choose the scatterer to be centered at the origin, i.e., $x_0=0$. \cor{For the choices of the other parameters we refer the reader to Appendix~\ref{sec:app_truncation}}. We solve the corresponding scattering problem following the steps described in Sec.~\ref{sec:FSM} to derive the Floquet scattering matrix $\SF$. This immediately enables us to apply the FWS approach to the given scattering problem.

Next we address the issue of optimal spatiotemporal focusing inside the target. Our goal is to identify input states that focus inside the target at a given instant in time. This can be realized with a FWS approach by choosing $\alpha=\delta \epsilon$, i.e., by performing a derivative of the scattering matrix with respect to the amplitude of the modulation. To understand this, consider that the eigenstates of $\QF{\delta\epsilon}$ are input states that fulfill Eq.~\eqref{eq:FWS_relation_general}, which reduces in this case to
\begin{equation}
    \evF{\delta\epsilon}=\frac{1}{2} \int_{-T/2}^{T/2} \frac{\dd{t}}{T} \int_{-L/2}^{L/2} \dd{x} \abs{\Es(x,t)}^2 \cos(\Omega t). \label{eq:FWS_relation_cos_V1}
\end{equation}
The eigenstate corresponding to the most positive eigenvalue maximizes the integral appearing on the right-hand side of the above equation. As one observes in Fig.~\ref{fig:bstH_cos_deltaeps}(b), this is achieved by a wave field consisting of two pulses approaching the target from the left and right, interfering inside the scatterer to create a strong intensity maximum at times when the weighting function $\cos(\Omega t)$ is positive, i.e., for $-\frac{1}{4} < t/T < \frac{1}{4}$. The largest amount of intensity is focused around $t \approx 0$ when the weighting function has its maximum. On the other hand, the eigenstate corresponding to the most negative eigenvalue shows opposite temporal behavior, applying a strong focus at the beginning and end of the period, respectively. This wave field is depicted in Fig.~\ref{fig:bstH_cos_deltaeps}(c). The eigenstate corresponding to an eigenvalue close to zero shows yet another behavior: To keep the integral of Eq.~\eqref{eq:FWS_relation_cos_V1} as close to zero as possible, this state applies a sharp intensity peak at times when the weighting function  $\cos(\Omega t)$ is approximately zero, thus giving no contribution to the integral. This demonstrates that by selecting the appropriate eigenstate, we can choose the specific moment in time when light gets focused on the target. We remark that the exact position of the peaks depends on the modulation strength $\delta \epsilon$. This can be observed, for example, in Fig.~\ref{fig:bstH_cos_deltaeps}(c), where the light field has its maximum slightly before the end of the period. This is because the wave gains energy from the modulated scatterer during times when the permittivity decreases, while energy is taken from the wave field when the permittivity rises. The FWS approach takes these amplifying and deamplifying effects into account and the resulting wave field thus possesses a slight asymmetry with respect to the temporal period.

\begin{figure*}
    \includegraphics[width=\textwidth]{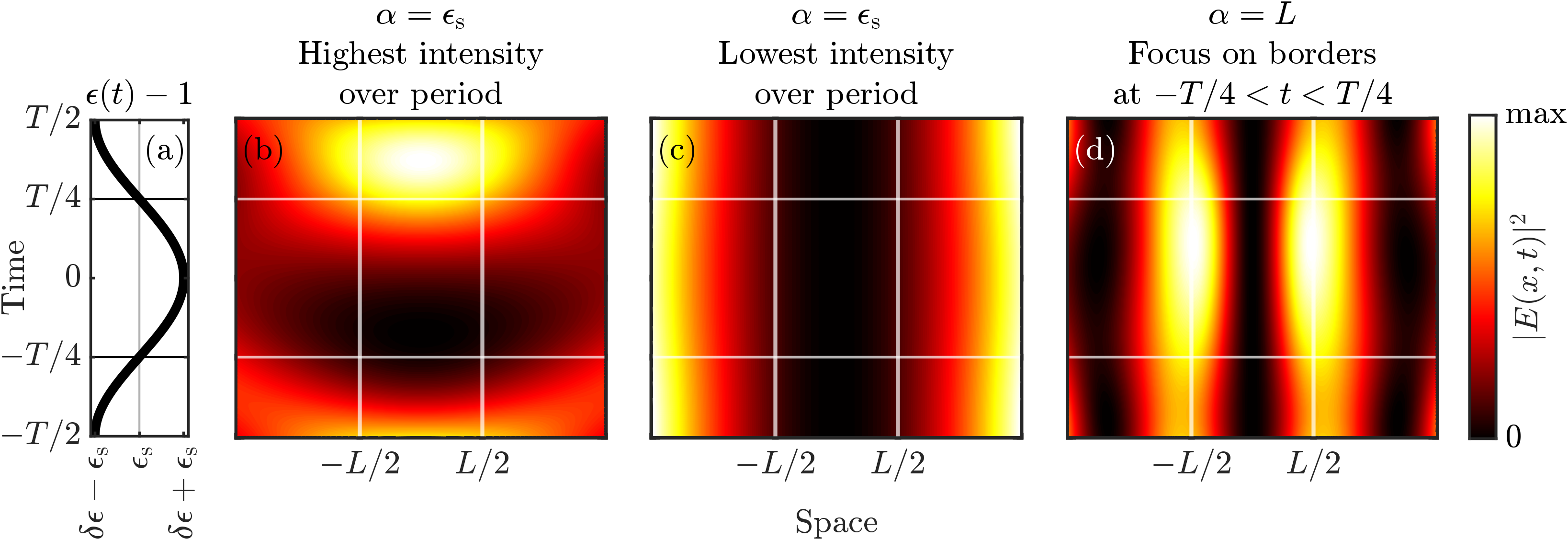}
    \caption{\label{fig:bstH_cos_L_eps} Focusing light inside and at the borders of a target scatterer. (a) One period of the harmonic modulation protocol $\epsilon(t)=1+\epss+\delta\epsilon \cos(\Omega t)$ of the target scatterer centered at the origin. (b) Intensity of the eigenstate corresponding to the largest eigenvalue of $Q_{\epss}$. This wave field maximizes the intensity built up inside the scatterer. (c) Intensity of an eigenstate to an eigenvalue close to zero of $Q_{\epss}$ has minimal intensity inside the target at all times. (d) The intensity of the eigenstate corresponding to the most positive eigenvalue of $Q_{L}$. This light field exhibits a focus at the borders of the scatterer during times when the permittivity $\epsilon(t)$ is high.}
\end{figure*}

Furthermore, we show that we can identify states that focus the maximum amount of intensity on the scatterer over the full period. Such light fields can be found with a FWS matrix involving a derivative with respect to the static part of the permittivity, $\alpha=\epss$. Applying Eq.~\eqref{eq:FWS_relation_general} in this case results in
\begin{equation}
    \evF{\epss}=\frac{1}{2} \int_{-T/2}^{T/2} \frac{\dd{t}}{T} \int_{-L/2}^{L/2} \dd{x} \abs{\Es(x,t)}^2. \label{eq:FWS_cos_epss}
\end{equation}
Figure~\ref{fig:bstH_cos_L_eps}(b) depicts the eigenstate corresponding to the maximal eigenvalue of $\QF{\epss}$. This optimal light field exhibits a single intensity maximum at the end of the period, which is built up at times when the permittivity is changing from a higher to a lower value. This can be understood as the wave gains energy from the modulated scatterer, which amplifies the intensity inside the target. Accordingly, the wave avoids times at which the permittivity rises, which would result in a deamplifying effect. In this way the wave field maximizes the intensity built up inside the scatterer during the period.

On the other hand, in Fig.~\ref{fig:bstH_cos_L_eps}(c) we show an eigenstate to an eigenvalue close to zero, which only deposits minimal intensity inside the target. In the one-dimensional scenario discussed here, this is achieved by destructive interference of the waves incoming from the left and right. In a higher-dimensional setup this eigenstate can be expected to constitute a light pulse that would completely avoid a given target by propagating around it. \corr{It would be interesting to explore how also other time-invariant schemes for optimal focusing \cite{Abboud,MaCheng,ZhouEdward,Sol} can be generalized to the Floquet regime.}

\begin{figure*} 
    \includegraphics[width=\textwidth]{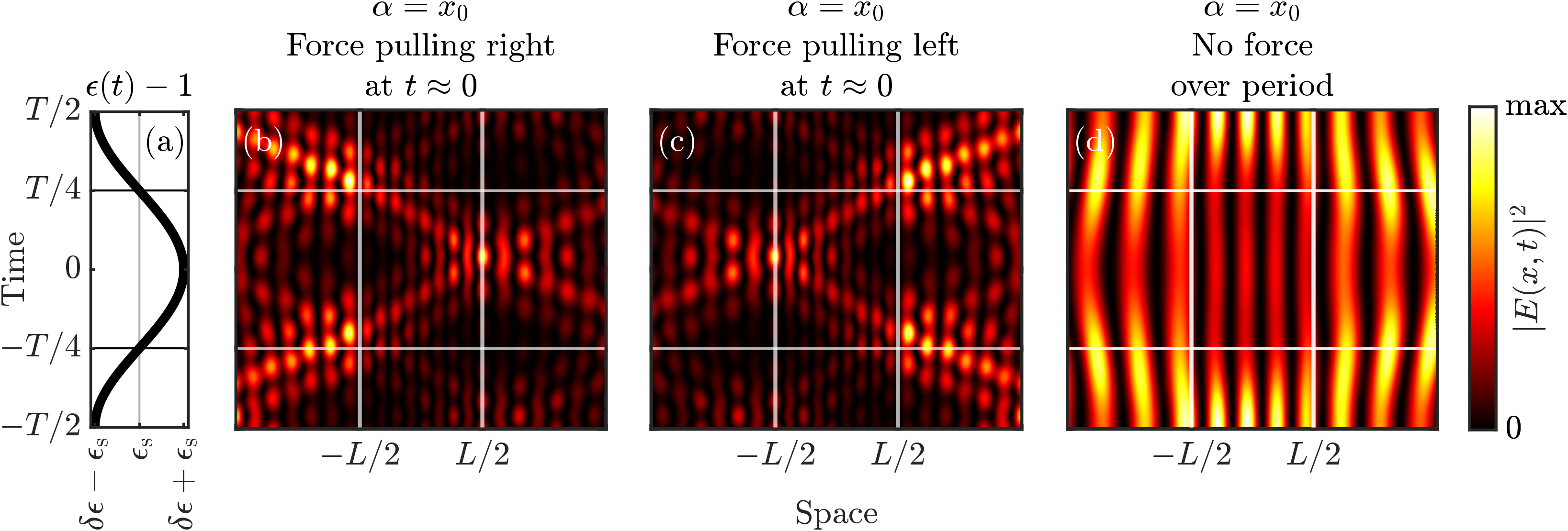}
    \caption{\label{fig:bstH_cos_x0} Optimally exerting force onto a target scatterer. (a) One period of the harmonic modulation protocol ${\epsilon(t)=1+\epss+\delta\epsilon \cos(\Omega t)}$ of the target scatterer centered at the origin. (b) Intensity of the eigenstate corresponding to the largest eigenvalue of $\QF{x_0}$. This optimal spatiotemporal state exhibits a focus on the right border of the target at times when the permittivity $\epsilon(t)$ is high. This light field thus pulls the target to the right. (c) Intensity of the eigenstate corresponding to the most negative eigenvalue. This wave field applies a force to the target, pulling it to the left during times when the permittivity is high. (d) Intensity of an eigenstate associated with an eigenvalue close to zero exerting nearly no force onto the target.}
\end{figure*}

Another useful task is to focus light not inside but at the surface of a given object. In the one-dimensional setup we consider here, this corresponds to focusing intensity at the borders $x=\pm L/2$ of the scatterer. As we demonstrate, this task can be optimally achieved with a FWS matrix involving a derivative with respect to the spatial extension $L$ of the target scatterer. The associated eigenvalues of $\QF{L}$ are proportional to the electric field at the edges of the scatterer as can be seen by applying Eq.~\eqref{eq:FWS_relation_general} again,
\begin{equation}
    \begin{split}
        \evF{L} =& \frac{1}{4} \int_{-T/2}^{T/2} \frac{\dd{t}}{T} \abs{\Es(-L/2,t)}^2 \left[\epss+ \delta\epsilon \cos(\Omega t) \right] \\
            +& \frac{1}{4} \int_{-T/2}^{T/2} \frac{\dd{t}}{T} \abs{\Es(L/2,t)}^2 \left[\epss+ \delta\epsilon \cos(\Omega t) \right].
    \end{split}
\label{eq:FWS_relation_cos_L}
\end{equation}
In Fig.~\ref{fig:bstH_cos_L_eps}(d) the light field corresponding to the largest eigenvalue $\evF{L}$ is shown. Indeed, it exhibits a peak at the borders of the scatterer. As stipulated by the time-dependent weighting function in Eq.~\eqref{eq:FWS_relation_cos_L}, the peak intensity occurs at times when the weighting function is most positive during the middle of the period.

Finally, we demonstrate the abilities of the FWS matrix involving a derivative with respect to the position of the center of the target $x_0$. The spatiotemporal behavior of the eigenstates of the matrix $\QF{x_0}$ can be understood when making use of Eq.~\eqref{eq:FWS_relation_general}, which reduces in the case of a target centered at the origin to
\begin{equation}
    \begin{split}
        \evF{x_0} =& -\frac{1}{2} \int_{-T/2}^{T/2} \frac{\dd{t}}{T} \abs{\Es(-L/2,t)}^2 \left[\epss+ \delta\epsilon \cos(\Omega t) \right] \\
            &+ \frac{1}{2} \int_{-\frac{T}{2}}^{\frac{T}{2}} \frac{\dd{t}}{T} \abs{\Es(L/2,t)}^2 \left[\epss+ \delta\epsilon \cos(\Omega t) \right].
    \end{split}
\label{eq:FWS_relation_cos_x0}
\end{equation}
The eigenvalues $\evF{x_0}$ of $\QF{x_0}$ are proportional to the difference between the intensities at the two borders of the target, weighted with the time-periodic modulation. If the electric field appearing in the above relation is real and $T$ periodic, the right hand-side of Eq.~\eqref{eq:FWS_relation_cos_x0} is proportional to the force exerted onto the target by the light field, including both the scattering and the gradient force \cite{Horodynski,HupflPRA}. A $T$-periodic electric field can be realized by choosing the input field to have vanishing quasienergy $\omega=0$. For an extension to complex electric fields with $\omega \neq 0$, we refer the reader to Appendix~\ref{sec:app_RealFields}.

In Fig.~\ref{fig:bstH_cos_x0} (b) the eigenstate of $\QF{x_0}$ corresponding to the most positive eigenvalue is depicted for a state with quasienergy $\omega=0$. A pronounced intensity peak is built up at the right border of the object at times when the permittivity is high while nearly no intensity is exhibited at the left edge of the object resulting in a force pulling the target to the right. Correspondingly, the eigenstate to the most negative eigenvalue exhibits an intensity peak on the left border [see Fig.~\ref{fig:bstH_cos_x0}(c)]. This wave field pulls the target to the left during times when the permittivity is high. Lastly, in Fig.~\ref{fig:bstH_cos_x0}(d) an eigenstate corresponding to an eigenvalue close to zero is shown. This wave field exhibits nearly no intensity at the borders of the target during times when the permittivity is high. We note that the colorscale in Fig.~\ref{fig:bstH_cos_x0} is normalized to the individual wave field in each panel. Therefore, we highlight that the intensity at the borders of the target at the beginning and end of the period is approximately three times smaller than the peak intensity at the respective borders in Fig.~\ref{fig:bstH_cos_x0}(b) and \ref{fig:bstH_cos_x0}(c). In this way the wave field corresponding to an eigenvalue of $\QF{x_0}$ exerts nearly no force onto the target at all times. A way to reduce the applied force even further is to use a superposition of eigenstates corresponding to two eigenvalues with opposite sign but similar magnitude (not explicitly shown here). A similar strategy has been employed for time-harmonic waves in \cite{Horodynski}.

With the above we demonstrate that by exploiting the scattering properties of a periodically modulated slab, we are able to access the spatial and temporal degrees of freedom for a wave control technique based on the Floquet Wigner-Smith matrix. The extremal eigenstates of this matrix correspond to pulsed light fields that are optimally shaped in their spatial and temporal degrees of freedom and accomplish specific micromanipulation tasks at the optimal level of efficiency.

\begin{figure*} 
    \includegraphics[width=\textwidth]{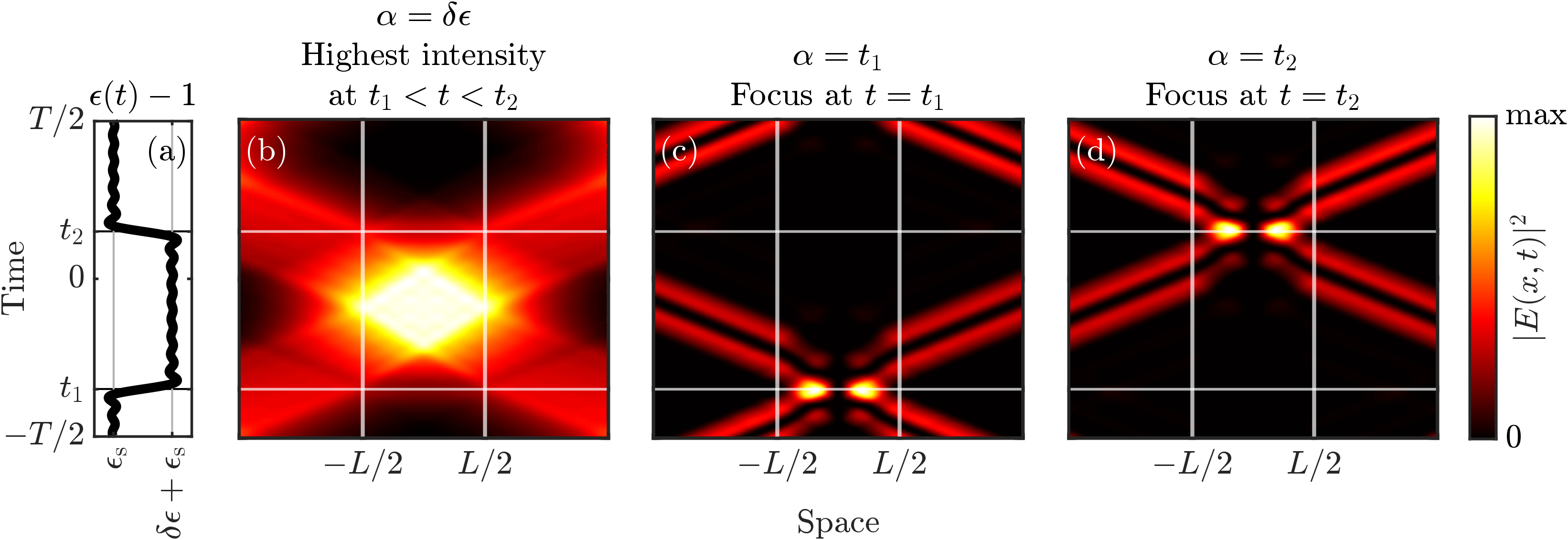}
    \caption{\label{fig:bstH_jump_all} Optimal spatiotemporal focusing of light on a target scatterer. (a) One period of the steplike modulation protocol $\epsilon(t)=1+\epss+ \delta \epsilon \Theta_T(t,t_1,t_2)$ of the target object centered at the origin ($t_1/T=-0.35$ and $t_2/T=\cor{0.15}$). (b) Intensity of the eigenstate corresponding to the largest eigenvalue of $\QF{\delta\epsilon}$. This light field exhibits a focus during times when the permittivity $\epsilon(t)$ is high. (c) Intensity of the eigenstate corresponding to the most negative eigenvalue of $\QF{t_1}$. This wave field builds up a strong focus at $t=t_1$. (d) Intensity of the eigenstate corresponding to the most positive eigenvalue of $\QF{t_2}$ building up a strong focus at $t=t_2$.}
\end{figure*}

In this section we assumed a time-harmonic modulation of the permittivity associated with the frequency $\Omega$. This results in eigenstates of the FWS matrix that are proportional to the intensity times a weighting function having the same time-harmonic variation as the permittivity itself (for $\alpha=\delta \epsilon, L$). In the next section we replace the time-harmonic variation of the permittivity with a periodically repeating steplike modulation. This allows us to adjust the temporal behavior of the optimal wave fields even further, giving rise, e.g., to states that focus inside the target during a customized time span of the period or only at a very short moment in time.

\subsection{Step Modulation \label{sec:step_mod}}
We now study a setting, in which the permittivity of the target switches periodically in a steplike manner between two values. Abrupt steplike changes of the permittivity have been realized experimentally in the microwave regime \cite{Moussa} and in the optical domain \cite{Lustig23}. We model the modulation by a dielectric function of the form
\begin{equation}
    \epsilon(x,t) = 1+\left[\epss + \delta \epsilon \, \Theta_T(t,t_1,t_2) \right] \Theta(L/2-\abs{x-x_0}). \label{eq:epsilon_jump}
\end{equation}
The time-periodic function we use here, ${\Theta_T(t,t_1,t_2)=\sum_n \Theta(t-t_1-nT)\Theta(t_2+nT-t)}$, implements the desired switching process: At the beginning of the period, the target object has permittivity $1+\epss$ until at time $t_1$ a jump to higher permittivity $1+\epss+\delta \epsilon$ occurs. At time $t_2$ the permittivity switches back to its original lower value. This temporal change in permittivity is repeated periodically [see Fig.~\ref{fig:bstH_jump_all}(a) for an illustration]. We can solve the scattering problem by means of Floquet theory and derive the Floquet scattering matrix following the discussion presented in Sec.~\ref{sec:FSM}. Again, the scatterer will be centered at the origin here, $x_0=0$, \cor{while for the other values of the parameters we refer the reader to Appendix~\ref{sec:app_truncation}}. In the following, we demonstrate that a steplike temporal modulation provides access to an even more detailed spatiotemporal control of light fields with the FWS approach.

First, we show how to maximize the intensity inside the target from time $t_1$ to time $t_2$ with the FWS matrix $\QF{\delta \epsilon}$, where the derivative is taken with respect to the modulation strength $\alpha=\delta \epsilon$. By exploiting Eq.~\eqref{eq:FWS_relation_general} we arrive at
\begin{equation}
    \evF{\delta \epsilon}=\frac{1}{2} \int_{t_1}^{t_2} \frac{\dd{t}}{T} \int_{-L/2}^{L/2} \dd{x} \abs{\Es(x,t)}^2. \label{eq:FWS_relation_jump_V1}
\end{equation}
Note, in particular, that the time integration does not span the full period $T$, but only times of high permittivity $t_1 < t < t_2$. This means that the eigenstate to the largest eigenvalue of $\QF{\delta\epsilon}$ is a light pulse that maximizes the intensity inside the scatterer during this time interval, as depicted in Fig.~\ref{fig:bstH_jump_all}(b).

Furthermore, the FWS matrix is capable of finding states that focus on the target at a single moment in time. More specifically, either of the two jump times $t_1$ or $t_2$ can serve as those specific moments in time when a pulse delivers its maximum focus onto the target. To emphasize this point in greater detail, we choose $\alpha=t_{1}$ ($\alpha=t_{2}$) and, following Eq.~\eqref{eq:FWS_relation_general}, the eigenvalues obey the relation
\begin{equation}
    \evF{t_{j}}= (-1)^j \frac{\delta \epsilon}{2T} \int_{-L/2}^{L/2} \dd{x} \abs{\Es(x,t_{j})}^2,
\end{equation}
where $j=\{1,2\}$. Note that the time integral completely disappears here and the eigenvalues are proportional to the intensity inside the scatterer at one instant of time $t=t_{1}$ ($t=t_{2}$). Thus, to extremize the right-hand side of the above equation we expect the wave state corresponding to the extremal eigenvalue (most negative for $t_1$ and most positive for $t_2$) to build up a pronounced intensity maximum at this moment in time. In Figs.~\ref{fig:bstH_jump_all}(c) and \ref{fig:bstH_jump_all}(d) the intensity distribution of these wave fields are plotted, having indeed the desired property of focusing at $t=t_{1}$ or $t_{2}$. Additional FWS matrices with respect to $\alpha=L$, $\alpha=\epss$, and $\alpha=x_0$ are discussed in Appendix~\ref{sec:app_FWS_jump}.

\cor{Finally, we highlight that for static media several matrix-based concepts for spatiotemporal shaping of light fields have already been developed and applied successfully \cite{McCabe,Mounaix,mounaix2020time,Bouchet23,Ferise} (see \cite{horstmeyer2015guidestar} and \cite{Cao22} for corresponding reviews). What distinguishes the current FWS approach presented in this paper from these earlier studies is the fact that it exploits the periodic time variation of the target itself to shape the spatial and temporal degrees of freedom of the light field. In this way it allows for not only optimal spatiotemporal focusing on a time-varying object but also for optimal micromanipulation of this object (even if the object is located inside a complex environment). }

\section{Conclusion}
To conclude, we have demonstrated that the Floquet scattering matrix fulfills a pseudounitary relation as its central algebraic property when expressed in a photon-flux normalized basis. This feature is a consequence of the conservation of the wave action (number of pseudophotons), for which we formulated here a source-free continuity equation. The pseudounitarity of the Floquet scattering matrix allowed us to generalize static wave control techniques to periodically time-varying scattering systems. In particular, we presented the Floquet Wigner-Smith matrix, which provides a toolbox for the micromanipulation of small particles and paves the way for the generation of tailor-made light fields for optimal interactions with designated targets in both the spatial and the temporal domain. Ultimately, we expect the FWS matrix to serve as a useful tool for accelerating, cooling, and manipulating particles with light pulses operating at the optimal level of efficiency. 

\begin{acknowledgments}
We thank Sh.~Amiranashvili, P.~Brouwer, M.~Kühmayer, U.~Leonhardt, F.~Schneider, S.~Yoshida, and M.~Zens for useful discussions. We acknowledge support from the Austrian Science Fund (FWF) under project P32300 (WAVELAND).
\end{acknowledgments}

\appendix
\section{Derivation of the Wave Field \label{sec:app_wavefield}}
We summarize the procedure of finding a scattering solution to the scalar wave equation with a time-periodic dielectric function
\begin{equation}
    \partial_x^2 \Es(x,t)-\partial_t^2 \left[ \epsilon(t) \Es(x,t) \right]=0, \label{eq:scalar_waveeq_E_app}
\end{equation}
which is Eq.~\eqref{eq:scalar_waveeq_E} of the main text. We assume a time-periodic permittivity function 
\begin{equation}
    \epsilon(x,t)=1+[\epss+\delta \epsilon \, g(t)]\Theta(L/2-\abs{x-x_0}),
\end{equation}
where we have $g(t)=g(t+T)$ and therefore $\epsilon(x,t)=\epsilon(x,t+T)$ with $T=2\pi/\Omega$. For brevity, we set $x_0=0$ here and follow closely \cite{Zurita} for the derivation of the wave field. First, we separate the scattering landscape into three regions: the left and right ($\mathrm{l}$ and $\mathrm{r}$) leads, which are the asymptotic regions $\abs{x}>L/2$ and the scattering region $\mathrm{m}$ such that
\begin{equation}
    \Es(x,t) = \begin{cases}
    \Es_\mathrm{l}(x,t), & x \leq -L/2 \\
    \Es_\mathrm{m}(x,t), & -L/2<x<L/2 \\
    \Es_\mathrm{r}(x,t), & x \geq L/2. \end{cases}
\end{equation}
In the asymptotic regions \corr{$\abs{x}>L/2$}, we assume \corr{free space} and therefore Eq.~\eqref{eq:scalar_waveeq_E_app} reduces to
\begin{equation}
    \partial_x^2  \Es_\lead (x,t)-\partial_t^2  \Es_\lead (x,t)=0,
\end{equation}
where we introduced the lead index $\lead \in \{ \mathrm{l}, \mathrm{r} \}$. A superposition of right- and left-propagating plane waves with frequencies $\omega_n=\omega+n\Omega$ with $n\in \mathbb{Z}$, and wavevector $k_\indexL = -k_\indexR = \omega_n$ is a solution to the wave equation. Therefore, we get
\begin{equation}
E_\lead (x,t)= \sum_\indexLR \mathcal{N}_\indexLR \left( \cin_\indexLR e^{i k_\indexLR x} + \cout_\indexL e^{-i k_\indexLR x} \right) e^{-i\omega_n t} \label{eq:Elr_full}
\end{equation}
with constant coefficients $\cin_\indexLR$ and $\cout_\indexLR$, as well as the normalization constant $\mathcal{N}_\indexLR$. The assignment of input and output coefficient holds true even for negative channels. This is because of the choice of the dispersion relation $k_\indexL=-k_\indexR=\omega_n$. Thus, the propagation direction of a negative channel is the same as for a corresponding positive channel since the sign of the phase velocity $v=\omega_{n}/k_\indexLR$ is independent of the sign of $n$.

In the region $-L/2<x<L/2$, i.e., inside the modulated scatterer, the dielectric function is periodically time dependent. A solution can be found by applying Floquet's theorem, which states that the wave field is of the form
\begin{equation}
    \Es_\mathrm{m}(x,t)=\Phi_\mathrm{m}(x)\xi(t)e^{-i\omega t},  
\end{equation}
with a periodic function $\xi(t)=\xi(t+T)$ having the same periodicity as the permittivity $\epsilon(t)=\epsilon(t+T)$. This allows for a Fourier decomposition as
\begin{align}
    \epsilon(t) &= \sum_n \epsilon_n e^{-i n\Omega t}, \label{eq:epsilon_Fourier}\\
    \xi(t) &= \sum_n \xi_n e^{-i n\Omega t}. \label{eq:xi_Fourier}
\end{align}
For the spatial part of the wave field we choose a plane-wave ansatz $\Phi_\mathrm{m}(x)=\alpha e^{iqx}+\beta e^{-iqx}$ with a wave vector $q$ and constant coefficients $\alpha$ and $\beta$. After a separation of variables, the wave equation inside the scattering region [Eq.~\eqref{eq:scalar_waveeq_E_app}] reduces to an eigenvalue problem of the form
\begin{equation}
    \sum_n \Gamma_{m,n} \xi_n = q^2 \xi_m,
\end{equation}
where we defined $\Gamma_{m,n}=(\omega+\Omega m)^2\epsilon_{m-n}$. We introduce the band index $p$ and allow for right- and left-propagating channels with wave vector $q_{p}$. Finally, we arrive at
\begin{equation}
    \Es_\mathrm{m}(x,t) = \sum_{n,p} \mathcal{N}_{n,p} \left( \alpha_{p} e^{iq_{p}x}+\beta_{p} e^{-iq_{p}x} \right) \xi_{n,p} e^{-i\omega_n t}. \label{eq:E2_full}
\end{equation}

\corr{We solve the scattering problem by expressing the outgoing coefficients $\cout_\indexLR$ in terms of the incoming coefficient $\cin_\indexLR$. This is achieved by demanding continuity of the electric  fields and their first spatial derivative (which corresponds in our case to the continuity of the associated magnetic fields) everywhere and especially at the boundaries of the target at $x=\pm L/2$ at all times. This gives rise to four linear coupled equations
\begin{widetext}
\begin{subequations}
\begin{align}
    \mathcal{N}_\indexL \left( \cin_\indexL e^{-i k_n L/2} + \cout_\indexL e^{i k_n L/2} \right) &= \sum_p  \mathcal{N}_{n,p} \left( \alpha_{p} e^{-iq_{p}L/2} + \beta_{p} e^{iq_{p}L/2} \right)\xi_{n,p}, \\
    \mathcal{N}_\indexL k_n \left( \cin_\indexL e^{-i k_n L/2} - \cout_\indexL e^{i k_n L/2} \right) &= \sum_p \mathcal{N}_{n,p} q_p \left( \alpha_{p} e^{-iq_{p}L/2} - \beta_{p} e^{iq_{p}L/2} \right)\xi_{n,p}, \\
    \sum_p \mathcal{N}_{n,p} \left( \alpha_{p} e^{iq_{p}L/2} + \beta_{p} e^{-iq_{p}L/2} \right)\xi_{n,p} &= \mathcal{N}_\indexR \left( \cin_\indexR e^{i k_n L/2} + \cout_\indexR e^{-i k_n L/2} \right), \\
    \sum_p \mathcal{N}_{n,p} q_p \left( \alpha_{p} e^{iq_{p}L/2} - \beta_{p} e^{-iq_{p}L/2} \right)\xi_{n,p} &= \mathcal{N}_\indexR k_n \left( \cin_\indexR e^{i k_n L/2} - \cout_\indexR e^{-i k_n L/2} \right).
\end{align}
\end{subequations}
\end{widetext}
Performing straightforward algebraic manipulations leads to the solution of the scattering problem (for details we again refer the reader to \cite{Zurita}). We cross-checked this analytical treatment of the problem by performing independent numerical simulations based on the finite difference time domain method (finding perfect agreement within the bounds of numerical precision).}

We use the same normalization for the asymptotic fields and the field inside the scatterer throughout the paper. As shown in the main text, the appropriate normalization for Floquet scattering problems is the photon-flux normalization, which is realized by the choice $\mathcal{N}_{n,p}=\mathcal{N}_{\sigma,n}=\sqrt{\abs{\omega_n}}$.

\section{Truncation of the Floquet Scattering Matrix} \label{sec:app_truncation}
In general, the Floquet scattering matrix is of infinite dimension as there are infinitely many Floquet channels~$n$. Clearly, for computational reasons one always has to truncate to a finite dimension. For the modulated permittivity $\epsilon(t)$ in Eq.~\eqref{eq:epsilon_Fourier} we use $n\in [-18,18]$ to model the steplike modulation $g(t)=\Theta_T(t,t_1,t_2)$ sufficiently accurately. Still, a finite number of Fourier channels implies that the switching takes place in a short but finite time. Apparently, for the harmonic modulation, $g(t)=\cos(\Omega t)$, only the $n=\pm 1$ terms are nonzero.

We use $n\in [-n_\mathrm{max}^{\mathrm{in}}, n_\mathrm{max}^{\mathrm{in}}]$ channels for the incoming part of the Fourier expansion of the electric field in both leads corresponding to the coefficients $\cin_\indexLR$ [see, e.g., in Eqs.~\eqref{eq:Elr_full} and \eqref{eq:E2_full}]. However, in order to account for the scattering into $n>n_\mathrm{max}^{\mathrm{in}}$ outgoing channels, we introduce a different cutoff for the outgoing channels as $n_\mathrm{max}^{\mathrm{out}} \geq n_\mathrm{max}^{\mathrm{in}}$. This results in $2n_\mathrm{max}^{\mathrm{out}}+1$ outgoing coefficients in each lead associated with the coefficients $\cout_\indexLR$. Therefore, the Floquet scattering matrix $\SF$ has dimension $2(2n_\mathrm{max}^{\mathrm{out}}+1) \times 2(2n_\mathrm{max}^{\mathrm{in}}+1)$. In the special case of $\omega=0$ the Floquet scattering matrix has dimensions $4n_\mathrm{max}^{\mathrm{out}} \times 4n_\mathrm{max}^{\mathrm{in}}$ as the mode $n=0$ is not a propagating mode and therefore is not considered within $\SF$.

Specifically, we choose $n_\mathrm{max}^{\mathrm{out}}=31$ for all data in this paper. We truncate $n_\mathrm{max}^{\mathrm{in}}$ such that the reflection or transmission probability (whichever is larger) of the highest input mode $n=n_\mathrm{max}^{\mathrm{in}}$ to the highest outgoing mode $n_\mathrm{max}^{\mathrm{out}}$ is less than $10^{-5}$. Furthermore, we use $\epsilon_s=1$, $\delta\epsilon=0.3$, and $\omega/\Omega=0.2$ (if not stated otherwise).

\section{Derivation of the Continuity Equation and FWS Relation} \label{sec:app_continuity}
Here we provide further details on the continuity equation~\eqref{eq:continuity}, the pseudounitary relation of the Floquet scattering matrix~\eqref{eq:pseudounitarity}, and the FWS relation~\eqref{eq:FWS_relation_general}. To begin with, we choose to work in the Coulomb gauge and therefore the vector potential $A$ can be introduced as $E=-\ptA$ \cite{Leonhardt}, where the overdot represents a derivative with respect to time. This implies a Fourier expansion in the leads of the form
\begin{equation}
    A_\lead (x,t)= \sum_n \frac{\mathcal{N}_\indexLR}{i \omega_n} \left( \cin_\indexLR e^{ik_\indexLR x} + \cout_\indexLR e^{-ik_\indexLR x} \right) e^{-i\omega_n t} \label{eq:A_Fourier}
\end{equation}
The vector potential satisfies the wave equation
\begin{equation}
    \partial_x^2 A - \partial_t ( \epsilon \ptA )=0. \label{eq:waveequation_A}
\end{equation}
Furthermore, it is useful to define a generalized density consisting of two fields $A_1$ and $A_2$ as
\begin{equation}
    q(A_1,A_2)= A_1^* \epsilon_2 \ptA_2 - A_2 \epsilon_1 \ptA_1^*. \label{eq:density_A12}
\end{equation}
Here, $A_1$ and $A_2$ are solutions of Eq.~\eqref{eq:waveequation_A} corresponding to time-periodic permittivities $\epsilon_1(t)$ and $\epsilon_2(t)$. As we show in the following, this generalized density gives rise to a generalized continuity equation. Specifically, taking the derivative with respect to time and using Eq.~\eqref{eq:waveequation_A} results in
\begin{equation}
    \begin{split}
        \dot{q}(A_1,A_2) =& A_1^* \pt (\epsilon_2 \ptA_2 ) \pt ( \epsilon_1 \ptA_1^* ) - A_2 \pt (\epsilon_1 \ptA_1^* ) \pt ( \epsilon_1 \ptA_1^* ) \\
         &+ \ptA_1^* \epsilon_2 \ptA_2 - \ptA_2 \epsilon_1 \ptA_1^* \\
         =& \px ( A_1^* \px A_2 - A_2 \px A_1^* ) \\
         &+ \ptA_1^* \epsilon_2 \ptA_2 - \ptA_2 \epsilon_1 \ptA_1^*. \label{eq:dtq}
    \end{split}
\end{equation}
We identify the terms on the right hand side as a generalized flux-density $j(A_1,A_2)$ and a source term $\eta(A_1,A_2)$ as
\begin{align}
    j(A_1,A_2) &= A_2 \px A_1^* - A_1^* \px A_2, \\
    \eta(A_1,A_2) &= \ptA_1^* \epsilon_2 \ptA_2 - \ptA_2 \epsilon_1 \ptA_1^*.
\end{align}
Rewriting Eq.~\eqref{eq:dtq} in the form of a generalized continuity equation yields
\begin{equation}
    \pt q(A_1,A_2) + \px j(A_1,A_2) = \eta(A_1,A_2). \label{eq:continuity_A12}
\end{equation}

\subsection{Continuity Equation and Pseudounitarity}
To arrive at a proper continuity equation, which is Eq.~\eqref{eq:pseudounitarity} of the main text, we choose $\epsilon_1=\epsilon_2=\epsilon$ and $A_1=A_2=A$. Thus, Eq.~\eqref{eq:continuity_A12} reduces to
\begin{equation}
    \pt q(A,A) + \px j(A,A) = 0, \label{eq:continuity_AA}
\end{equation}
since the source term vanishes in this case, i.e., $\eta(A,A)=0$. The density and flux reduce to the form used in the main text
\begin{align}
    q(A,A) &= A^* \epsilon \ptA - A \epsilon \ptA^*, \\
    j(A,A) &= A \px A^* - A^* \px A. \label{eq:flux_AA}
\end{align}
From this continuity equation we can now derive the pseudounitarity of the Floquet scattering matrix. We start by integrating Eq.~\eqref{eq:continuity_AA} over a temporal period and the spatial extension of the scatterer ($x_0=0$). We observe that
\begin{equation}
\begin{split}
    &\int_{-T/2}^{T/2} \frac{\dd{t}}{T} \int_{-L/2}^{L/2} \dd{x} \pt q(A,A) \\
    &= \frac{1}{T} \int_{-L/2}^{L/2} \dd{x} q(A,A)\Big|_{-T/2}^{T/2} = 0,
\end{split}
\end{equation}
due to the periodicity of $q(A,A)$. Therefore, we are left with
\begin{equation}
\begin{split}
    &\int_{-T/2}^{T/2} \frac{\dd{t}}{T} \int_{-L/2}^{L/2} \dd{x} \px j(A,A) \\
    &=  \int_{-T/2}^{T/2} \frac{\dd{t}}{T} j(A,A)\Big|_{-L/2}^{L/2} = 0.    
\end{split}
\end{equation}
By making use of the Fourier expansion of the field, Eq.~\eqref{eq:A_Fourier}, we see that
\begin{equation}
    \begin{split}
        0 =& \int_{-T/2}^{T/2} \frac{\dd{t}}{T} j(A,A)\Big|_{-L/2}^{L/2} \\
         =& 2\sum_n \frac{\abs{\mathcal{N}_\indexR}^2}{\omega_n^2} \left(-i k_\indexR \right) \left( \abs{\cin_\indexR}^2 - \abs{\cout_\indexR}^2 \right) \\
         -& 2\sum_n \frac{\abs{\mathcal{N}_\indexL}^2}{\omega_n^2} \left(-i k_\indexL \right) \left( \abs{\cin_\indexL}^2 - \abs{\cout_\indexL}^2 \right) \\
         =& 2i \sum_\indexLR \frac{\abs{\mathcal{N}_\indexLR}^2}{\omega_n }\left( \abs{\cin_\indexLR}^2 - \abs{\cout_\indexLR}^2 \right). 
    \end{split}
\end{equation}
In the last line we used that $k_\indexL = - k_\indexR = \omega_n$. This is Eq.~\eqref{eq:continuity_integral} of the main text.

\subsection{FWS Relation}
To derive the connection between the eigenvalues of the FWS matrix and the near field of the target scatterer [Eq.~\eqref{eq:FWS_relation_general}], we again make use of Eq.~\eqref{eq:continuity_A12}. Now we set $\epsilon_1=\epsilon$ and $\epsilon_2=\epsilon + \Delta \alpha \pa \epsilon + O(\Delta \alpha^2)$ and therefore $A_1=A$ and $A_2=A+\Delta \alpha \pa A + O(\Delta \alpha^2)$. Here we introduce the parameter $\alpha$, on which the scattering landscape may depend parametrically. To first order in $\Delta \alpha$, Eq.~\eqref{eq:continuity_A12} reads
\begin{equation}
    \pt q^{(1)} + \px j^{(1)} = \eta^{(1)}, \label{eq:continuity_first}
\end{equation}
where we have
\begin{align}
    q^{(1)} &= A^* \left( \pa \epsilon \right) \ptA +A^* \epsilon \pa \ptA - \left( \pa A \right) \epsilon \ptA, \\
    j^{(1)} &= \left( \pa A \right) \px A^* - A^* \px \pa A, \\
    \eta^{(1)} &= \abs{E}^2 \pa \epsilon .
\end{align}
Now we integrate Eq.~\eqref{eq:continuity_first} over a temporal period and the spatial extension of the scatterer ($x_0=0$). Also, here we use the periodicity of $q^{(1)}$ to arrive at
\begin{equation}
    \int_{-T/2}^{T/2} \frac{\dd{t}}{T} j^{(1)} \Big|_{-L/2}^{L/2} = \int_{-T/2}^{T/2} \frac{\dd{t}}{T} \int_{-L/2}^{L/2} \dd{x} \eta^{(1)}
\end{equation}
After inserting the Fourier expansion of the vector field, Eq.~\eqref{eq:A_Fourier} with a photon-flux normalization, and some straightforward algebraic manipulations (note that only $\coutv$ depends on $\alpha$) we end up with
\begin{equation}
    \begin{split}
        -2i\sum_\indexLR& \sign(\omega_n) \left( \cout_\indexLR \right)^* \pa \cout_\indexLR \\
        &= \int_{-T/2}^{T/2} \frac{\dd{t}}{T} \int_{-L/2}^{L/2} \dd{x} \abs{E}^2 \pa \epsilon.
    \end{split}
\end{equation}
Finally, we can rewrite the above equation in a matrix form. We use $\SF \cinv=\coutv$ and introduce the matrix $V$ defined in the main text, which leads to
\begin{equation}
\begin{split}
    &\left( \cinv \right)^\dagger \left( -i \SF^\dagger V \pa \SF \right) \cinv \\
    &= \frac{1}{2} \int_{-T/2}^{T/2} \frac{\dd{t}}{T} \int_{-L/2}^{L/2} \dd{x} \abs{E}^2 \pa \epsilon.
\end{split}
\end{equation}
This is Eq.~\eqref{eq:FWS_relation_general} of the main text.

\section{Real Fields} \label{sec:app_RealFields}
\begin{figure}
    \includegraphics[width=\columnwidth]{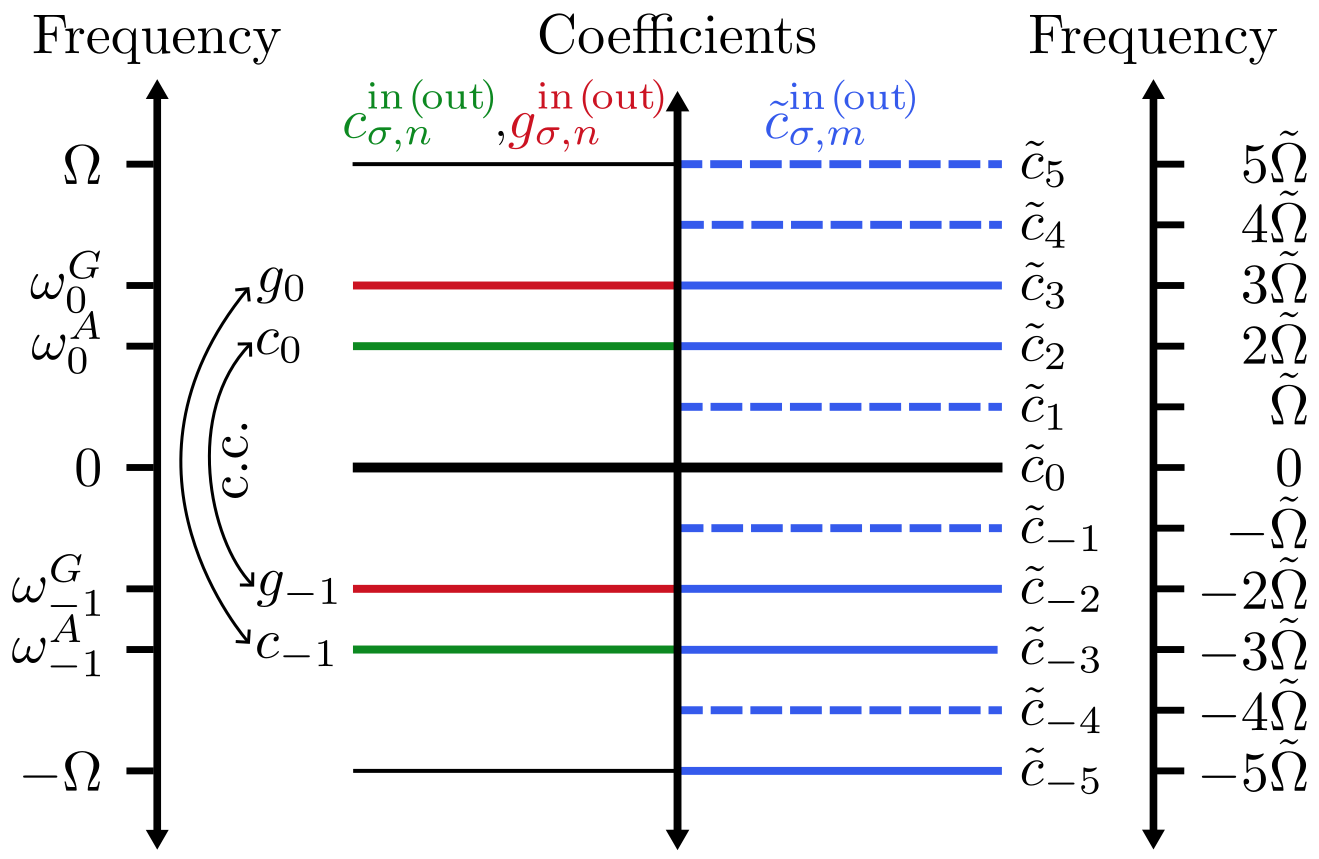}
    \caption{\label{fig:realfield} Schematic representation of the construction of the physical solution $A_\mathrm{p}=\Re(A)$ from the complex field $A$. The coefficients $\cinout_\indexLR$ (green) are associated with the field $A$ with quasifrequency $\omega^{A}$. The coefficients $\ginout_\indexLR$ (red) correspond to the complex-conjugate field $G=A^*$ with quasifrequency $\omega^{G}$. The physical field $A_\mathrm{p}$ is described as a part of a larger complex field with vanishing quasifrequency and period $\tilde{T}=2\pi/\tilde{\Omega}$. The expansion coefficients of $A_\mathrm{p}$ are $\ctildeinout_{\sigma,m}$ (blue), which we choose according to Eq.~\eqref{eq:coef_real_relation}. Only those coefficients that have a counterpart in $A$ or $G$ are nonzero (blue solid line), while the other ones are zero (blue dashed line). Here we choose $r=2$ and $s=5$.}
\end{figure}
In the main text we considered the general case of complex-valued electric fields $E(x,t)$ and vector potentials $A(x,t)$. The physical fields are however given by the corresponding real-valued fields. This section serves the purpose of a double-check, ensuring the Floquet scattering theory presented in the main text covers physical (real) fields as well.

We start by considering the special case of $\omega=0$. Here a real electric field is simply implemented by demanding that the coefficients obey
\begin{equation}
    \left( \cinout_\indexLR \right)^* = \cinout_{\sigma,-n}.
\end{equation}
We again note here that for $\omega=0$ the channel $n=0$ is not considered in the Floquet scattering matrix, since it does not represent a propagating electric-field channel.

We proceed with the case $\omega=\Omega/2$. Similarly, here a real field is realized by requiring
\begin{equation}
    (\cinout_\indexLR)^*= \cinout_{\sigma,-n-1}.
\end{equation}
In both cases, Eq.~\eqref{eq:FWS_relation_general} is applicable such that the results for complex electric fields can be one-to-one transferred to real fields.

For the general case of $0 \leq \omega < \Omega$ (but $\omega\neq 0$ or $\Omega/2$) we observe that given a complex field $A$, the physical field can be extracted as $A_\mathrm{p}=\Re(A)=(A+A^*)/2$. Therefore, the physical field $A_\mathrm{p}$ is an equal superposition of two complex fields $A$ and $G=A^*$. However, we note that both fields are associated with different quasifrequencies $\omega^{\mathrm{A}}=\omega$ and $\omega^{\mathrm{G}}=\Omega-\omega$, such that their expansion inside the leads read
\begin{align}
    A_\lead (x,t) &= \sum_n \frac{\mathcal{N}_\indexLR^{A}}{i \omega_n^{A}} \left( \cin_\indexLR e^{ik_\indexLR^{A} x} + \cout_\indexLR e^{-ik_\indexLR^{A} x} \right) e^{-i\omega_n^{A} t} \\
    G_\lead (x,t) &= \sum_n \frac{\mathcal{N}_\indexLR^{G}}{i \omega_n^{G}} \left( \gin_\indexLR e^{ik_\indexLR^{G} x} + \gout_\indexLR e^{-ik_\indexLR^{G} x} \right) e^{-i\omega_n^{G} t}
\end{align}
Since both field are complex conjugate to each other we have the connection
\begin{equation}
    \left( \ginout_\indexLR \right)^*=\cinout_{\sigma,-(n+1)}. \label{eq:connection_gc}
\end{equation}
The idea here is to embed the physical solution $A_\mathrm{p}$ into a field hosting a denser set of channel coefficients in frequency space compared to $A$ and $G$. Specifically, we construct this field in such a way that it contains the coefficients of both fields $A$ and $G$ as depicted in Fig.~\ref{fig:realfield}. We assume that the quasifrequency $\omega^{\mathrm{A}}$ and the oscillation frequency $\Omega$ are commensurable, such that there exists an $\tilde{\Omega}$ for which $\omega^{\mathrm{A}}=r\Tilde{\Omega}$ and $\Omega=s\Tilde{\Omega}$ hold with $r,s\in\mathbb{N}$. Since incommensurable frequencies can be approximated to arbitrary precision by commensurable ones, this assumption holds true for all practical purposes. The field $A_\mathrm{p}$ is now constructed such that it has a vanishing quasifrequency and a period $\Tilde{T}=2\pi/\Tilde{\Omega}$. This new field is then larger in the sense that the associated frequency $\tilde{\Omega}$ is smaller than $\Omega$ and thus hosts a denser set of channels for a given frequency range. The channel expansion of the field $A_\mathrm{p}$ reads
\begin{equation}
    A_\mathrm{p} = \sum_{\lead,m} \frac{\mathcal{N}_{\lead,m}}{i m \Tilde{\Omega}} \left( \ctildein_{\lead,m} e^{i \Tilde{k}_{\lead,m} x} + \ctildeout_{\lead,m} e^{-i \Tilde{k}_{\lead,m} x} \right)  e^{-i m \Tilde{\Omega} t}, 
\end{equation}
where accordingly $\Tilde{k}_{\mathrm{l},m}=-\Tilde{k}_{\mathrm{r},m}=m \Tilde{\Omega}$. For the channel coefficients of the physical field $A_\mathrm{p}$ we demand that (cf.\ Fig.~\ref{fig:realfield})
\begin{equation}
    \ctildeinout_m = \begin{cases}
    \frac{1}{2} \cinout_n \qif & m=r+ns, \\
    \frac{1}{2} \ginout_n \qif & m=s-r+ns, \\
    0 & \mathrm{else}. \label{eq:coef_real_relation}
    \end{cases}
\end{equation}
The time-periodic dielectric function only couples channels with frequency differences of $n\Omega$. Hence, we can treat the incoming and outgoing coefficients of the $A$ and $G$ fields independently. Using Eq.~\eqref{eq:continuity_first} with an extended integration range $[ -\Tilde{T}/2,\Tilde{T}/2]$ according to the periodicity of the field $A_\mathrm{p}$ and using a photon-flux normalization $\mathcal{N}_{\lead,m}=\sqrt{\abs*{m\Tilde{\Omega}}}$, we get
\begin{equation}
    \begin{split}
        &-2i\sum_{\lead,m} \sign(m) \left( \ctildeout_{\lead,m} \right)^{*} \partial_\alpha \ctildeout_{\lead,m} \\
        &= \int_{-\Tilde{T}/2}^{\Tilde{T}/2} \frac{\dd{t}}{\Tilde{T}} \int \dd{x} \abs{E_\mathrm{p}}^2 \partial_{\alpha} \epsilon,
    \end{split}
\end{equation}
where $E_\mathrm{p}=-\partial_t A_\mathrm{p}$
Furthermore, we make use of Eq.~\eqref{eq:connection_gc} and are left with
\begin{equation}
    \begin{split}
        -2i \sum_{\lead,m} \sign(m)& \left( \ctildeout_{\lead,m} \right)^{*} \partial_\alpha \ctildeout_{\lead,m} \\
        =&-\frac{i}{2} \sum_\indexLR \sign(\omega_n^{A}) \left( \cout_\indexLR \right)^{*} \partial_\alpha \cout_\indexLR \\
        &-\frac{i}{2} \sum_\indexLR \sign(\omega_n^{G}) \left( \gout_\indexLR \right)^{*} \partial_\alpha \gout_\indexLR \\
        =& -\frac{i}{2} \sum_\indexLR \sign(\omega_n^{A}) \left( \cout_\indexLR \right)^{*} \partial_\alpha \cout_\indexLR \\
        &+\frac{i}{2} \sum_\indexLR \sign(\omega_n^{A}) \cout_\indexLR \partial_\alpha \left( \cout_\indexLR \right)^{*} \\
        =& \left( \cinv \right)^\dagger \left(-i \SF^\dagger V \pa \SF \right) \cinv,
    \end{split}
\end{equation}
where in the last line we used the Hermiticity of the FWS matrix. Altogether, we arrive at
\begin{equation}
     \left(\cinv \right)^\dagger \left( -i \SF^\dagger V \partial_\alpha \SF \right) \cinv = \int_{-\Tilde{T}/2}^{\Tilde{T}/2} \frac{\dd{t}}{\Tilde{T}} \int \dd{x} E_\mathrm{p}^2 \partial_{\alpha} \epsilon,
\end{equation}
which states the connection between the FWS operator and the corresponding real electric field. This equation tells us that the eigenvalue of the FWS matrix is connected to the physical field in the same way as the complex field [cf.\ Eq.~\eqref{eq:FWS_relation_general}] with, apart from a factor $\frac{1}{2}$, the essential difference that the temporal integration range is extended from $T$ to $\tilde{T}=sT$.

\section{Further FWS Matrices for Step-Like Modulation} \label{sec:app_FWS_jump}
\begin{figure*}
    \includegraphics[width=\textwidth]{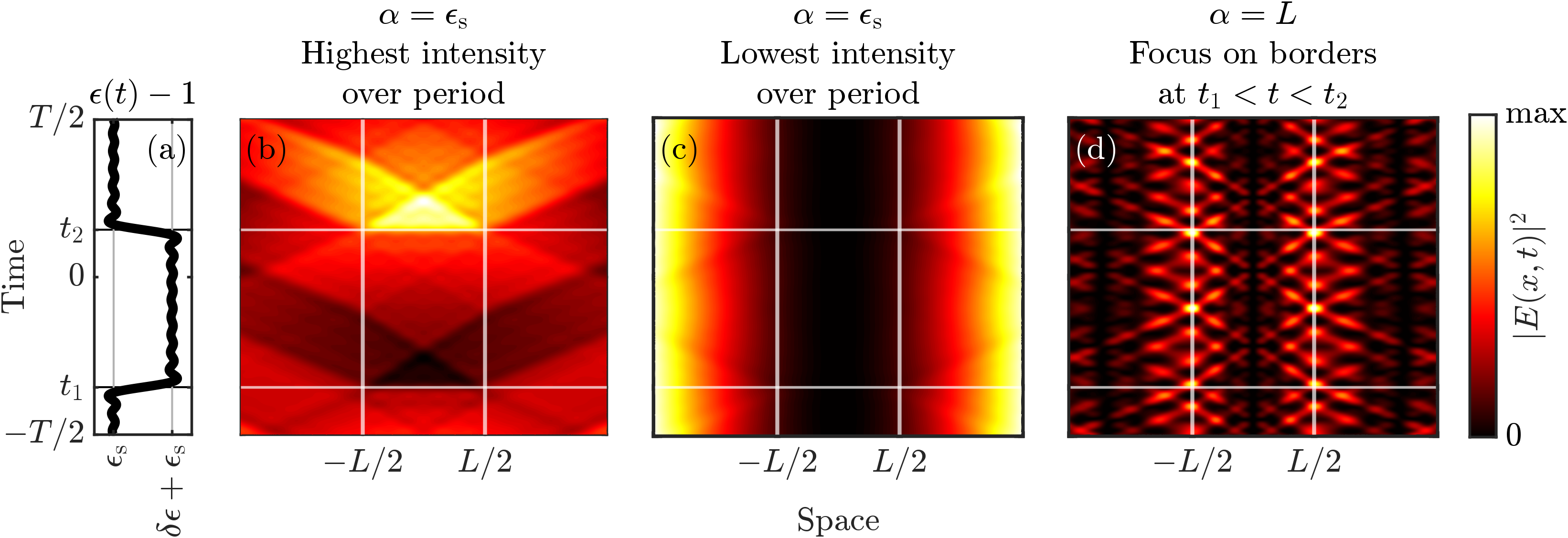}
    \caption{\label{fig:bstH_jump_L_epss} Focusing light inside and at the borders of a target scatterer. (a) Steplike modulation protocol $\epsilon(t)=1+\epss+ \delta \epsilon \Theta_T(t,t_1,t_2)$ of the target scatterer centered at the origin plotted for one period ($t_1/T=-0.35$ and $t_2/T=\cor{0.15}$). (b) Intensity of the eigenstate corresponding to the most positive eigenvalue of $\QF{\epss}$. This wave field maximizes the intensity built up inside the scatterer during the entire period $T$. (c) Intensity of an eigenstate to an eigenvalue close to zero of $\QF{\epss}$ having minimal intensity inside the target at all times. (d) Intensity of the eigenstate corresponding to the largest eigenvalue of $\QF{L}$. This light field exhibits several foci at the borders of the scatterer during times when the permittivity $\epsilon(t)$ is high.}
\end{figure*}
\begin{figure*}
    \includegraphics[width=\textwidth]{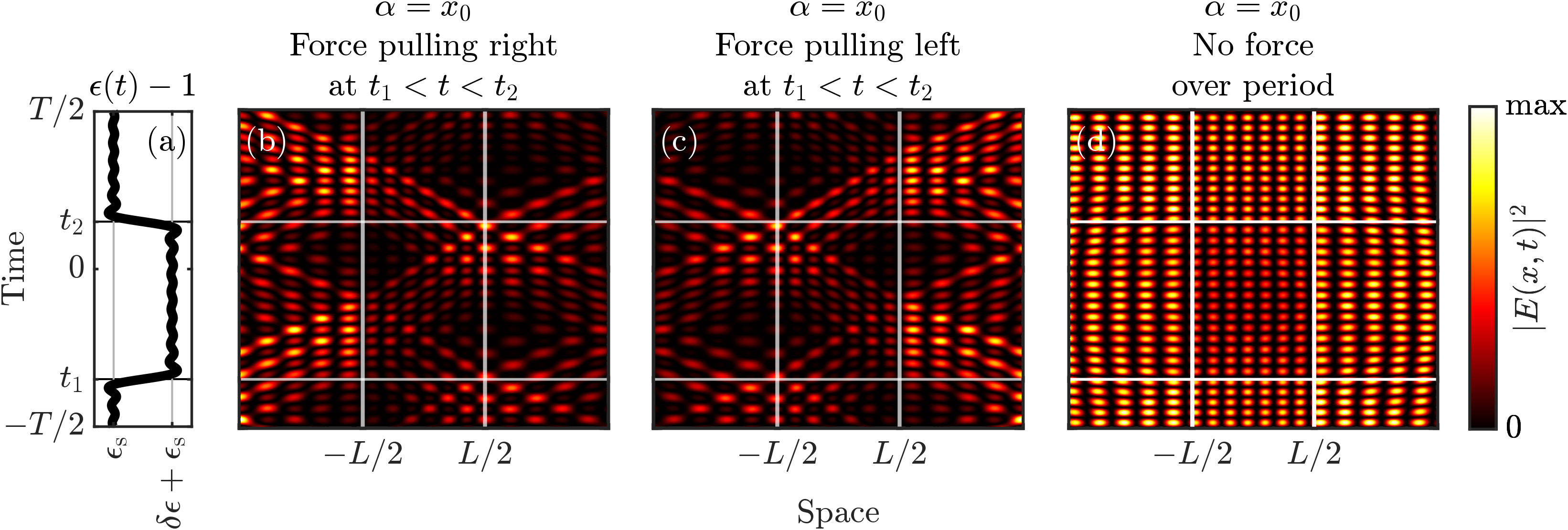}
    \caption{\label{fig:bstH_jump_x0} Optimally exerting force onto a target scatterer. (a) One period of the steplike modulation protocol ${\epsilon(t)=1+\epss+ \delta \epsilon \Theta_T(t,t_1,t_2)}$ of the target object centered at the origin ($t_1/T=-0.35$ and $t_2/T=\cor{0.15}$). (b) Intensity of the eigenstate corresponding to the largest eigenvalue of $\QF{x_0}$. This optimal spatiotemporal state exhibits several foci on the right border of the target at times when the permittivity $\epsilon(t)$ is high. This light field thus pulls the target to the right. (c) Intensity of the eigenstate corresponding to the most negative eigenvalue. This wave field applies a force to the target, pulling it to the left during times when the permittivity is high. (d) Intensity of an eigenstate associated with an eigenvalue close to zero exerting nearly no force onto the target.}
\end{figure*}

Here we discuss the spatio-temporal behavior of eigenstates of the FWS operator involving a derivative with respect to the static part of the permittivity ($\alpha=\epss$), the length of the scatterer ($\alpha=L$), and the central position ($\alpha=x_0$) for a steplike change of the refractive index [cf.\ Figs.~\eqref{fig:bstH_jump_L_epss}(a) and \eqref{fig:bstH_jump_x0}(a)]. Starting with the FWS matrix $\QF{\epss}$, its eigenvalues $\evF{\epss}$ are connected to the near field of the scatterer centered at the origin as
\begin{equation}
    \evF{\epss}=\frac{1}{2} \int_{-T/2}^{T/2} \frac{\dd{t}}{T} \int_{-L/2}^{L/2} \dd{x} \abs{\Es(x,t)}^2.
\end{equation}
The eigenstate corresponding to the most positive eigenvalue of $\QF{\epss}$ is depicted in Fig.~\ref{fig:bstH_jump_L_epss}(b). Similar to the case of a harmonic time modulation [cf.\ Fig.~\ref{fig:bstH_cos_L_eps}(b)], the intensity of the wave field inside the scatterer is maximized by exhibiting a focus while the jump from high to low permittivity takes place. In Fig.~\ref{fig:bstH_jump_L_epss}(c) an eigenstate of an eigenvalue close to zero is shown. This light field only has minimal intensity inside the scatterer because of destructive interference.

On the other hand, the eigenvalues $\evF{L}$ of the FWS matrix $\QF{L}$ corresponding to the length $L$ of the scatterer are proportional to the intensity of the light field at the edges of the scatterer,
\begin{equation}
    \begin{split}
    \evF{L} =& \frac{1}{4} \int_{-T/2}^{T/2} \frac{\dd{t}}{T} \abs{\Es(-L/2,t)}^2 \left[\epss+ \delta\epsilon \, \Theta_T(t,t_1,t_2) \right] \\
    +& \frac{1}{4} \int_{-T/2}^{T/2} \frac{\dd{t}}{T} \abs{\Es(L/2,t)}^2 \left[\epss+ \delta\epsilon \, \Theta_T(t,t_1,t_2) \right].
    \end{split}
\end{equation}
In Fig.~\ref{fig:bstH_jump_L_epss}(b) the eigenstate corresponding to the largest eigenvalue of $\QF{L}$ is depicted. Clearly, pronounced intensity maxima are built up at the edges of the scatterer predominantly during times when the permittivity is high.

Finally, the eigenvalues of the FWS matrix $\QF{x_0}$ involving a derivative with respect to the position of the center of the target $x_0$ are related to the near field of the scatterer centered at the origin as
\begin{equation}
    \begin{split}
    \evF{x_0} = -&\int_{-T/2}^{T/2} \frac{\dd{t}}{2T} \abs{\Es\left(-L/2,t\right)}^2 \left[\epss + \delta\epsilon \Theta_T(t,t_1,t_2) \right] \\
    +& \int_{-T/2}^{T/2} \frac{\dd{t}}{2T} \abs{\Es \left( L/2,t\right)}^2 \left[ \epss \!+ \! \delta\epsilon \Theta_T(t,t_1,t_2) \right].   
    \end{split}
\end{equation}
In Fig.~\ref{fig:bstH_jump_x0}(b) the eigenstate corresponding to the most positive eigenvalue $\evF{x_0}$ is depicted. This light field pulls the target scatterer to the right at times when the permittivity is high by exhibiting several intensity maxima at the right border of the target. Correspondingly, the eigenstate to the most negative eigenvalue applies a pulling force to left [see Fig.~\ref{fig:bstH_jump_x0}(c)]. An eigenstate to an eigenvalue close to zero applies nearly no force onto the target by exhibiting intensity maxima simultaneously on both borders of the target [see panel Fig.~\ref{fig:bstH_jump_x0}(d)]. We note that we choose $\omega=0$ for all wave fields in Fig.~\ref{fig:bstH_jump_x0}. \\ \\ \\ \\

\bibliography{FWS_bib}

\begin{thebibliography}{92}%
\makeatletter
\providecommand \@ifxundefined [1]{%
 \@ifx{#1\undefined}
}%
\providecommand \@ifnum [1]{%
 \ifnum #1\expandafter \@firstoftwo
 \else \expandafter \@secondoftwo
 \fi
}%
\providecommand \@ifx [1]{%
 \ifx #1\expandafter \@firstoftwo
 \else \expandafter \@secondoftwo
 \fi
}%
\providecommand \natexlab [1]{#1}%
\providecommand \enquote  [1]{``#1''}%
\providecommand \bibnamefont  [1]{#1}%
\providecommand \bibfnamefont [1]{#1}%
\providecommand \citenamefont [1]{#1}%
\providecommand \href@noop [0]{\@secondoftwo}%
\providecommand \href [0]{\begingroup \@sanitize@url \@href}%
\providecommand \@href[1]{\@@startlink{#1}\@@href}%
\providecommand \@@href[1]{\endgroup#1\@@endlink}%
\providecommand \@sanitize@url [0]{\catcode `\\12\catcode `\$12\catcode
  `\&12\catcode `\#12\catcode `\^12\catcode `\_12\catcode `\%12\relax}%
\providecommand \@@startlink[1]{}%
\providecommand \@@endlink[0]{}%
\providecommand \url  [0]{\begingroup\@sanitize@url \@url }%
\providecommand \@url [1]{\endgroup\@href {#1}{\urlprefix }}%
\providecommand \urlprefix  [0]{URL }%
\providecommand \Eprint [0]{\href }%
\providecommand \doibase [0]{https://doi.org/}%
\providecommand \selectlanguage [0]{\@gobble}%
\providecommand \bibinfo  [0]{\@secondoftwo}%
\providecommand \bibfield  [0]{\@secondoftwo}%
\providecommand \translation [1]{[#1]}%
\providecommand \BibitemOpen [0]{}%
\providecommand \bibitemStop [0]{}%
\providecommand \bibitemNoStop [0]{.\EOS\space}%
\providecommand \EOS [0]{\spacefactor3000\relax}%
\providecommand \BibitemShut  [1]{\csname bibitem#1\endcsname}%
\let\auto@bib@innerbib\@empty
\bibitem [{\citenamefont {Mosk}\ \emph {et~al.}(2012)\citenamefont {Mosk},
  \citenamefont {Lagendijk}, \citenamefont {Lerosey},\ and\ \citenamefont
  {Fink}}]{Mosk}%
  \BibitemOpen
  \bibfield  {author} {\bibinfo {author} {\bibfnamefont {A.~P.}\ \bibnamefont
  {Mosk}}, \bibinfo {author} {\bibfnamefont {A.}~\bibnamefont {Lagendijk}},
  \bibinfo {author} {\bibfnamefont {G.}~\bibnamefont {Lerosey}},\ and\ \bibinfo
  {author} {\bibfnamefont {M.}~\bibnamefont {Fink}},\ }\bibfield  {title}
  {\bibinfo {title} {Controlling waves in space and time for imaging and
  focusing in complex media},\ }\href {https://doi.org/10.1038/nphoton.2012.88}
  {\bibfield  {journal} {\bibinfo  {journal} {Nature Photonics}\ }\textbf
  {\bibinfo {volume} {6}},\ \bibinfo {pages} {283} (\bibinfo {year}
  {2012})}\BibitemShut {NoStop}%
\bibitem [{\citenamefont {Kim}\ \emph {et~al.}(2015)\citenamefont {Kim},
  \citenamefont {Choi}, \citenamefont {Choi}, \citenamefont {Yoon},\ and\
  \citenamefont {Choi}}]{Kim}%
  \BibitemOpen
  \bibfield  {author} {\bibinfo {author} {\bibfnamefont {M.}~\bibnamefont
  {Kim}}, \bibinfo {author} {\bibfnamefont {W.}~\bibnamefont {Choi}}, \bibinfo
  {author} {\bibfnamefont {Y.}~\bibnamefont {Choi}}, \bibinfo {author}
  {\bibfnamefont {C.}~\bibnamefont {Yoon}},\ and\ \bibinfo {author}
  {\bibfnamefont {W.}~\bibnamefont {Choi}},\ }\bibfield  {title} {\bibinfo
  {title} {Transmission matrix of a scattering medium and its applications in
  biophotonics},\ }\href {https://doi.org/10.1364/OE.23.012648} {\bibfield
  {journal} {\bibinfo  {journal} {Opt. Express}\ }\textbf {\bibinfo {volume}
  {23}},\ \bibinfo {pages} {12648} (\bibinfo {year} {2015})}\BibitemShut
  {NoStop}%
\bibitem [{\citenamefont {Rotter}\ and\ \citenamefont {Gigan}(2017)}]{Rotter}%
  \BibitemOpen
  \bibfield  {author} {\bibinfo {author} {\bibfnamefont {S.}~\bibnamefont
  {Rotter}}\ and\ \bibinfo {author} {\bibfnamefont {S.}~\bibnamefont {Gigan}},\
  }\bibfield  {title} {\bibinfo {title} {Light fields in complex media:
  Mesoscopic scattering meets wave control},\ }\href
  {https://doi.org/10.1103/RevModPhys.89.015005} {\bibfield  {journal}
  {\bibinfo  {journal} {Rev. Mod. Phys.}\ }\textbf {\bibinfo {volume} {89}},\
  \bibinfo {pages} {015005} (\bibinfo {year} {2017})}\BibitemShut {NoStop}%
\bibitem [{\citenamefont {Yoon}\ \emph {et~al.}(2020)\citenamefont {Yoon},
  \citenamefont {Kim}, \citenamefont {Jang}, \citenamefont {Choi},
  \citenamefont {Choi}, \citenamefont {Kang},\ and\ \citenamefont
  {Choi}}]{Yoon}%
  \BibitemOpen
  \bibfield  {author} {\bibinfo {author} {\bibfnamefont {S.}~\bibnamefont
  {Yoon}}, \bibinfo {author} {\bibfnamefont {M.}~\bibnamefont {Kim}}, \bibinfo
  {author} {\bibfnamefont {M.}~\bibnamefont {Jang}}, \bibinfo {author}
  {\bibfnamefont {Y.}~\bibnamefont {Choi}}, \bibinfo {author} {\bibfnamefont
  {W.}~\bibnamefont {Choi}}, \bibinfo {author} {\bibfnamefont {S.}~\bibnamefont
  {Kang}},\ and\ \bibinfo {author} {\bibfnamefont {W.}~\bibnamefont {Choi}},\
  }\bibfield  {title} {\bibinfo {title} {Deep optical imaging within complex
  scattering media},\ }\href {https://doi.org/10.1038/s42254-019-0143-2}
  {\bibfield  {journal} {\bibinfo  {journal} {Nature Reviews Physics}\ }\textbf
  {\bibinfo {volume} {2}},\ \bibinfo {pages} {141} (\bibinfo {year}
  {2020})}\BibitemShut {NoStop}%
\bibitem [{\citenamefont {Gigan}\ \emph {et~al.}(2022)\citenamefont {Gigan},
  \citenamefont {Katz}, \citenamefont {de~Aguiar}, \citenamefont {Andresen},
  \citenamefont {Aubry}, \citenamefont {Bertolotti}, \citenamefont {Bossy},
  \citenamefont {Bouchet}, \citenamefont {Brake}, \citenamefont {Brasselet},
  \citenamefont {Bromberg}, \citenamefont {Cao}, \citenamefont {Chaigne},
  \citenamefont {Cheng}, \citenamefont {Choi}, \citenamefont {Čižmár},
  \citenamefont {Cui}, \citenamefont {Curtis}, \citenamefont {Defienne},
  \citenamefont {Hofer}, \citenamefont {Horisaki}, \citenamefont {Horstmeyer},
  \citenamefont {Ji}, \citenamefont {LaViolette}, \citenamefont {Mertz},
  \citenamefont {Moser}, \citenamefont {Mosk}, \citenamefont {Pégard},
  \citenamefont {Piestun}, \citenamefont {Popoff}, \citenamefont {Phillips},
  \citenamefont {Psaltis}, \citenamefont {Rahmani}, \citenamefont {Rigneault},
  \citenamefont {Rotter}, \citenamefont {Tian}, \citenamefont {Vellekoop},
  \citenamefont {Waller}, \citenamefont {Wang}, \citenamefont {Weber},
  \citenamefont {Xiao}, \citenamefont {Xu}, \citenamefont {Yamilov},
  \citenamefont {Yang},\ and\ \citenamefont {Yılmaz}}]{Gigan}%
  \BibitemOpen
  \bibfield  {author} {\bibinfo {author} {\bibfnamefont {S.}~\bibnamefont
  {Gigan}}, \bibinfo {author} {\bibfnamefont {O.}~\bibnamefont {Katz}},
  \bibinfo {author} {\bibfnamefont {H.~B.}\ \bibnamefont {de~Aguiar}}, \bibinfo
  {author} {\bibfnamefont {E.~R.}\ \bibnamefont {Andresen}}, \bibinfo {author}
  {\bibfnamefont {A.}~\bibnamefont {Aubry}}, \bibinfo {author} {\bibfnamefont
  {J.}~\bibnamefont {Bertolotti}}, \bibinfo {author} {\bibfnamefont
  {E.}~\bibnamefont {Bossy}}, \bibinfo {author} {\bibfnamefont
  {D.}~\bibnamefont {Bouchet}}, \bibinfo {author} {\bibfnamefont
  {J.}~\bibnamefont {Brake}}, \bibinfo {author} {\bibfnamefont
  {S.}~\bibnamefont {Brasselet}}, \bibinfo {author} {\bibfnamefont
  {Y.}~\bibnamefont {Bromberg}}, \bibinfo {author} {\bibfnamefont
  {H.}~\bibnamefont {Cao}}, \bibinfo {author} {\bibfnamefont {T.}~\bibnamefont
  {Chaigne}}, \bibinfo {author} {\bibfnamefont {Z.}~\bibnamefont {Cheng}},
  \bibinfo {author} {\bibfnamefont {W.}~\bibnamefont {Choi}}, \bibinfo {author}
  {\bibfnamefont {T.}~\bibnamefont {Čižmár}}, \bibinfo {author}
  {\bibfnamefont {M.}~\bibnamefont {Cui}}, \bibinfo {author} {\bibfnamefont
  {V.~R.}\ \bibnamefont {Curtis}}, \bibinfo {author} {\bibfnamefont
  {H.}~\bibnamefont {Defienne}}, \bibinfo {author} {\bibfnamefont
  {M.}~\bibnamefont {Hofer}}, \bibinfo {author} {\bibfnamefont
  {R.}~\bibnamefont {Horisaki}}, \bibinfo {author} {\bibfnamefont
  {R.}~\bibnamefont {Horstmeyer}}, \bibinfo {author} {\bibfnamefont
  {N.}~\bibnamefont {Ji}}, \bibinfo {author} {\bibfnamefont {A.~K.}\
  \bibnamefont {LaViolette}}, \bibinfo {author} {\bibfnamefont
  {J.}~\bibnamefont {Mertz}}, \bibinfo {author} {\bibfnamefont
  {C.}~\bibnamefont {Moser}}, \bibinfo {author} {\bibfnamefont {A.~P.}\
  \bibnamefont {Mosk}}, \bibinfo {author} {\bibfnamefont {N.~C.}\ \bibnamefont
  {Pégard}}, \bibinfo {author} {\bibfnamefont {R.}~\bibnamefont {Piestun}},
  \bibinfo {author} {\bibfnamefont {S.}~\bibnamefont {Popoff}}, \bibinfo
  {author} {\bibfnamefont {D.~B.}\ \bibnamefont {Phillips}}, \bibinfo {author}
  {\bibfnamefont {D.}~\bibnamefont {Psaltis}}, \bibinfo {author} {\bibfnamefont
  {B.}~\bibnamefont {Rahmani}}, \bibinfo {author} {\bibfnamefont
  {H.}~\bibnamefont {Rigneault}}, \bibinfo {author} {\bibfnamefont
  {S.}~\bibnamefont {Rotter}}, \bibinfo {author} {\bibfnamefont
  {L.}~\bibnamefont {Tian}}, \bibinfo {author} {\bibfnamefont {I.~M.}\
  \bibnamefont {Vellekoop}}, \bibinfo {author} {\bibfnamefont {L.}~\bibnamefont
  {Waller}}, \bibinfo {author} {\bibfnamefont {L.}~\bibnamefont {Wang}},
  \bibinfo {author} {\bibfnamefont {T.}~\bibnamefont {Weber}}, \bibinfo
  {author} {\bibfnamefont {S.}~\bibnamefont {Xiao}}, \bibinfo {author}
  {\bibfnamefont {C.}~\bibnamefont {Xu}}, \bibinfo {author} {\bibfnamefont
  {A.}~\bibnamefont {Yamilov}}, \bibinfo {author} {\bibfnamefont
  {C.}~\bibnamefont {Yang}},\ and\ \bibinfo {author} {\bibfnamefont
  {H.}~\bibnamefont {Yılmaz}},\ }\bibfield  {title} {\bibinfo {title} {Roadmap
  on wavefront shaping and deep imaging in complex media},\ }\href
  {https://doi.org/10.1088/2515-7647/ac76f9} {\bibfield  {journal} {\bibinfo
  {journal} {Journal of Physics: Photonics}\ }\textbf {\bibinfo {volume} {4}},\
  \bibinfo {pages} {042501} (\bibinfo {year} {2022})}\BibitemShut {NoStop}%
\bibitem [{\citenamefont {Bertolotti}\ and\ \citenamefont
  {Katz}(2022)}]{Bertolotti}%
  \BibitemOpen
  \bibfield  {author} {\bibinfo {author} {\bibfnamefont {J.}~\bibnamefont
  {Bertolotti}}\ and\ \bibinfo {author} {\bibfnamefont {O.}~\bibnamefont
  {Katz}},\ }\bibfield  {title} {\bibinfo {title} {Imaging in complex media},\
  }\href {https://doi.org/10.1038/s41567-022-01723-8} {\bibfield  {journal}
  {\bibinfo  {journal} {Nature Physics}\ }\textbf {\bibinfo {volume} {18}},\
  \bibinfo {pages} {1008} (\bibinfo {year} {2022})}\BibitemShut {NoStop}%
\bibitem [{\citenamefont {Cao}\ \emph {et~al.}(2022)\citenamefont {Cao},
  \citenamefont {Mosk},\ and\ \citenamefont {Rotter}}]{Cao22}%
  \BibitemOpen
  \bibfield  {author} {\bibinfo {author} {\bibfnamefont {H.}~\bibnamefont
  {Cao}}, \bibinfo {author} {\bibfnamefont {A.~P.}\ \bibnamefont {Mosk}},\ and\
  \bibinfo {author} {\bibfnamefont {S.}~\bibnamefont {Rotter}},\ }\bibfield
  {title} {\bibinfo {title} {Shaping the propagation of light in complex
  media},\ }\href {https://doi.org/10.1038/s41567-022-01677-x} {\bibfield
  {journal} {\bibinfo  {journal} {Nature Physics}\ }\textbf {\bibinfo {volume}
  {18}},\ \bibinfo {pages} {994} (\bibinfo {year} {2022})}\BibitemShut
  {NoStop}%
\bibitem [{\citenamefont {Gigan}(2022)}]{Gigan22}%
  \BibitemOpen
  \bibfield  {author} {\bibinfo {author} {\bibfnamefont {S.}~\bibnamefont
  {Gigan}},\ }\bibfield  {title} {\bibinfo {title} {Imaging and computing with
  disorder},\ }\href {https://doi.org/10.1038/s41567-022-01681-1} {\bibfield
  {journal} {\bibinfo  {journal} {Nature Physics}\ }\textbf {\bibinfo {volume}
  {18}},\ \bibinfo {pages} {980} (\bibinfo {year} {2022})}\BibitemShut
  {NoStop}%
\bibitem [{\citenamefont {Patel}\ \emph {et~al.}(2023)\citenamefont {Patel},
  \citenamefont {Mao},\ and\ \citenamefont {Michielssen}}]{Patel}%
  \BibitemOpen
  \bibfield  {author} {\bibinfo {author} {\bibfnamefont {U.~R.}\ \bibnamefont
  {Patel}}, \bibinfo {author} {\bibfnamefont {Y.}~\bibnamefont {Mao}},\ and\
  \bibinfo {author} {\bibfnamefont {E.}~\bibnamefont {Michielssen}},\
  }\bibfield  {title} {\bibinfo {title} {Wigner–{Smith} time delay matrix for
  acoustic scattering: {Theory} and phenomenology},\ }\href
  {https://doi.org/10.1121/10.0017826} {\bibfield  {journal} {\bibinfo
  {journal} {The Journal of the Acoustical Society of America}\ }\textbf
  {\bibinfo {volume} {153}},\ \bibinfo {pages} {2769} (\bibinfo {year}
  {2023})}\BibitemShut {NoStop}%
\bibitem [{\citenamefont {Cao}\ \emph {et~al.}(2023)\citenamefont {Cao},
  \citenamefont {\v{C}i\v{z}m\'{a}r}, \citenamefont {Turtaev}, \citenamefont
  {Tyc},\ and\ \citenamefont {Rotter}}]{Cao23}%
  \BibitemOpen
  \bibfield  {author} {\bibinfo {author} {\bibfnamefont {H.}~\bibnamefont
  {Cao}}, \bibinfo {author} {\bibfnamefont {T.}~\bibnamefont
  {\v{C}i\v{z}m\'{a}r}}, \bibinfo {author} {\bibfnamefont {S.}~\bibnamefont
  {Turtaev}}, \bibinfo {author} {\bibfnamefont {T.}~\bibnamefont {Tyc}},\ and\
  \bibinfo {author} {\bibfnamefont {S.}~\bibnamefont {Rotter}},\ }\bibfield
  {title} {\bibinfo {title} {Controlling light propagation in multimode fibers
  for imaging, spectroscopy, and beyond},\ }\href
  {https://doi.org/10.1364/AOP.484298} {\bibfield  {journal} {\bibinfo
  {journal} {Adv. Opt. Photon.}\ }\textbf {\bibinfo {volume} {15}},\ \bibinfo
  {pages} {524} (\bibinfo {year} {2023})}\BibitemShut {NoStop}%
\bibitem [{\citenamefont {Asadova}\ \emph {et~al.}(2024)\citenamefont
  {Asadova}, \citenamefont {Achouri}, \citenamefont {Arjas}, \citenamefont
  {Auguié}, \citenamefont {Aydin}, \citenamefont {Baron}, \citenamefont
  {Beutel}, \citenamefont {Bodermann}, \citenamefont {Boussaoud}, \citenamefont
  {Burger}, \citenamefont {Choi}, \citenamefont {Czajkowski}, \citenamefont
  {Evlyukhin}, \citenamefont {Fazel-Najafabadi}, \citenamefont
  {Fernandez-Corbaton}, \citenamefont {Garg}, \citenamefont {Globosits},
  \citenamefont {Hohenester}, \citenamefont {Kim}, \citenamefont {Kim},
  \citenamefont {Lalanne}, \citenamefont {Ru}, \citenamefont {Meyer},
  \citenamefont {Mun}, \citenamefont {Pattelli}, \citenamefont {Pflug},
  \citenamefont {Rockstuhl}, \citenamefont {Rho}, \citenamefont {Rotter},
  \citenamefont {Stout}, \citenamefont {Törmä}, \citenamefont {Trigo},
  \citenamefont {Tristram}, \citenamefont {Tsitsas}, \citenamefont {Vallée},
  \citenamefont {Vynck}, \citenamefont {Weiss}, \citenamefont {Wiecha},
  \citenamefont {Wriedt}, \citenamefont {Yannopapas}, \citenamefont {Yurkin},\
  and\ \citenamefont {Zouros}}]{Asadova}%
  \BibitemOpen
  \bibfield  {author} {\bibinfo {author} {\bibfnamefont {N.}~\bibnamefont
  {Asadova}}, \bibinfo {author} {\bibfnamefont {K.}~\bibnamefont {Achouri}},
  \bibinfo {author} {\bibfnamefont {K.}~\bibnamefont {Arjas}}, \bibinfo
  {author} {\bibfnamefont {B.}~\bibnamefont {Auguié}}, \bibinfo {author}
  {\bibfnamefont {R.}~\bibnamefont {Aydin}}, \bibinfo {author} {\bibfnamefont
  {A.}~\bibnamefont {Baron}}, \bibinfo {author} {\bibfnamefont
  {D.}~\bibnamefont {Beutel}}, \bibinfo {author} {\bibfnamefont
  {B.}~\bibnamefont {Bodermann}}, \bibinfo {author} {\bibfnamefont
  {K.}~\bibnamefont {Boussaoud}}, \bibinfo {author} {\bibfnamefont
  {S.}~\bibnamefont {Burger}}, \bibinfo {author} {\bibfnamefont
  {M.}~\bibnamefont {Choi}}, \bibinfo {author} {\bibfnamefont {K.~M.}\
  \bibnamefont {Czajkowski}}, \bibinfo {author} {\bibfnamefont {A.~B.}\
  \bibnamefont {Evlyukhin}}, \bibinfo {author} {\bibfnamefont {A.}~\bibnamefont
  {Fazel-Najafabadi}}, \bibinfo {author} {\bibfnamefont {I.}~\bibnamefont
  {Fernandez-Corbaton}}, \bibinfo {author} {\bibfnamefont {P.}~\bibnamefont
  {Garg}}, \bibinfo {author} {\bibfnamefont {D.}~\bibnamefont {Globosits}},
  \bibinfo {author} {\bibfnamefont {U.}~\bibnamefont {Hohenester}}, \bibinfo
  {author} {\bibfnamefont {H.}~\bibnamefont {Kim}}, \bibinfo {author}
  {\bibfnamefont {S.}~\bibnamefont {Kim}}, \bibinfo {author} {\bibfnamefont
  {P.}~\bibnamefont {Lalanne}}, \bibinfo {author} {\bibfnamefont {E.~C.~L.}\
  \bibnamefont {Ru}}, \bibinfo {author} {\bibfnamefont {J.}~\bibnamefont
  {Meyer}}, \bibinfo {author} {\bibfnamefont {J.}~\bibnamefont {Mun}}, \bibinfo
  {author} {\bibfnamefont {L.}~\bibnamefont {Pattelli}}, \bibinfo {author}
  {\bibfnamefont {L.}~\bibnamefont {Pflug}}, \bibinfo {author} {\bibfnamefont
  {C.}~\bibnamefont {Rockstuhl}}, \bibinfo {author} {\bibfnamefont
  {J.}~\bibnamefont {Rho}}, \bibinfo {author} {\bibfnamefont {S.}~\bibnamefont
  {Rotter}}, \bibinfo {author} {\bibfnamefont {B.}~\bibnamefont {Stout}},
  \bibinfo {author} {\bibfnamefont {P.}~\bibnamefont {Törmä}}, \bibinfo
  {author} {\bibfnamefont {J.~O.}\ \bibnamefont {Trigo}}, \bibinfo {author}
  {\bibfnamefont {F.}~\bibnamefont {Tristram}}, \bibinfo {author}
  {\bibfnamefont {N.~L.}\ \bibnamefont {Tsitsas}}, \bibinfo {author}
  {\bibfnamefont {R.}~\bibnamefont {Vallée}}, \bibinfo {author} {\bibfnamefont
  {K.}~\bibnamefont {Vynck}}, \bibinfo {author} {\bibfnamefont
  {T.}~\bibnamefont {Weiss}}, \bibinfo {author} {\bibfnamefont
  {P.}~\bibnamefont {Wiecha}}, \bibinfo {author} {\bibfnamefont
  {T.}~\bibnamefont {Wriedt}}, \bibinfo {author} {\bibfnamefont
  {V.}~\bibnamefont {Yannopapas}}, \bibinfo {author} {\bibfnamefont {M.~A.}\
  \bibnamefont {Yurkin}},\ and\ \bibinfo {author} {\bibfnamefont {G.~P.}\
  \bibnamefont {Zouros}},\ }\href {https://arxiv.org/abs/2408.10727} {\bibinfo
  {title} {T-matrix representation of optical scattering response: Suggestion
  for a data format}} (\bibinfo {year} {2024}),\ \Eprint
  {https://arxiv.org/abs/2408.10727} {arXiv:2408.10727 [physics.optics]}
  \BibitemShut {NoStop}%
\bibitem [{\citenamefont {Popoff}\ \emph {et~al.}(2010)\citenamefont {Popoff},
  \citenamefont {Lerosey}, \citenamefont {Carminati}, \citenamefont {Fink},
  \citenamefont {Boccara},\ and\ \citenamefont {Gigan}}]{Popoff}%
  \BibitemOpen
  \bibfield  {author} {\bibinfo {author} {\bibfnamefont {S.~M.}\ \bibnamefont
  {Popoff}}, \bibinfo {author} {\bibfnamefont {G.}~\bibnamefont {Lerosey}},
  \bibinfo {author} {\bibfnamefont {R.}~\bibnamefont {Carminati}}, \bibinfo
  {author} {\bibfnamefont {M.}~\bibnamefont {Fink}}, \bibinfo {author}
  {\bibfnamefont {A.~C.}\ \bibnamefont {Boccara}},\ and\ \bibinfo {author}
  {\bibfnamefont {S.}~\bibnamefont {Gigan}},\ }\bibfield  {title} {\bibinfo
  {title} {Measuring the transmission matrix in optics: An approach to the
  study and control of light propagation in disordered media},\ }\href
  {https://doi.org/10.1103/PhysRevLett.104.100601} {\bibfield  {journal}
  {\bibinfo  {journal} {Phys. Rev. Lett.}\ }\textbf {\bibinfo {volume} {104}},\
  \bibinfo {pages} {100601} (\bibinfo {year} {2010})}\BibitemShut {NoStop}%
\bibitem [{\citenamefont {Shi}\ \emph {et~al.}(2013)\citenamefont {Shi},
  \citenamefont {Davy}, \citenamefont {Wang},\ and\ \citenamefont
  {Genack}}]{Shi}%
  \BibitemOpen
  \bibfield  {author} {\bibinfo {author} {\bibfnamefont {Z.}~\bibnamefont
  {Shi}}, \bibinfo {author} {\bibfnamefont {M.}~\bibnamefont {Davy}}, \bibinfo
  {author} {\bibfnamefont {J.}~\bibnamefont {Wang}},\ and\ \bibinfo {author}
  {\bibfnamefont {A.~Z.}\ \bibnamefont {Genack}},\ }\bibfield  {title}
  {\bibinfo {title} {Focusing through random media in space and time: a
  transmission matrix approach},\ }\href {https://doi.org/10.1364/OL.38.002714}
  {\bibfield  {journal} {\bibinfo  {journal} {Opt. Lett.}\ }\textbf {\bibinfo
  {volume} {38}},\ \bibinfo {pages} {2714} (\bibinfo {year}
  {2013})}\BibitemShut {NoStop}%
\bibitem [{\citenamefont {Yu}\ \emph {et~al.}(2013)\citenamefont {Yu},
  \citenamefont {Hillman}, \citenamefont {Choi}, \citenamefont {Lee},
  \citenamefont {Feld}, \citenamefont {Dasari},\ and\ \citenamefont
  {Park}}]{Yu}%
  \BibitemOpen
  \bibfield  {author} {\bibinfo {author} {\bibfnamefont {H.}~\bibnamefont
  {Yu}}, \bibinfo {author} {\bibfnamefont {T.~R.}\ \bibnamefont {Hillman}},
  \bibinfo {author} {\bibfnamefont {W.}~\bibnamefont {Choi}}, \bibinfo {author}
  {\bibfnamefont {J.~O.}\ \bibnamefont {Lee}}, \bibinfo {author} {\bibfnamefont
  {M.~S.}\ \bibnamefont {Feld}}, \bibinfo {author} {\bibfnamefont {R.~R.}\
  \bibnamefont {Dasari}},\ and\ \bibinfo {author} {\bibfnamefont
  {Y.}~\bibnamefont {Park}},\ }\bibfield  {title} {\bibinfo {title} {Measuring
  large optical transmission matrices of disordered media},\ }\href
  {https://doi.org/10.1103/PhysRevLett.111.153902} {\bibfield  {journal}
  {\bibinfo  {journal} {Phys. Rev. Lett.}\ }\textbf {\bibinfo {volume} {111}},\
  \bibinfo {pages} {153902} (\bibinfo {year} {2013})}\BibitemShut {NoStop}%
\bibitem [{\citenamefont {Chaigne}\ \emph {et~al.}(2014)\citenamefont
  {Chaigne}, \citenamefont {Katz}, \citenamefont {Boccara}, \citenamefont
  {Fink}, \citenamefont {Bossy},\ and\ \citenamefont {Gigan}}]{Chaigne}%
  \BibitemOpen
  \bibfield  {author} {\bibinfo {author} {\bibfnamefont {T.}~\bibnamefont
  {Chaigne}}, \bibinfo {author} {\bibfnamefont {O.}~\bibnamefont {Katz}},
  \bibinfo {author} {\bibfnamefont {A.~C.}\ \bibnamefont {Boccara}}, \bibinfo
  {author} {\bibfnamefont {M.}~\bibnamefont {Fink}}, \bibinfo {author}
  {\bibfnamefont {E.}~\bibnamefont {Bossy}},\ and\ \bibinfo {author}
  {\bibfnamefont {S.}~\bibnamefont {Gigan}},\ }\bibfield  {title} {\bibinfo
  {title} {Controlling light in scattering media non-invasively using the
  photoacoustic transmission matrix},\ }\href
  {https://doi.org/10.1038/nphoton.2013.307} {\bibfield  {journal} {\bibinfo
  {journal} {Nature Photonics}\ }\textbf {\bibinfo {volume} {8}},\ \bibinfo
  {pages} {58} (\bibinfo {year} {2014})}\BibitemShut {NoStop}%
\bibitem [{\citenamefont {Andreoli}\ \emph {et~al.}(2015)\citenamefont
  {Andreoli}, \citenamefont {Volpe}, \citenamefont {Popoff}, \citenamefont
  {Katz}, \citenamefont {Gr{\'e}sillon},\ and\ \citenamefont
  {Gigan}}]{Andreoli}%
  \BibitemOpen
  \bibfield  {author} {\bibinfo {author} {\bibfnamefont {D.}~\bibnamefont
  {Andreoli}}, \bibinfo {author} {\bibfnamefont {G.}~\bibnamefont {Volpe}},
  \bibinfo {author} {\bibfnamefont {S.}~\bibnamefont {Popoff}}, \bibinfo
  {author} {\bibfnamefont {O.}~\bibnamefont {Katz}}, \bibinfo {author}
  {\bibfnamefont {S.}~\bibnamefont {Gr{\'e}sillon}},\ and\ \bibinfo {author}
  {\bibfnamefont {S.}~\bibnamefont {Gigan}},\ }\bibfield  {title} {\bibinfo
  {title} {Deterministic control of broadband light through a multiply
  scattering medium via the multispectral transmission matrix},\ }\href
  {https://doi.org/10.1038/srep10347} {\bibfield  {journal} {\bibinfo
  {journal} {Scientific Reports}\ }\textbf {\bibinfo {volume} {5}},\ \bibinfo
  {pages} {10347} (\bibinfo {year} {2015})}\BibitemShut {NoStop}%
\bibitem [{\citenamefont {Dr\'{e}meau}\ \emph {et~al.}(2015)\citenamefont
  {Dr\'{e}meau}, \citenamefont {Liutkus}, \citenamefont {Martina},
  \citenamefont {Katz}, \citenamefont {Sch\"{u}lke}, \citenamefont {Krzakala},
  \citenamefont {Gigan},\ and\ \citenamefont {Daudet}}]{Dremeau}%
  \BibitemOpen
  \bibfield  {author} {\bibinfo {author} {\bibfnamefont {A.}~\bibnamefont
  {Dr\'{e}meau}}, \bibinfo {author} {\bibfnamefont {A.}~\bibnamefont
  {Liutkus}}, \bibinfo {author} {\bibfnamefont {D.}~\bibnamefont {Martina}},
  \bibinfo {author} {\bibfnamefont {O.}~\bibnamefont {Katz}}, \bibinfo {author}
  {\bibfnamefont {C.}~\bibnamefont {Sch\"{u}lke}}, \bibinfo {author}
  {\bibfnamefont {F.}~\bibnamefont {Krzakala}}, \bibinfo {author}
  {\bibfnamefont {S.}~\bibnamefont {Gigan}},\ and\ \bibinfo {author}
  {\bibfnamefont {L.}~\bibnamefont {Daudet}},\ }\bibfield  {title} {\bibinfo
  {title} {Reference-less measurement of the transmission matrix of a highly
  scattering material using a dmd and phase retrieval techniques},\ }\href
  {https://doi.org/10.1364/OE.23.011898} {\bibfield  {journal} {\bibinfo
  {journal} {Opt. Express}\ }\textbf {\bibinfo {volume} {23}},\ \bibinfo
  {pages} {11898} (\bibinfo {year} {2015})}\BibitemShut {NoStop}%
\bibitem [{\citenamefont {Mounaix}\ \emph {et~al.}(2016)\citenamefont
  {Mounaix}, \citenamefont {Defienne},\ and\ \citenamefont {Gigan}}]{Mounaix}%
  \BibitemOpen
  \bibfield  {author} {\bibinfo {author} {\bibfnamefont {M.}~\bibnamefont
  {Mounaix}}, \bibinfo {author} {\bibfnamefont {H.}~\bibnamefont {Defienne}},\
  and\ \bibinfo {author} {\bibfnamefont {S.}~\bibnamefont {Gigan}},\ }\bibfield
   {title} {\bibinfo {title} {Deterministic light focusing in space and time
  through multiple scattering media with a time-resolved transmission matrix
  approach},\ }\href {https://doi.org/10.1103/PhysRevA.94.041802} {\bibfield
  {journal} {\bibinfo  {journal} {Phys. Rev. A}\ }\textbf {\bibinfo {volume}
  {94}},\ \bibinfo {pages} {041802} (\bibinfo {year} {2016})}\BibitemShut
  {NoStop}%
\bibitem [{\citenamefont {Horodynski}\ \emph {et~al.}(2020)\citenamefont
  {Horodynski}, \citenamefont {Kühmayer}, \citenamefont {Brandstötter},
  \citenamefont {Pichler}, \citenamefont {Fyodorov}, \citenamefont {Kuhl},\
  and\ \citenamefont {Rotter}}]{Horodynski}%
  \BibitemOpen
  \bibfield  {author} {\bibinfo {author} {\bibfnamefont {M.}~\bibnamefont
  {Horodynski}}, \bibinfo {author} {\bibfnamefont {M.}~\bibnamefont
  {Kühmayer}}, \bibinfo {author} {\bibfnamefont {A.}~\bibnamefont
  {Brandstötter}}, \bibinfo {author} {\bibfnamefont {K.}~\bibnamefont
  {Pichler}}, \bibinfo {author} {\bibfnamefont {Y.~V.}\ \bibnamefont
  {Fyodorov}}, \bibinfo {author} {\bibfnamefont {U.}~\bibnamefont {Kuhl}},\
  and\ \bibinfo {author} {\bibfnamefont {S.}~\bibnamefont {Rotter}},\
  }\bibfield  {title} {\bibinfo {title} {Optimal wave fields for
  micromanipulation in complex scattering environments},\ }\href
  {https://doi.org/10.1038/s41566-019-0550-z} {\bibfield  {journal} {\bibinfo
  {journal} {Nature Photonics}\ }\textbf {\bibinfo {volume} {14}},\ \bibinfo
  {pages} {149} (\bibinfo {year} {2020})}\BibitemShut {NoStop}%
\bibitem [{\citenamefont {Boniface}\ \emph {et~al.}(2020)\citenamefont
  {Boniface}, \citenamefont {Dong},\ and\ \citenamefont {Gigan}}]{Boniface}%
  \BibitemOpen
  \bibfield  {author} {\bibinfo {author} {\bibfnamefont {A.}~\bibnamefont
  {Boniface}}, \bibinfo {author} {\bibfnamefont {J.}~\bibnamefont {Dong}},\
  and\ \bibinfo {author} {\bibfnamefont {S.}~\bibnamefont {Gigan}},\ }\bibfield
   {title} {\bibinfo {title} {Non-invasive focusing and imaging in scattering
  media with a fluorescence-based transmission matrix},\ }\href
  {https://doi.org/10.1038/s41467-020-19696-8} {\bibfield  {journal} {\bibinfo
  {journal} {Nature Communications}\ }\textbf {\bibinfo {volume} {11}},\
  \bibinfo {pages} {6154} (\bibinfo {year} {2020})}\BibitemShut {NoStop}%
\bibitem [{\citenamefont {Li}\ \emph {et~al.}(2021{\natexlab{a}})\citenamefont
  {Li}, \citenamefont {Saunders}, \citenamefont {Lum}, \citenamefont
  {Murray-Bruce}, \citenamefont {Goyal}, \citenamefont {{\v C}i{\v z}m{\'a}r},\
  and\ \citenamefont {Phillips}}]{Li21}%
  \BibitemOpen
  \bibfield  {author} {\bibinfo {author} {\bibfnamefont {S.}~\bibnamefont
  {Li}}, \bibinfo {author} {\bibfnamefont {C.}~\bibnamefont {Saunders}},
  \bibinfo {author} {\bibfnamefont {D.~J.}\ \bibnamefont {Lum}}, \bibinfo
  {author} {\bibfnamefont {J.}~\bibnamefont {Murray-Bruce}}, \bibinfo {author}
  {\bibfnamefont {V.~K.}\ \bibnamefont {Goyal}}, \bibinfo {author}
  {\bibfnamefont {T.}~\bibnamefont {{\v C}i{\v z}m{\'a}r}},\ and\ \bibinfo
  {author} {\bibfnamefont {D.~B.}\ \bibnamefont {Phillips}},\ }\bibfield
  {title} {\bibinfo {title} {Compressively sampling the optical transmission
  matrix of a multimode fibre},\ }\href
  {https://doi.org/10.1038/s41377-021-00514-9} {\bibfield  {journal} {\bibinfo
  {journal} {Light: Science \& Applications}\ }\textbf {\bibinfo {volume}
  {10}},\ \bibinfo {pages} {88} (\bibinfo {year}
  {2021}{\natexlab{a}})}\BibitemShut {NoStop}%
\bibitem [{\citenamefont {Horodynski}\ \emph {et~al.}(2022)\citenamefont
  {Horodynski}, \citenamefont {K{\"u}hmayer}, \citenamefont {Ferise},
  \citenamefont {Rotter},\ and\ \citenamefont {Davy}}]{Horodynski22}%
  \BibitemOpen
  \bibfield  {author} {\bibinfo {author} {\bibfnamefont {M.}~\bibnamefont
  {Horodynski}}, \bibinfo {author} {\bibfnamefont {M.}~\bibnamefont
  {K{\"u}hmayer}}, \bibinfo {author} {\bibfnamefont {C.}~\bibnamefont
  {Ferise}}, \bibinfo {author} {\bibfnamefont {S.}~\bibnamefont {Rotter}},\
  and\ \bibinfo {author} {\bibfnamefont {M.}~\bibnamefont {Davy}},\ }\bibfield
  {title} {\bibinfo {title} {Anti-reflection structure for perfect transmission
  through complex media},\ }\href {https://doi.org/10.1038/s41586-022-04843-6}
  {\bibfield  {journal} {\bibinfo  {journal} {Nature}\ }\textbf {\bibinfo
  {volume} {607}},\ \bibinfo {pages} {281} (\bibinfo {year}
  {2022})}\BibitemShut {NoStop}%
\bibitem [{\citenamefont {Choi}\ \emph {et~al.}(2013)\citenamefont {Choi},
  \citenamefont {Hillman}, \citenamefont {Choi}, \citenamefont {Lue},
  \citenamefont {Dasari}, \citenamefont {So}, \citenamefont {Choi},\ and\
  \citenamefont {Yaqoob}}]{Choi}%
  \BibitemOpen
  \bibfield  {author} {\bibinfo {author} {\bibfnamefont {Y.}~\bibnamefont
  {Choi}}, \bibinfo {author} {\bibfnamefont {T.~R.}\ \bibnamefont {Hillman}},
  \bibinfo {author} {\bibfnamefont {W.}~\bibnamefont {Choi}}, \bibinfo {author}
  {\bibfnamefont {N.}~\bibnamefont {Lue}}, \bibinfo {author} {\bibfnamefont
  {R.~R.}\ \bibnamefont {Dasari}}, \bibinfo {author} {\bibfnamefont {P.~T.~C.}\
  \bibnamefont {So}}, \bibinfo {author} {\bibfnamefont {W.}~\bibnamefont
  {Choi}},\ and\ \bibinfo {author} {\bibfnamefont {Z.}~\bibnamefont {Yaqoob}},\
  }\bibfield  {title} {\bibinfo {title} {Measurement of the time-resolved
  reflection matrix for enhancing light energy delivery into a scattering
  medium},\ }\href {https://doi.org/10.1103/PhysRevLett.111.243901} {\bibfield
  {journal} {\bibinfo  {journal} {Phys. Rev. Lett.}\ }\textbf {\bibinfo
  {volume} {111}},\ \bibinfo {pages} {243901} (\bibinfo {year}
  {2013})}\BibitemShut {NoStop}%
\bibitem [{\citenamefont {Galiffi}\ \emph {et~al.}(2022)\citenamefont
  {Galiffi}, \citenamefont {Tirole}, \citenamefont {Yin}, \citenamefont {Li},
  \citenamefont {Vezzoli}, \citenamefont {Huidobro}, \citenamefont
  {Silveirinha}, \citenamefont {Sapienza}, \citenamefont {Al{\`u}},\ and\
  \citenamefont {Pendry}}]{Galiffi22}%
  \BibitemOpen
  \bibfield  {author} {\bibinfo {author} {\bibfnamefont {E.}~\bibnamefont
  {Galiffi}}, \bibinfo {author} {\bibfnamefont {R.}~\bibnamefont {Tirole}},
  \bibinfo {author} {\bibfnamefont {S.}~\bibnamefont {Yin}}, \bibinfo {author}
  {\bibfnamefont {H.}~\bibnamefont {Li}}, \bibinfo {author} {\bibfnamefont
  {S.}~\bibnamefont {Vezzoli}}, \bibinfo {author} {\bibfnamefont {P.~A.}\
  \bibnamefont {Huidobro}}, \bibinfo {author} {\bibfnamefont {M.~G.}\
  \bibnamefont {Silveirinha}}, \bibinfo {author} {\bibfnamefont
  {R.}~\bibnamefont {Sapienza}}, \bibinfo {author} {\bibfnamefont
  {A.}~\bibnamefont {Al{\`u}}},\ and\ \bibinfo {author} {\bibfnamefont {J.~B.}\
  \bibnamefont {Pendry}},\ }\bibfield  {title} {\bibinfo {title} {{Photonics of
  time-varying media}},\ }\href {https://doi.org/10.1117/1.AP.4.1.014002}
  {\bibfield  {journal} {\bibinfo  {journal} {Advanced Photonics}\ }\textbf
  {\bibinfo {volume} {4}},\ \bibinfo {pages} {014002} (\bibinfo {year}
  {2022})}\BibitemShut {NoStop}%
\bibitem [{\citenamefont {Pacheco-Peña}\ \emph {et~al.}(2022)\citenamefont
  {Pacheco-Peña}, \citenamefont {Sol\'{i}s},\ and\ \citenamefont
  {Engheta}}]{Pacheco}%
  \BibitemOpen
  \bibfield  {author} {\bibinfo {author} {\bibfnamefont {V.}~\bibnamefont
  {Pacheco-Peña}}, \bibinfo {author} {\bibfnamefont {D.~M.}\ \bibnamefont
  {Sol\'{i}s}},\ and\ \bibinfo {author} {\bibfnamefont {N.}~\bibnamefont
  {Engheta}},\ }\bibfield  {title} {\bibinfo {title} {Time-varying
  electromagnetic media: opinion},\ }\href {https://doi.org/10.1364/OME.471007}
  {\bibfield  {journal} {\bibinfo  {journal} {Opt. Mater. Express}\ }\textbf
  {\bibinfo {volume} {12}},\ \bibinfo {pages} {3829} (\bibinfo {year}
  {2022})}\BibitemShut {NoStop}%
\bibitem [{\citenamefont {Won}(2023)}]{Won}%
  \BibitemOpen
  \bibfield  {author} {\bibinfo {author} {\bibfnamefont {R.}~\bibnamefont
  {Won}},\ }\bibfield  {title} {\bibinfo {title} {It's a matter of time},\
  }\href {https://doi.org/10.1038/s41566-023-01163-7} {\bibfield  {journal}
  {\bibinfo  {journal} {Nature Photonics}\ }\textbf {\bibinfo {volume} {17}},\
  \bibinfo {pages} {209} (\bibinfo {year} {2023})}\BibitemShut {NoStop}%
\bibitem [{\citenamefont {Biancalana}\ \emph {et~al.}(2007)\citenamefont
  {Biancalana}, \citenamefont {Amann}, \citenamefont {Uskov},\ and\
  \citenamefont {O'Reilly}}]{Biancalana}%
  \BibitemOpen
  \bibfield  {author} {\bibinfo {author} {\bibfnamefont {F.}~\bibnamefont
  {Biancalana}}, \bibinfo {author} {\bibfnamefont {A.}~\bibnamefont {Amann}},
  \bibinfo {author} {\bibfnamefont {A.~V.}\ \bibnamefont {Uskov}},\ and\
  \bibinfo {author} {\bibfnamefont {E.~P.}\ \bibnamefont {O'Reilly}},\
  }\bibfield  {title} {\bibinfo {title} {Dynamics of light propagation in
  spatiotemporal dielectric structures},\ }\href
  {https://doi.org/10.1103/PhysRevE.75.046607} {\bibfield  {journal} {\bibinfo
  {journal} {Phys. Rev. E}\ }\textbf {\bibinfo {volume} {75}},\ \bibinfo
  {pages} {046607} (\bibinfo {year} {2007})}\BibitemShut {NoStop}%
\bibitem [{\citenamefont {Xiao}\ \emph {et~al.}(2014)\citenamefont {Xiao},
  \citenamefont {Maywar},\ and\ \citenamefont {Agrawal}}]{Xiao14}%
  \BibitemOpen
  \bibfield  {author} {\bibinfo {author} {\bibfnamefont {Y.}~\bibnamefont
  {Xiao}}, \bibinfo {author} {\bibfnamefont {D.~N.}\ \bibnamefont {Maywar}},\
  and\ \bibinfo {author} {\bibfnamefont {G.~P.}\ \bibnamefont {Agrawal}},\
  }\bibfield  {title} {\bibinfo {title} {Reflection and transmission of
  electromagnetic waves at a temporal boundary},\ }\href
  {https://doi.org/10.1364/OL.39.000574} {\bibfield  {journal} {\bibinfo
  {journal} {Opt. Lett.}\ }\textbf {\bibinfo {volume} {39}},\ \bibinfo {pages}
  {574} (\bibinfo {year} {2014})}\BibitemShut {NoStop}%
\bibitem [{\citenamefont {Hayrapetyan}\ \emph {et~al.}(2016)\citenamefont
  {Hayrapetyan}, \citenamefont {Götte}, \citenamefont {Grigoryan},
  \citenamefont {Fritzsche},\ and\ \citenamefont {Petrosyan}}]{Hayrapetyan}%
  \BibitemOpen
  \bibfield  {author} {\bibinfo {author} {\bibfnamefont {A.~G.}\ \bibnamefont
  {Hayrapetyan}}, \bibinfo {author} {\bibfnamefont {J.~B.}\ \bibnamefont
  {Götte}}, \bibinfo {author} {\bibfnamefont {K.~K.}\ \bibnamefont
  {Grigoryan}}, \bibinfo {author} {\bibfnamefont {S.}~\bibnamefont
  {Fritzsche}},\ and\ \bibinfo {author} {\bibfnamefont {R.~G.}\ \bibnamefont
  {Petrosyan}},\ }\bibfield  {title} {\bibinfo {title} {Electromagnetic wave
  propagation in spatially homogeneous yet smoothly time-varying dielectric
  media},\ }\href {https://doi.org/https://doi.org/10.1016/j.jqsrt.2015.12.007}
  {\bibfield  {journal} {\bibinfo  {journal} {Journal of Quantitative
  Spectroscopy and Radiative Transfer}\ }\textbf {\bibinfo {volume} {178}},\
  \bibinfo {pages} {158} (\bibinfo {year} {2016})}\BibitemShut {NoStop}%
\bibitem [{\citenamefont {Koutserimpas}\ and\ \citenamefont
  {Fleury}(2018{\natexlab{a}})}]{Koutserimpas18}%
  \BibitemOpen
  \bibfield  {author} {\bibinfo {author} {\bibfnamefont {T.~T.}\ \bibnamefont
  {Koutserimpas}}\ and\ \bibinfo {author} {\bibfnamefont {R.}~\bibnamefont
  {Fleury}},\ }\bibfield  {title} {\bibinfo {title} {Electromagnetic waves in a
  time periodic medium with step-varying refractive index},\ }\href
  {https://doi.org/10.1109/TAP.2018.2858200} {\bibfield  {journal} {\bibinfo
  {journal} {IEEE Transactions on Antennas and Propagation}\ }\textbf {\bibinfo
  {volume} {66}},\ \bibinfo {pages} {5300} (\bibinfo {year}
  {2018}{\natexlab{a}})}\BibitemShut {NoStop}%
\bibitem [{\citenamefont {Bacot}\ \emph {et~al.}(2016)\citenamefont {Bacot},
  \citenamefont {Labousse}, \citenamefont {Eddi}, \citenamefont {Fink},\ and\
  \citenamefont {Fort}}]{Bacot}%
  \BibitemOpen
  \bibfield  {author} {\bibinfo {author} {\bibfnamefont {V.}~\bibnamefont
  {Bacot}}, \bibinfo {author} {\bibfnamefont {M.}~\bibnamefont {Labousse}},
  \bibinfo {author} {\bibfnamefont {A.}~\bibnamefont {Eddi}}, \bibinfo {author}
  {\bibfnamefont {M.}~\bibnamefont {Fink}},\ and\ \bibinfo {author}
  {\bibfnamefont {E.}~\bibnamefont {Fort}},\ }\bibfield  {title} {\bibinfo
  {title} {Time reversal and holography with spacetime transformations},\
  }\href {https://doi.org/10.1038/nphys3810} {\bibfield  {journal} {\bibinfo
  {journal} {Nature Physics}\ }\textbf {\bibinfo {volume} {12}},\ \bibinfo
  {pages} {972} (\bibinfo {year} {2016})}\BibitemShut {NoStop}%
\bibitem [{\citenamefont {Apffel}\ and\ \citenamefont {Fort}(2022)}]{Apffel}%
  \BibitemOpen
  \bibfield  {author} {\bibinfo {author} {\bibfnamefont {B.}~\bibnamefont
  {Apffel}}\ and\ \bibinfo {author} {\bibfnamefont {E.}~\bibnamefont {Fort}},\
  }\bibfield  {title} {\bibinfo {title} {Frequency conversion cascade by
  crossing multiple space and time interfaces},\ }\href
  {https://doi.org/10.1103/PhysRevLett.128.064501} {\bibfield  {journal}
  {\bibinfo  {journal} {Phys. Rev. Lett.}\ }\textbf {\bibinfo {volume} {128}},\
  \bibinfo {pages} {064501} (\bibinfo {year} {2022})}\BibitemShut {NoStop}%
\bibitem [{\citenamefont {Moussa}\ \emph {et~al.}(2023)\citenamefont {Moussa},
  \citenamefont {Xu}, \citenamefont {Yin}, \citenamefont {Galiffi},
  \citenamefont {Ra'di},\ and\ \citenamefont {Al{\`u}}}]{Moussa}%
  \BibitemOpen
  \bibfield  {author} {\bibinfo {author} {\bibfnamefont {H.}~\bibnamefont
  {Moussa}}, \bibinfo {author} {\bibfnamefont {G.}~\bibnamefont {Xu}}, \bibinfo
  {author} {\bibfnamefont {S.}~\bibnamefont {Yin}}, \bibinfo {author}
  {\bibfnamefont {E.}~\bibnamefont {Galiffi}}, \bibinfo {author} {\bibfnamefont
  {Y.}~\bibnamefont {Ra'di}},\ and\ \bibinfo {author} {\bibfnamefont
  {A.}~\bibnamefont {Al{\`u}}},\ }\bibfield  {title} {\bibinfo {title}
  {Observation of temporal reflection and broadband frequency translation at
  photonic time interfaces},\ }\href
  {https://doi.org/10.1038/s41567-023-01975-y} {\bibfield  {journal} {\bibinfo
  {journal} {Nature Physics}\ }\textbf {\bibinfo {volume} {19}},\ \bibinfo
  {pages} {863} (\bibinfo {year} {2023})}\BibitemShut {NoStop}%
\bibitem [{\citenamefont {Lustig}\ \emph
  {et~al.}(2023{\natexlab{a}})\citenamefont {Lustig}, \citenamefont {Segal},
  \citenamefont {Saha}, \citenamefont {Bordo}, \citenamefont {Chowdhury},
  \citenamefont {Sharabi}, \citenamefont {Fleischer}, \citenamefont
  {Boltasseva}, \citenamefont {Cohen}, \citenamefont {Shalaev},\ and\
  \citenamefont {Segev}}]{Lustig23}%
  \BibitemOpen
  \bibfield  {author} {\bibinfo {author} {\bibfnamefont {E.}~\bibnamefont
  {Lustig}}, \bibinfo {author} {\bibfnamefont {O.}~\bibnamefont {Segal}},
  \bibinfo {author} {\bibfnamefont {S.}~\bibnamefont {Saha}}, \bibinfo {author}
  {\bibfnamefont {E.}~\bibnamefont {Bordo}}, \bibinfo {author} {\bibfnamefont
  {S.~N.}\ \bibnamefont {Chowdhury}}, \bibinfo {author} {\bibfnamefont
  {Y.}~\bibnamefont {Sharabi}}, \bibinfo {author} {\bibfnamefont
  {A.}~\bibnamefont {Fleischer}}, \bibinfo {author} {\bibfnamefont
  {A.}~\bibnamefont {Boltasseva}}, \bibinfo {author} {\bibfnamefont
  {O.}~\bibnamefont {Cohen}}, \bibinfo {author} {\bibfnamefont {V.~M.}\
  \bibnamefont {Shalaev}},\ and\ \bibinfo {author} {\bibfnamefont
  {M.}~\bibnamefont {Segev}},\ }\bibfield  {title} {\bibinfo {title}
  {Time-refraction optics with single cycle modulation},\ }\href
  {https://doi.org/doi:10.1515/nanoph-2023-0126} {\bibfield  {journal}
  {\bibinfo  {journal} {Nanophotonics}\ }\textbf {\bibinfo {volume} {12}},\
  \bibinfo {pages} {2221} (\bibinfo {year} {2023}{\natexlab{a}})}\BibitemShut
  {NoStop}%
\bibitem [{\citenamefont {Tirole}\ \emph {et~al.}(2023)\citenamefont {Tirole},
  \citenamefont {Vezzoli}, \citenamefont {Galiffi}, \citenamefont {Robertson},
  \citenamefont {Maurice}, \citenamefont {Tilmann}, \citenamefont {Maier},
  \citenamefont {Pendry},\ and\ \citenamefont {Sapienza}}]{Tirole}%
  \BibitemOpen
  \bibfield  {author} {\bibinfo {author} {\bibfnamefont {R.}~\bibnamefont
  {Tirole}}, \bibinfo {author} {\bibfnamefont {S.}~\bibnamefont {Vezzoli}},
  \bibinfo {author} {\bibfnamefont {E.}~\bibnamefont {Galiffi}}, \bibinfo
  {author} {\bibfnamefont {I.}~\bibnamefont {Robertson}}, \bibinfo {author}
  {\bibfnamefont {D.}~\bibnamefont {Maurice}}, \bibinfo {author} {\bibfnamefont
  {B.}~\bibnamefont {Tilmann}}, \bibinfo {author} {\bibfnamefont {S.~A.}\
  \bibnamefont {Maier}}, \bibinfo {author} {\bibfnamefont {J.~B.}\ \bibnamefont
  {Pendry}},\ and\ \bibinfo {author} {\bibfnamefont {R.}~\bibnamefont
  {Sapienza}},\ }\bibfield  {title} {\bibinfo {title} {Double-slit time
  diffraction at optical frequencies},\ }\href
  {https://doi.org/10.1038/s41567-023-01993-w} {\bibfield  {journal} {\bibinfo
  {journal} {Nature Physics}\ }\textbf {\bibinfo {volume} {19}},\ \bibinfo
  {pages} {999} (\bibinfo {year} {2023})}\BibitemShut {NoStop}%
\bibitem [{\citenamefont {Lustig}\ \emph
  {et~al.}(2023{\natexlab{b}})\citenamefont {Lustig}, \citenamefont {Segal},
  \citenamefont {Saha}, \citenamefont {Fruhling}, \citenamefont {Shalaev},
  \citenamefont {Boltasseva},\ and\ \citenamefont {Segev}}]{Lustig232}%
  \BibitemOpen
  \bibfield  {author} {\bibinfo {author} {\bibfnamefont {E.}~\bibnamefont
  {Lustig}}, \bibinfo {author} {\bibfnamefont {O.}~\bibnamefont {Segal}},
  \bibinfo {author} {\bibfnamefont {S.}~\bibnamefont {Saha}}, \bibinfo {author}
  {\bibfnamefont {C.}~\bibnamefont {Fruhling}}, \bibinfo {author}
  {\bibfnamefont {V.~M.}\ \bibnamefont {Shalaev}}, \bibinfo {author}
  {\bibfnamefont {A.}~\bibnamefont {Boltasseva}},\ and\ \bibinfo {author}
  {\bibfnamefont {M.}~\bibnamefont {Segev}},\ }\bibfield  {title} {\bibinfo
  {title} {Photonic time-crystals - fundamental concepts [invited]},\ }\href
  {https://doi.org/10.1364/OE.479367} {\bibfield  {journal} {\bibinfo
  {journal} {Opt. Express}\ }\textbf {\bibinfo {volume} {31}},\ \bibinfo
  {pages} {9165} (\bibinfo {year} {2023}{\natexlab{b}})}\BibitemShut {NoStop}%
\bibitem [{\citenamefont {Wang}\ \emph {et~al.}(2023)\citenamefont {Wang},
  \citenamefont {Mirmoosa}, \citenamefont {Asadchy}, \citenamefont {Rockstuhl},
  \citenamefont {Fan},\ and\ \citenamefont {Tretyakov}}]{Wang23}%
  \BibitemOpen
  \bibfield  {author} {\bibinfo {author} {\bibfnamefont {X.}~\bibnamefont
  {Wang}}, \bibinfo {author} {\bibfnamefont {M.~S.}\ \bibnamefont {Mirmoosa}},
  \bibinfo {author} {\bibfnamefont {V.~S.}\ \bibnamefont {Asadchy}}, \bibinfo
  {author} {\bibfnamefont {C.}~\bibnamefont {Rockstuhl}}, \bibinfo {author}
  {\bibfnamefont {S.}~\bibnamefont {Fan}},\ and\ \bibinfo {author}
  {\bibfnamefont {S.~A.}\ \bibnamefont {Tretyakov}},\ }\bibfield  {title}
  {\bibinfo {title} {Metasurface-based realization of photonic time crystals},\
  }\href {https://doi.org/10.1126/sciadv.adg7541} {\bibfield  {journal}
  {\bibinfo  {journal} {Science Advances}\ }\textbf {\bibinfo {volume} {9}},\
  \bibinfo {pages} {eadg7541} (\bibinfo {year} {2023})}\BibitemShut {NoStop}%
\bibitem [{\citenamefont {Li}\ and\ \citenamefont {Reichl}(1999)}]{Reichl99}%
  \BibitemOpen
  \bibfield  {author} {\bibinfo {author} {\bibfnamefont {W.}~\bibnamefont
  {Li}}\ and\ \bibinfo {author} {\bibfnamefont {L.~E.}\ \bibnamefont
  {Reichl}},\ }\bibfield  {title} {\bibinfo {title} {Floquet scattering through
  a time-periodic potential},\ }\href
  {https://doi.org/10.1103/PhysRevB.60.15732} {\bibfield  {journal} {\bibinfo
  {journal} {Phys. Rev. B}\ }\textbf {\bibinfo {volume} {60}},\ \bibinfo
  {pages} {15732} (\bibinfo {year} {1999})}\BibitemShut {NoStop}%
\bibitem [{\citenamefont {Martinez}\ and\ \citenamefont
  {Reichl}(2001)}]{Reichl01}%
  \BibitemOpen
  \bibfield  {author} {\bibinfo {author} {\bibfnamefont {D.~F.}\ \bibnamefont
  {Martinez}}\ and\ \bibinfo {author} {\bibfnamefont {L.~E.}\ \bibnamefont
  {Reichl}},\ }\bibfield  {title} {\bibinfo {title} {Transmission properties of
  the oscillating \ensuremath{\delta}-function potential},\ }\href
  {https://doi.org/10.1103/PhysRevB.64.245315} {\bibfield  {journal} {\bibinfo
  {journal} {Phys. Rev. B}\ }\textbf {\bibinfo {volume} {64}},\ \bibinfo
  {pages} {245315} (\bibinfo {year} {2001})}\BibitemShut {NoStop}%
\bibitem [{\citenamefont {Emmanouilidou}\ and\ \citenamefont
  {Reichl}(2002)}]{Reichl02}%
  \BibitemOpen
  \bibfield  {author} {\bibinfo {author} {\bibfnamefont {A.}~\bibnamefont
  {Emmanouilidou}}\ and\ \bibinfo {author} {\bibfnamefont {L.~E.}\ \bibnamefont
  {Reichl}},\ }\bibfield  {title} {\bibinfo {title} {Floquet scattering and
  classical-quantum correspondence in strong time-periodic fields},\ }\href
  {https://doi.org/10.1103/PhysRevA.65.033405} {\bibfield  {journal} {\bibinfo
  {journal} {Phys. Rev. A}\ }\textbf {\bibinfo {volume} {65}},\ \bibinfo
  {pages} {033405} (\bibinfo {year} {2002})}\BibitemShut {NoStop}%
\bibitem [{\citenamefont {Chegnizadeh}\ \emph {et~al.}(2018)\citenamefont
  {Chegnizadeh}, \citenamefont {Mehrany},\ and\ \citenamefont
  {Memarian}}]{Chegnizadeh}%
  \BibitemOpen
  \bibfield  {author} {\bibinfo {author} {\bibfnamefont {M.}~\bibnamefont
  {Chegnizadeh}}, \bibinfo {author} {\bibfnamefont {K.}~\bibnamefont
  {Mehrany}},\ and\ \bibinfo {author} {\bibfnamefont {M.}~\bibnamefont
  {Memarian}},\ }\bibfield  {title} {\bibinfo {title} {General solution to wave
  propagation in media undergoing arbitrary transient or periodic temporal
  variations of permittivity},\ }\href
  {https://doi.org/10.1364/JOSAB.35.002923} {\bibfield  {journal} {\bibinfo
  {journal} {J. Opt. Soc. Am. B}\ }\textbf {\bibinfo {volume} {35}},\ \bibinfo
  {pages} {2923} (\bibinfo {year} {2018})}\BibitemShut {NoStop}%
\bibitem [{\citenamefont {Li}\ \emph {et~al.}(2021{\natexlab{b}})\citenamefont
  {Li}, \citenamefont {Yin}, \citenamefont {Galiffi},\ and\ \citenamefont
  {Al\`u}}]{Li}%
  \BibitemOpen
  \bibfield  {author} {\bibinfo {author} {\bibfnamefont {H.}~\bibnamefont
  {Li}}, \bibinfo {author} {\bibfnamefont {S.}~\bibnamefont {Yin}}, \bibinfo
  {author} {\bibfnamefont {E.}~\bibnamefont {Galiffi}},\ and\ \bibinfo {author}
  {\bibfnamefont {A.}~\bibnamefont {Al\`u}},\ }\bibfield  {title} {\bibinfo
  {title} {Temporal parity-time symmetry for extreme energy transformations},\
  }\href {https://doi.org/10.1103/PhysRevLett.127.153903} {\bibfield  {journal}
  {\bibinfo  {journal} {Phys. Rev. Lett.}\ }\textbf {\bibinfo {volume} {127}},\
  \bibinfo {pages} {153903} (\bibinfo {year} {2021}{\natexlab{b}})}\BibitemShut
  {NoStop}%
\bibitem [{\citenamefont {Yin}\ \emph {et~al.}(2023)\citenamefont {Yin},
  \citenamefont {Galiffi}, \citenamefont {Xu},\ and\ \citenamefont
  {Alù}}]{Yin}%
  \BibitemOpen
  \bibfield  {author} {\bibinfo {author} {\bibfnamefont {S.}~\bibnamefont
  {Yin}}, \bibinfo {author} {\bibfnamefont {E.}~\bibnamefont {Galiffi}},
  \bibinfo {author} {\bibfnamefont {G.}~\bibnamefont {Xu}},\ and\ \bibinfo
  {author} {\bibfnamefont {A.}~\bibnamefont {Alù}},\ }\bibfield  {title}
  {\bibinfo {title} {Scattering at temporal interfaces: An overview from an
  antennas and propagation engineering perspective.},\ }\href
  {https://doi.org/10.1109/MAP.2023.3254486} {\bibfield  {journal} {\bibinfo
  {journal} {IEEE Antennas and Propagation Magazine}\ ,\ \bibinfo {pages} {2}}
  (\bibinfo {year} {2023})}\BibitemShut {NoStop}%
\bibitem [{\citenamefont {Barnett}(2010)}]{PhysRevLett.104.070401}%
  \BibitemOpen
  \bibfield  {author} {\bibinfo {author} {\bibfnamefont {S.~M.}\ \bibnamefont
  {Barnett}},\ }\bibfield  {title} {\bibinfo {title} {Resolution of the
  {A}braham-{M}inkowski dilemma},\ }\href
  {https://doi.org/10.1103/PhysRevLett.104.070401} {\bibfield  {journal}
  {\bibinfo  {journal} {Phys. Rev. Lett.}\ }\textbf {\bibinfo {volume} {104}},\
  \bibinfo {pages} {070401} (\bibinfo {year} {2010})}\BibitemShut {NoStop}%
\bibitem [{\citenamefont {Zurita-S\'anchez}\ \emph {et~al.}(2009)\citenamefont
  {Zurita-S\'anchez}, \citenamefont {Halevi},\ and\ \citenamefont
  {Cervantes-Gonz\'alez}}]{Zurita}%
  \BibitemOpen
  \bibfield  {author} {\bibinfo {author} {\bibfnamefont {J.~R.}\ \bibnamefont
  {Zurita-S\'anchez}}, \bibinfo {author} {\bibfnamefont {P.}~\bibnamefont
  {Halevi}},\ and\ \bibinfo {author} {\bibfnamefont {J.~C.}\ \bibnamefont
  {Cervantes-Gonz\'alez}},\ }\bibfield  {title} {\bibinfo {title} {Reflection
  and transmission of a wave incident on a slab with a time-periodic dielectric
  function},\ }\href {https://doi.org/10.1103/PhysRevA.79.053821} {\bibfield
  {journal} {\bibinfo  {journal} {Phys. Rev. A}\ }\textbf {\bibinfo {volume}
  {79}},\ \bibinfo {pages} {053821} (\bibinfo {year} {2009})}\BibitemShut
  {NoStop}%
\bibitem [{\citenamefont {Mart\'{\i}nez-Romero}\ \emph
  {et~al.}(2016)\citenamefont {Mart\'{\i}nez-Romero}, \citenamefont
  {Becerra-Fuentes},\ and\ \citenamefont {Halevi}}]{Martinez16}%
  \BibitemOpen
  \bibfield  {author} {\bibinfo {author} {\bibfnamefont {J.~S.}\ \bibnamefont
  {Mart\'{\i}nez-Romero}}, \bibinfo {author} {\bibfnamefont {O.~M.}\
  \bibnamefont {Becerra-Fuentes}},\ and\ \bibinfo {author} {\bibfnamefont
  {P.}~\bibnamefont {Halevi}},\ }\bibfield  {title} {\bibinfo {title} {Temporal
  photonic crystals with modulations of both permittivity and permeability},\
  }\href {https://doi.org/10.1103/PhysRevA.93.063813} {\bibfield  {journal}
  {\bibinfo  {journal} {Phys. Rev. A}\ }\textbf {\bibinfo {volume} {93}},\
  \bibinfo {pages} {063813} (\bibinfo {year} {2016})}\BibitemShut {NoStop}%
\bibitem [{\citenamefont {Li}\ \emph {et~al.}(2018)\citenamefont {Li},
  \citenamefont {Shapiro},\ and\ \citenamefont {Kottos}}]{Huanan}%
  \BibitemOpen
  \bibfield  {author} {\bibinfo {author} {\bibfnamefont {H.}~\bibnamefont
  {Li}}, \bibinfo {author} {\bibfnamefont {B.}~\bibnamefont {Shapiro}},\ and\
  \bibinfo {author} {\bibfnamefont {T.}~\bibnamefont {Kottos}},\ }\bibfield
  {title} {\bibinfo {title} {Floquet scattering theory based on effective
  {H}amiltonians of driven systems},\ }\href
  {https://doi.org/10.1103/PhysRevB.98.121101} {\bibfield  {journal} {\bibinfo
  {journal} {Phys. Rev. B}\ }\textbf {\bibinfo {volume} {98}},\ \bibinfo
  {pages} {121101} (\bibinfo {year} {2018})}\BibitemShut {NoStop}%
\bibitem [{\citenamefont {Mart\'{\i}nez-Romero}\ and\ \citenamefont
  {Halevi}(2018)}]{Martinez18}%
  \BibitemOpen
  \bibfield  {author} {\bibinfo {author} {\bibfnamefont {J.~S.}\ \bibnamefont
  {Mart\'{\i}nez-Romero}}\ and\ \bibinfo {author} {\bibfnamefont
  {P.}~\bibnamefont {Halevi}},\ }\bibfield  {title} {\bibinfo {title}
  {Parametric resonances in a temporal photonic crystal slab},\ }\href
  {https://doi.org/10.1103/PhysRevA.98.053852} {\bibfield  {journal} {\bibinfo
  {journal} {Phys. Rev. A}\ }\textbf {\bibinfo {volume} {98}},\ \bibinfo
  {pages} {053852} (\bibinfo {year} {2018})}\BibitemShut {NoStop}%
\bibitem [{\citenamefont {Pantazopoulos}\ and\ \citenamefont
  {Stefanou}(2019)}]{Pantazopoulos}%
  \BibitemOpen
  \bibfield  {author} {\bibinfo {author} {\bibfnamefont {P.~A.}\ \bibnamefont
  {Pantazopoulos}}\ and\ \bibinfo {author} {\bibfnamefont {N.}~\bibnamefont
  {Stefanou}},\ }\bibfield  {title} {\bibinfo {title} {Layered optomagnonic
  structures: Time {F}loquet scattering-matrix approach},\ }\href
  {https://doi.org/10.1103/PhysRevB.99.144415} {\bibfield  {journal} {\bibinfo
  {journal} {Phys. Rev. B}\ }\textbf {\bibinfo {volume} {99}},\ \bibinfo
  {pages} {144415} (\bibinfo {year} {2019})}\BibitemShut {NoStop}%
\bibitem [{\citenamefont {Stefanou}\ \emph {et~al.}(2021)\citenamefont
  {Stefanou}, \citenamefont {Pantazopoulos},\ and\ \citenamefont
  {Stefanou}}]{Stefanou}%
  \BibitemOpen
  \bibfield  {author} {\bibinfo {author} {\bibfnamefont {I.}~\bibnamefont
  {Stefanou}}, \bibinfo {author} {\bibfnamefont {P.~A.}\ \bibnamefont
  {Pantazopoulos}},\ and\ \bibinfo {author} {\bibfnamefont {N.}~\bibnamefont
  {Stefanou}},\ }\bibfield  {title} {\bibinfo {title} {Light scattering by a
  spherical particle with a time-periodicrefractive index},\ }\href
  {https://doi.org/10.1364/JOSAB.408559} {\bibfield  {journal} {\bibinfo
  {journal} {J. Opt. Soc. Am. B}\ }\textbf {\bibinfo {volume} {38}},\ \bibinfo
  {pages} {407} (\bibinfo {year} {2021})}\BibitemShut {NoStop}%
\bibitem [{\citenamefont {Stefanou}\ \emph {et~al.}(2023)\citenamefont
  {Stefanou}, \citenamefont {Stefanou}, \citenamefont {Almpanis}, \citenamefont
  {Papanikolaou}, \citenamefont {Garg},\ and\ \citenamefont
  {Rockstuhl}}]{Stefanou23}%
  \BibitemOpen
  \bibfield  {author} {\bibinfo {author} {\bibfnamefont {N.}~\bibnamefont
  {Stefanou}}, \bibinfo {author} {\bibfnamefont {I.}~\bibnamefont {Stefanou}},
  \bibinfo {author} {\bibfnamefont {E.}~\bibnamefont {Almpanis}}, \bibinfo
  {author} {\bibfnamefont {N.}~\bibnamefont {Papanikolaou}}, \bibinfo {author}
  {\bibfnamefont {P.}~\bibnamefont {Garg}},\ and\ \bibinfo {author}
  {\bibfnamefont {C.}~\bibnamefont {Rockstuhl}},\ }\bibfield  {title} {\bibinfo
  {title} {Light scattering by a periodically time-modulated object of
  arbitrary shape: {T}he extended boundary condition method},\ }\href
  {https://doi.org/10.1364/JOSAB.502171} {\bibfield  {journal} {\bibinfo
  {journal} {J. Opt. Soc. Am. B}\ }\textbf {\bibinfo {volume} {40}},\ \bibinfo
  {pages} {2842} (\bibinfo {year} {2023})}\BibitemShut {NoStop}%
\bibitem [{\citenamefont {Ptitcyn}\ \emph {et~al.}(2023)\citenamefont
  {Ptitcyn}, \citenamefont {Lamprianidis}, \citenamefont {Karamanos},
  \citenamefont {Asadchy}, \citenamefont {Alaee}, \citenamefont {Müller},
  \citenamefont {Albooyeh}, \citenamefont {Mirmoosa}, \citenamefont {Fan},
  \citenamefont {Tretyakov},\ and\ \citenamefont {Rockstuhl}}]{Ptitcyn}%
  \BibitemOpen
  \bibfield  {author} {\bibinfo {author} {\bibfnamefont {G.}~\bibnamefont
  {Ptitcyn}}, \bibinfo {author} {\bibfnamefont {A.}~\bibnamefont
  {Lamprianidis}}, \bibinfo {author} {\bibfnamefont {T.}~\bibnamefont
  {Karamanos}}, \bibinfo {author} {\bibfnamefont {V.}~\bibnamefont {Asadchy}},
  \bibinfo {author} {\bibfnamefont {R.}~\bibnamefont {Alaee}}, \bibinfo
  {author} {\bibfnamefont {M.}~\bibnamefont {Müller}}, \bibinfo {author}
  {\bibfnamefont {M.}~\bibnamefont {Albooyeh}}, \bibinfo {author}
  {\bibfnamefont {M.~S.}\ \bibnamefont {Mirmoosa}}, \bibinfo {author}
  {\bibfnamefont {S.}~\bibnamefont {Fan}}, \bibinfo {author} {\bibfnamefont
  {S.}~\bibnamefont {Tretyakov}},\ and\ \bibinfo {author} {\bibfnamefont
  {C.}~\bibnamefont {Rockstuhl}},\ }\bibfield  {title} {\bibinfo {title}
  {Floquet–{M}ie theory for time-varying dispersive spheres},\ }\href
  {https://doi.org/https://doi.org/10.1002/lpor.202100683} {\bibfield
  {journal} {\bibinfo  {journal} {Laser \& Photonics Reviews}\ }\textbf
  {\bibinfo {volume} {17}},\ \bibinfo {pages} {2100683} (\bibinfo {year}
  {2023})}\BibitemShut {NoStop}%
\bibitem [{\citenamefont {Zurita-S\'anchez}\ and\ \citenamefont
  {Halevi}(2010)}]{Zurita10}%
  \BibitemOpen
  \bibfield  {author} {\bibinfo {author} {\bibfnamefont {J.~R.}\ \bibnamefont
  {Zurita-S\'anchez}}\ and\ \bibinfo {author} {\bibfnamefont {P.}~\bibnamefont
  {Halevi}},\ }\bibfield  {title} {\bibinfo {title} {Resonances in the optical
  response of a slab with time-periodic dielectric function
  $\ensuremath{\epsilon}(t)$},\ }\href
  {https://doi.org/10.1103/PhysRevA.81.053834} {\bibfield  {journal} {\bibinfo
  {journal} {Phys. Rev. A}\ }\textbf {\bibinfo {volume} {81}},\ \bibinfo
  {pages} {053834} (\bibinfo {year} {2010})}\BibitemShut {NoStop}%
\bibitem [{\citenamefont {Brizard}\ and\ \citenamefont
  {Kaufman}(1995)}]{Brizard}%
  \BibitemOpen
  \bibfield  {author} {\bibinfo {author} {\bibfnamefont {A.~J.}\ \bibnamefont
  {Brizard}}\ and\ \bibinfo {author} {\bibfnamefont {A.~N.}\ \bibnamefont
  {Kaufman}},\ }\bibfield  {title} {\bibinfo {title} {Local {M}anley-{R}owe
  relations for noneikonal wave fields},\ }\href
  {https://doi.org/10.1103/PhysRevLett.74.4567} {\bibfield  {journal} {\bibinfo
   {journal} {Phys. Rev. Lett.}\ }\textbf {\bibinfo {volume} {74}},\ \bibinfo
  {pages} {4567} (\bibinfo {year} {1995})}\BibitemShut {NoStop}%
\bibitem [{\citenamefont {Bellotti}\ \emph {et~al.}(1997)\citenamefont
  {Bellotti}, \citenamefont {Bornatici},\ and\ \citenamefont
  {Engelmann}}]{Bellotti}%
  \BibitemOpen
  \bibfield  {author} {\bibinfo {author} {\bibfnamefont {U.}~\bibnamefont
  {Bellotti}}, \bibinfo {author} {\bibfnamefont {M.}~\bibnamefont
  {Bornatici}},\ and\ \bibinfo {author} {\bibfnamefont {F.}~\bibnamefont
  {Engelmann}},\ }\bibfield  {title} {\bibinfo {title} {Radiative energy
  transfer in anisotropic, spatially dispersive, weakly inhomogeneous and
  dissipative media with embedded sources},\ }\href
  {https://doi.org/10.1007/BF02897900} {\bibfield  {journal} {\bibinfo
  {journal} {La Rivista del Nuovo Cimento (1978-1999)}\ }\textbf {\bibinfo
  {volume} {20}},\ \bibinfo {pages} {1} (\bibinfo {year} {1997})}\BibitemShut
  {NoStop}%
\bibitem [{\citenamefont {Buddhiraju}\ \emph {et~al.}(2021)\citenamefont
  {Buddhiraju}, \citenamefont {Dutt}, \citenamefont {Minkov}, \citenamefont
  {Williamson},\ and\ \citenamefont {Fan}}]{Buddhiraju}%
  \BibitemOpen
  \bibfield  {author} {\bibinfo {author} {\bibfnamefont {S.}~\bibnamefont
  {Buddhiraju}}, \bibinfo {author} {\bibfnamefont {A.}~\bibnamefont {Dutt}},
  \bibinfo {author} {\bibfnamefont {M.}~\bibnamefont {Minkov}}, \bibinfo
  {author} {\bibfnamefont {I.~A.~D.}\ \bibnamefont {Williamson}},\ and\
  \bibinfo {author} {\bibfnamefont {S.}~\bibnamefont {Fan}},\ }\bibfield
  {title} {\bibinfo {title} {Arbitrary linear transformations for photons in
  the frequency synthetic dimension},\ }\href
  {https://doi.org/10.1038/s41467-021-22670-7} {\bibfield  {journal} {\bibinfo
  {journal} {Nature Communications}\ }\textbf {\bibinfo {volume} {12}},\
  \bibinfo {pages} {2401} (\bibinfo {year} {2021})}\BibitemShut {NoStop}%
\bibitem [{\citenamefont {Fan}\ \emph {et~al.}(2022)\citenamefont {Fan},
  \citenamefont {Zhao}, \citenamefont {Wang}, \citenamefont {Dutt},
  \citenamefont {Wang}, \citenamefont {Buddhiraju}, \citenamefont {Wojcik},\
  and\ \citenamefont {Fan}}]{Fan22}%
  \BibitemOpen
  \bibfield  {author} {\bibinfo {author} {\bibfnamefont {L.}~\bibnamefont
  {Fan}}, \bibinfo {author} {\bibfnamefont {Z.}~\bibnamefont {Zhao}}, \bibinfo
  {author} {\bibfnamefont {K.}~\bibnamefont {Wang}}, \bibinfo {author}
  {\bibfnamefont {A.}~\bibnamefont {Dutt}}, \bibinfo {author} {\bibfnamefont
  {J.}~\bibnamefont {Wang}}, \bibinfo {author} {\bibfnamefont {S.}~\bibnamefont
  {Buddhiraju}}, \bibinfo {author} {\bibfnamefont {C.~C.}\ \bibnamefont
  {Wojcik}},\ and\ \bibinfo {author} {\bibfnamefont {S.}~\bibnamefont {Fan}},\
  }\bibfield  {title} {\bibinfo {title} {Multidimensional convolution operation
  with synthetic frequency dimensions in photonics},\ }\href
  {https://doi.org/10.1103/PhysRevApplied.18.034088} {\bibfield  {journal}
  {\bibinfo  {journal} {Phys. Rev. Appl.}\ }\textbf {\bibinfo {volume} {18}},\
  \bibinfo {pages} {034088} (\bibinfo {year} {2022})}\BibitemShut {NoStop}%
\bibitem [{\citenamefont {Joachain}\ \emph {et~al.}(2011)\citenamefont
  {Joachain}, \citenamefont {Kylstra},\ and\ \citenamefont
  {Potvliege}}]{Joachain}%
  \BibitemOpen
  \bibfield  {author} {\bibinfo {author} {\bibfnamefont {C.~J.}\ \bibnamefont
  {Joachain}}, \bibinfo {author} {\bibfnamefont {N.~J.}\ \bibnamefont
  {Kylstra}},\ and\ \bibinfo {author} {\bibfnamefont {R.~M.}\ \bibnamefont
  {Potvliege}},\ }\href@noop {} {\emph {\bibinfo {title} {Atoms in Intense
  Laser Fields}}}\ (\bibinfo  {publisher} {Cambridge University Press},\
  \bibinfo {year} {2011})\BibitemShut {NoStop}%
\bibitem [{\citenamefont {Leonhardt}(2003)}]{Leonhardt}%
  \BibitemOpen
  \bibfield  {author} {\bibinfo {author} {\bibfnamefont {U.}~\bibnamefont
  {Leonhardt}},\ }\bibfield  {title} {\bibinfo {title} {Quantum physics of
  simple optical instruments},\ }\href
  {https://doi.org/10.1088/0034-4885/66/7/203} {\bibfield  {journal} {\bibinfo
  {journal} {Reports on Progress in Physics}\ }\textbf {\bibinfo {volume}
  {66}},\ \bibinfo {pages} {1207} (\bibinfo {year} {2003})}\BibitemShut
  {NoStop}%
\bibitem [{\citenamefont {Birrell}\ and\ \citenamefont
  {Davies}(1982)}]{Birrell}%
  \BibitemOpen
  \bibfield  {author} {\bibinfo {author} {\bibfnamefont {N.~D.}\ \bibnamefont
  {Birrell}}\ and\ \bibinfo {author} {\bibfnamefont {P.~C.~W.}\ \bibnamefont
  {Davies}},\ }\href {https://doi.org/10.1017/CBO9780511622632} {\emph
  {\bibinfo {title} {Quantum Fields in Curved Space}}},\ Cambridge Monographs
  on Mathematical Physics\ (\bibinfo  {publisher} {Cambridge University
  Press},\ \bibinfo {year} {1982})\BibitemShut {NoStop}%
\bibitem [{\citenamefont {Cohen-Tannoudji}\ \emph {et~al.}(2004)\citenamefont
  {Cohen-Tannoudji}, \citenamefont {Dupont-Roc},\ and\ \citenamefont
  {Grynberg}}]{Cohen}%
  \BibitemOpen
  \bibfield  {author} {\bibinfo {author} {\bibfnamefont {C.}~\bibnamefont
  {Cohen-Tannoudji}}, \bibinfo {author} {\bibfnamefont {J.}~\bibnamefont
  {Dupont-Roc}},\ and\ \bibinfo {author} {\bibfnamefont {G.}~\bibnamefont
  {Grynberg}},\ }\href@noop {} {\emph {\bibinfo {title} {Photons and atoms:
  introduction to quantum electrodynamics}}}\ (\bibinfo  {publisher}
  {Wiley-VCH},\ \bibinfo {address} {Weinheim},\ \bibinfo {year}
  {2004})\BibitemShut {NoStop}%
\bibitem [{\citenamefont {New}(2011)}]{New}%
  \BibitemOpen
  \bibfield  {author} {\bibinfo {author} {\bibfnamefont {G.}~\bibnamefont
  {New}},\ }\href {https://doi.org/10.1017/CBO9780511975851} {\emph {\bibinfo
  {title} {Introduction to Nonlinear Optics}}}\ (\bibinfo  {publisher}
  {Cambridge University Press},\ \bibinfo {year} {2011})\BibitemShut {NoStop}%
\bibitem [{\citenamefont {Xiao}\ \emph {et~al.}(2011)\citenamefont {Xiao},
  \citenamefont {Agrawal},\ and\ \citenamefont {Maywar}}]{Xiao11}%
  \BibitemOpen
  \bibfield  {author} {\bibinfo {author} {\bibfnamefont {Y.}~\bibnamefont
  {Xiao}}, \bibinfo {author} {\bibfnamefont {G.~P.}\ \bibnamefont {Agrawal}},\
  and\ \bibinfo {author} {\bibfnamefont {D.~N.}\ \bibnamefont {Maywar}},\
  }\bibfield  {title} {\bibinfo {title} {Spectral and temporal changes of
  optical pulses propagating through time-varying linear media},\ }\href
  {https://doi.org/10.1364/OL.36.000505} {\bibfield  {journal} {\bibinfo
  {journal} {Opt. Lett.}\ }\textbf {\bibinfo {volume} {36}},\ \bibinfo {pages}
  {505} (\bibinfo {year} {2011})}\BibitemShut {NoStop}%
\bibitem [{\citenamefont {Pendry}(2023)}]{Pendry23}%
  \BibitemOpen
  \bibfield  {author} {\bibinfo {author} {\bibfnamefont {J.~B.}\ \bibnamefont
  {Pendry}},\ }\bibfield  {title} {\bibinfo {title} {Photon number conservation
  in time dependent systems},\ }\href {https://doi.org/10.1364/OE.476961}
  {\bibfield  {journal} {\bibinfo  {journal} {Opt. Express}\ }\textbf {\bibinfo
  {volume} {31}},\ \bibinfo {pages} {452} (\bibinfo {year} {2023})}\BibitemShut
  {NoStop}%
\bibitem [{\citenamefont {Mostafazadeh}(2004)}]{Mostafazadeh}%
  \BibitemOpen
  \bibfield  {author} {\bibinfo {author} {\bibfnamefont {A.}~\bibnamefont
  {Mostafazadeh}},\ }\bibfield  {title} {\bibinfo {title} {Pseudounitary
  operators and pseudounitary quantum dynamics},\ }\href
  {https://doi.org/10.1063/1.1646448} {\bibfield  {journal} {\bibinfo
  {journal} {Journal of Mathematical Physics}\ }\textbf {\bibinfo {volume}
  {45}},\ \bibinfo {pages} {932–946} (\bibinfo {year} {2004})}\BibitemShut
  {NoStop}%
\bibitem [{\citenamefont {Serra}\ \emph {et~al.}(2024)\citenamefont {Serra},
  \citenamefont {Galiffi}, \citenamefont {Huidobro}, \citenamefont {Pendry},\
  and\ \citenamefont {Silveirinha}}]{Serra}%
  \BibitemOpen
  \bibfield  {author} {\bibinfo {author} {\bibfnamefont {J.~C.}\ \bibnamefont
  {Serra}}, \bibinfo {author} {\bibfnamefont {E.}~\bibnamefont {Galiffi}},
  \bibinfo {author} {\bibfnamefont {P.~A.}\ \bibnamefont {Huidobro}}, \bibinfo
  {author} {\bibfnamefont {J.~B.}\ \bibnamefont {Pendry}},\ and\ \bibinfo
  {author} {\bibfnamefont {M.~G.}\ \bibnamefont {Silveirinha}},\ }\bibfield
  {title} {\bibinfo {title} {Particle-hole instabilities in photonic
  time-varying systems},\ }\href {https://doi.org/10.1364/OME.521571}
  {\bibfield  {journal} {\bibinfo  {journal} {Opt. Mater. Express}\ }\textbf
  {\bibinfo {volume} {14}},\ \bibinfo {pages} {1459} (\bibinfo {year}
  {2024})}\BibitemShut {NoStop}%
\bibitem [{\citenamefont {Jalas}\ \emph {et~al.}(2013)\citenamefont {Jalas},
  \citenamefont {Petrov}, \citenamefont {Eich}, \citenamefont {Freude},
  \citenamefont {Fan}, \citenamefont {Yu}, \citenamefont {Baets}, \citenamefont
  {Popović}, \citenamefont {Melloni}, \citenamefont {Joannopoulos},
  \citenamefont {Vanwolleghem}, \citenamefont {Doerr},\ and\ \citenamefont
  {Renner}}]{Jalas}%
  \BibitemOpen
  \bibfield  {author} {\bibinfo {author} {\bibfnamefont {D.}~\bibnamefont
  {Jalas}}, \bibinfo {author} {\bibfnamefont {A.}~\bibnamefont {Petrov}},
  \bibinfo {author} {\bibfnamefont {M.}~\bibnamefont {Eich}}, \bibinfo {author}
  {\bibfnamefont {W.}~\bibnamefont {Freude}}, \bibinfo {author} {\bibfnamefont
  {S.}~\bibnamefont {Fan}}, \bibinfo {author} {\bibfnamefont {Z.}~\bibnamefont
  {Yu}}, \bibinfo {author} {\bibfnamefont {R.}~\bibnamefont {Baets}}, \bibinfo
  {author} {\bibfnamefont {M.}~\bibnamefont {Popović}}, \bibinfo {author}
  {\bibfnamefont {A.}~\bibnamefont {Melloni}}, \bibinfo {author} {\bibfnamefont
  {J.~D.}\ \bibnamefont {Joannopoulos}}, \bibinfo {author} {\bibfnamefont
  {M.}~\bibnamefont {Vanwolleghem}}, \bibinfo {author} {\bibfnamefont {C.~R.}\
  \bibnamefont {Doerr}},\ and\ \bibinfo {author} {\bibfnamefont
  {H.}~\bibnamefont {Renner}},\ }\bibfield  {title} {\bibinfo {title} {What is
  — and what is not — an optical isolator},\ }\href
  {https://doi.org/10.1038/nphoton.2013.185} {\bibfield  {journal} {\bibinfo
  {journal} {Nature Photonics}\ }\textbf {\bibinfo {volume} {7}},\ \bibinfo
  {pages} {579} (\bibinfo {year} {2013})}\BibitemShut {NoStop}%
\bibitem [{\citenamefont {Sounas}\ and\ \citenamefont
  {Al{\`u}}(2017)}]{Sounas}%
  \BibitemOpen
  \bibfield  {author} {\bibinfo {author} {\bibfnamefont {D.~L.}\ \bibnamefont
  {Sounas}}\ and\ \bibinfo {author} {\bibfnamefont {A.}~\bibnamefont
  {Al{\`u}}},\ }\bibfield  {title} {\bibinfo {title} {Non-reciprocal photonics
  based on time modulation},\ }\href
  {https://doi.org/10.1038/s41566-017-0051-x} {\bibfield  {journal} {\bibinfo
  {journal} {Nature Photonics}\ }\textbf {\bibinfo {volume} {11}},\ \bibinfo
  {pages} {774} (\bibinfo {year} {2017})}\BibitemShut {NoStop}%
\bibitem [{\citenamefont {Koutserimpas}\ and\ \citenamefont
  {Fleury}(2018{\natexlab{b}})}]{Koutserimpas182}%
  \BibitemOpen
  \bibfield  {author} {\bibinfo {author} {\bibfnamefont {T.~T.}\ \bibnamefont
  {Koutserimpas}}\ and\ \bibinfo {author} {\bibfnamefont {R.}~\bibnamefont
  {Fleury}},\ }\bibfield  {title} {\bibinfo {title} {Nonreciprocal gain in
  non-{H}ermitian time-{F}loquet systems},\ }\href
  {https://doi.org/10.1103/PhysRevLett.120.087401} {\bibfield  {journal}
  {\bibinfo  {journal} {Phys. Rev. Lett.}\ }\textbf {\bibinfo {volume} {120}},\
  \bibinfo {pages} {087401} (\bibinfo {year} {2018}{\natexlab{b}})}\BibitemShut
  {NoStop}%
\bibitem [{\citenamefont {Wang}\ \emph {et~al.}(2020)\citenamefont {Wang},
  \citenamefont {Ptitcyn}, \citenamefont {Asadchy}, \citenamefont
  {D\'{\i}az-Rubio}, \citenamefont {Mirmoosa}, \citenamefont {Fan},\ and\
  \citenamefont {Tretyakov}}]{Wang20}%
  \BibitemOpen
  \bibfield  {author} {\bibinfo {author} {\bibfnamefont {X.}~\bibnamefont
  {Wang}}, \bibinfo {author} {\bibfnamefont {G.}~\bibnamefont {Ptitcyn}},
  \bibinfo {author} {\bibfnamefont {V.~S.}\ \bibnamefont {Asadchy}}, \bibinfo
  {author} {\bibfnamefont {A.}~\bibnamefont {D\'{\i}az-Rubio}}, \bibinfo
  {author} {\bibfnamefont {M.~S.}\ \bibnamefont {Mirmoosa}}, \bibinfo {author}
  {\bibfnamefont {S.}~\bibnamefont {Fan}},\ and\ \bibinfo {author}
  {\bibfnamefont {S.~A.}\ \bibnamefont {Tretyakov}},\ }\bibfield  {title}
  {\bibinfo {title} {Nonreciprocity in bianisotropic systems with uniform time
  modulation},\ }\href {https://doi.org/10.1103/PhysRevLett.125.266102}
  {\bibfield  {journal} {\bibinfo  {journal} {Phys. Rev. Lett.}\ }\textbf
  {\bibinfo {volume} {125}},\ \bibinfo {pages} {266102} (\bibinfo {year}
  {2020})}\BibitemShut {NoStop}%
\bibitem [{\citenamefont {Wang}\ \emph {et~al.}(2021)\citenamefont {Wang},
  \citenamefont {Herrmann}, \citenamefont {Witmer}, \citenamefont
  {Safavi-Naeini},\ and\ \citenamefont {Fan}}]{Wang21}%
  \BibitemOpen
  \bibfield  {author} {\bibinfo {author} {\bibfnamefont {J.}~\bibnamefont
  {Wang}}, \bibinfo {author} {\bibfnamefont {J.~F.}\ \bibnamefont {Herrmann}},
  \bibinfo {author} {\bibfnamefont {J.~D.}\ \bibnamefont {Witmer}}, \bibinfo
  {author} {\bibfnamefont {A.~H.}\ \bibnamefont {Safavi-Naeini}},\ and\
  \bibinfo {author} {\bibfnamefont {S.}~\bibnamefont {Fan}},\ }\bibfield
  {title} {\bibinfo {title} {Photonic modal circulator using temporal
  refractive-index modulation with spatial inversion symmetry},\ }\href
  {https://doi.org/10.1103/PhysRevLett.126.193901} {\bibfield  {journal}
  {\bibinfo  {journal} {Phys. Rev. Lett.}\ }\textbf {\bibinfo {volume} {126}},\
  \bibinfo {pages} {193901} (\bibinfo {year} {2021})}\BibitemShut {NoStop}%
\bibitem [{\citenamefont {Abboud}\ \emph {et~al.}(2013)\citenamefont {Abboud},
  \citenamefont {Cozza},\ and\ \citenamefont {Pichon}}]{Abboud}%
  \BibitemOpen
  \bibfield  {author} {\bibinfo {author} {\bibfnamefont {L.}~\bibnamefont
  {Abboud}}, \bibinfo {author} {\bibfnamefont {A.}~\bibnamefont {Cozza}},\ and\
  \bibinfo {author} {\bibfnamefont {L.}~\bibnamefont {Pichon}},\ }\bibfield
  {title} {\bibinfo {title} {A noniterative method for locating soft faults in
  complex wire networks},\ }\href {https://doi.org/10.1109/TVT.2013.2237796}
  {\bibfield  {journal} {\bibinfo  {journal} {IEEE Transactions on Vehicular
  Technology}\ }\textbf {\bibinfo {volume} {62}},\ \bibinfo {pages} {1010}
  (\bibinfo {year} {2013})}\BibitemShut {NoStop}%
\bibitem [{\citenamefont {Ma}\ \emph {et~al.}(2014)\citenamefont {Ma},
  \citenamefont {Xu}, \citenamefont {Liu},\ and\ \citenamefont
  {Wang}}]{MaCheng}%
  \BibitemOpen
  \bibfield  {author} {\bibinfo {author} {\bibfnamefont {C.}~\bibnamefont
  {Ma}}, \bibinfo {author} {\bibfnamefont {X.}~\bibnamefont {Xu}}, \bibinfo
  {author} {\bibfnamefont {Y.}~\bibnamefont {Liu}},\ and\ \bibinfo {author}
  {\bibfnamefont {L.~V.}\ \bibnamefont {Wang}},\ }\bibfield  {title} {\bibinfo
  {title} {Time-reversed adapted-perturbation ({{TRAP}}) optical focusing onto
  dynamic objects inside scattering media},\ }\href
  {https://doi.org/10.1038/nphoton.2014.251} {\bibfield  {journal} {\bibinfo
  {journal} {Nature Photonics}\ }\textbf {\bibinfo {volume} {8}},\ \bibinfo
  {pages} {931} (\bibinfo {year} {2014})}\BibitemShut {NoStop}%
\bibitem [{\citenamefont {Zhou}\ \emph {et~al.}(2014)\citenamefont {Zhou},
  \citenamefont {Ruan}, \citenamefont {Yang},\ and\ \citenamefont
  {Judkewitz}}]{ZhouEdward}%
  \BibitemOpen
  \bibfield  {author} {\bibinfo {author} {\bibfnamefont {E.~H.}\ \bibnamefont
  {Zhou}}, \bibinfo {author} {\bibfnamefont {H.}~\bibnamefont {Ruan}}, \bibinfo
  {author} {\bibfnamefont {C.}~\bibnamefont {Yang}},\ and\ \bibinfo {author}
  {\bibfnamefont {B.}~\bibnamefont {Judkewitz}},\ }\bibfield  {title} {\bibinfo
  {title} {Focusing on moving targets through scattering samples},\ }\href
  {https://doi.org/10.1364/OPTICA.1.000227} {\bibfield  {journal} {\bibinfo
  {journal} {Optica}\ }\textbf {\bibinfo {volume} {1}},\ \bibinfo {pages} {227}
  (\bibinfo {year} {2014})}\BibitemShut {NoStop}%
\bibitem [{\citenamefont {Sol}\ \emph {et~al.}(2024)\citenamefont {Sol},
  \citenamefont {Le~Magoarou},\ and\ \citenamefont {del Hougne}}]{Sol}%
  \BibitemOpen
  \bibfield  {author} {\bibinfo {author} {\bibfnamefont {J.}~\bibnamefont
  {Sol}}, \bibinfo {author} {\bibfnamefont {L.}~\bibnamefont {Le~Magoarou}},\
  and\ \bibinfo {author} {\bibfnamefont {P.}~\bibnamefont {del Hougne}},\
  }\bibfield  {title} {\bibinfo {title} {Optimal blind focusing on
  perturbation-inducing targets in sub-unitary complex media},\ }\href
  {https://doi.org/https://doi.org/10.1002/lpor.202400619} {\bibfield
  {journal} {\bibinfo  {journal} {Laser \& Photonics Reviews}\ }\textbf
  {\bibinfo {volume} {2024}},\ \bibinfo {pages} {2400619} (\bibinfo {year}
  {2024})}\BibitemShut {NoStop}%
\bibitem [{\citenamefont {Brouwer}\ \emph {et~al.}(1997)\citenamefont
  {Brouwer}, \citenamefont {Frahm},\ and\ \citenamefont
  {Beenakker}}]{Brouwer97}%
  \BibitemOpen
  \bibfield  {author} {\bibinfo {author} {\bibfnamefont {P.~W.}\ \bibnamefont
  {Brouwer}}, \bibinfo {author} {\bibfnamefont {K.~M.}\ \bibnamefont {Frahm}},\
  and\ \bibinfo {author} {\bibfnamefont {C.~W.~J.}\ \bibnamefont {Beenakker}},\
  }\bibfield  {title} {\bibinfo {title} {Quantum mechanical time-delay matrix
  in chaotic scattering},\ }\href {https://doi.org/10.1103/PhysRevLett.78.4737}
  {\bibfield  {journal} {\bibinfo  {journal} {Phys. Rev. Lett.}\ }\textbf
  {\bibinfo {volume} {78}},\ \bibinfo {pages} {4737} (\bibinfo {year}
  {1997})}\BibitemShut {NoStop}%
\bibitem [{\citenamefont {Brouwer}\ \emph {et~al.}(1999)\citenamefont
  {Brouwer}, \citenamefont {Frahm},\ and\ \citenamefont
  {Beenakker}}]{Brouwer99}%
  \BibitemOpen
  \bibfield  {author} {\bibinfo {author} {\bibfnamefont {P.~W.}\ \bibnamefont
  {Brouwer}}, \bibinfo {author} {\bibfnamefont {K.~M.}\ \bibnamefont {Frahm}},\
  and\ \bibinfo {author} {\bibfnamefont {C.~W.~J.}\ \bibnamefont {Beenakker}},\
  }\bibfield  {title} {\bibinfo {title} {Distribution of the quantum mechanical
  time-delay matrix for a chaotic cavity},\ }\href
  {https://doi.org/10.1088/0959-7174/9/2/303} {\bibfield  {journal} {\bibinfo
  {journal} {Waves in Random Media}\ }\textbf {\bibinfo {volume} {9}},\
  \bibinfo {pages} {91} (\bibinfo {year} {1999})}\BibitemShut {NoStop}%
\bibitem [{\citenamefont {Ambichl}\ \emph {et~al.}(2017)\citenamefont
  {Ambichl}, \citenamefont {Brandst\"otter}, \citenamefont {B\"ohm},
  \citenamefont {K\"uhmayer}, \citenamefont {Kuhl},\ and\ \citenamefont
  {Rotter}}]{Ambichl}%
  \BibitemOpen
  \bibfield  {author} {\bibinfo {author} {\bibfnamefont {P.}~\bibnamefont
  {Ambichl}}, \bibinfo {author} {\bibfnamefont {A.}~\bibnamefont
  {Brandst\"otter}}, \bibinfo {author} {\bibfnamefont {J.}~\bibnamefont
  {B\"ohm}}, \bibinfo {author} {\bibfnamefont {M.}~\bibnamefont {K\"uhmayer}},
  \bibinfo {author} {\bibfnamefont {U.}~\bibnamefont {Kuhl}},\ and\ \bibinfo
  {author} {\bibfnamefont {S.}~\bibnamefont {Rotter}},\ }\bibfield  {title}
  {\bibinfo {title} {Focusing inside disordered media with the generalized
  {W}igner-{S}mith operator},\ }\href
  {https://doi.org/10.1103/PhysRevLett.119.033903} {\bibfield  {journal}
  {\bibinfo  {journal} {Phys. Rev. Lett.}\ }\textbf {\bibinfo {volume} {119}},\
  \bibinfo {pages} {033903} (\bibinfo {year} {2017})}\BibitemShut {NoStop}%
\bibitem [{\citenamefont {Wigner}(1955)}]{Wigner}%
  \BibitemOpen
  \bibfield  {author} {\bibinfo {author} {\bibfnamefont {E.~P.}\ \bibnamefont
  {Wigner}},\ }\bibfield  {title} {\bibinfo {title} {Lower limit for the energy
  derivative of the scattering phase shift},\ }\href
  {https://doi.org/10.1103/PhysRev.98.145} {\bibfield  {journal} {\bibinfo
  {journal} {Phys. Rev.}\ }\textbf {\bibinfo {volume} {98}},\ \bibinfo {pages}
  {145} (\bibinfo {year} {1955})}\BibitemShut {NoStop}%
\bibitem [{\citenamefont {Smith}(1960)}]{Smith}%
  \BibitemOpen
  \bibfield  {author} {\bibinfo {author} {\bibfnamefont {F.~T.}\ \bibnamefont
  {Smith}},\ }\bibfield  {title} {\bibinfo {title} {Lifetime matrix in
  collision theory},\ }\href {https://doi.org/10.1103/PhysRev.118.349}
  {\bibfield  {journal} {\bibinfo  {journal} {Phys. Rev.}\ }\textbf {\bibinfo
  {volume} {118}},\ \bibinfo {pages} {349} (\bibinfo {year}
  {1960})}\BibitemShut {NoStop}%
\bibitem [{\citenamefont {Horodynski}\ \emph {et~al.}(2023)\citenamefont
  {Horodynski}, \citenamefont {Reiter}, \citenamefont {K\"uhmayer},\ and\
  \citenamefont {Rotter}}]{Horodynski23PRA}%
  \BibitemOpen
  \bibfield  {author} {\bibinfo {author} {\bibfnamefont {M.}~\bibnamefont
  {Horodynski}}, \bibinfo {author} {\bibfnamefont {T.}~\bibnamefont {Reiter}},
  \bibinfo {author} {\bibfnamefont {M.}~\bibnamefont {K\"uhmayer}},\ and\
  \bibinfo {author} {\bibfnamefont {S.}~\bibnamefont {Rotter}},\ }\bibfield
  {title} {\bibinfo {title} {Tractor beams with optimal pulling force using
  structured waves},\ }\href {https://doi.org/10.1103/PhysRevA.108.023504}
  {\bibfield  {journal} {\bibinfo  {journal} {Phys. Rev. A}\ }\textbf {\bibinfo
  {volume} {108}},\ \bibinfo {pages} {023504} (\bibinfo {year}
  {2023})}\BibitemShut {NoStop}%
\bibitem [{\citenamefont {Būtaitė}\ \emph {et~al.}(2024)\citenamefont
  {Būtaitė}, \citenamefont {Sharp}, \citenamefont {Horodynski}, \citenamefont
  {Gibson}, \citenamefont {Padgett}, \citenamefont {Rotter}, \citenamefont
  {Taylor},\ and\ \citenamefont {Phillips}}]{Butaite}%
  \BibitemOpen
  \bibfield  {author} {\bibinfo {author} {\bibfnamefont {U.~G.}\ \bibnamefont
  {Būtaitė}}, \bibinfo {author} {\bibfnamefont {C.}~\bibnamefont {Sharp}},
  \bibinfo {author} {\bibfnamefont {M.}~\bibnamefont {Horodynski}}, \bibinfo
  {author} {\bibfnamefont {G.~M.}\ \bibnamefont {Gibson}}, \bibinfo {author}
  {\bibfnamefont {M.~J.}\ \bibnamefont {Padgett}}, \bibinfo {author}
  {\bibfnamefont {S.}~\bibnamefont {Rotter}}, \bibinfo {author} {\bibfnamefont
  {J.~M.}\ \bibnamefont {Taylor}},\ and\ \bibinfo {author} {\bibfnamefont
  {D.~B.}\ \bibnamefont {Phillips}},\ }\bibfield  {title} {\bibinfo {title}
  {Photon-efficient optical tweezers via wavefront shaping},\ }\href
  {https://doi.org/10.1126/sciadv.adi7792} {\bibfield  {journal} {\bibinfo
  {journal} {Science Advances}\ }\textbf {\bibinfo {volume} {10}},\ \bibinfo
  {pages} {eadi7792} (\bibinfo {year} {2024})},\ \Eprint
  {https://arxiv.org/abs/https://www.science.org/doi/pdf/10.1126/sciadv.adi7792}
  {https://www.science.org/doi/pdf/10.1126/sciadv.adi7792} \BibitemShut
  {NoStop}%
\bibitem [{\citenamefont {Matth\`es}\ \emph {et~al.}(2021)\citenamefont
  {Matth\`es}, \citenamefont {Bromberg}, \citenamefont {de~Rosny},\ and\
  \citenamefont {Popoff}}]{Matthes}%
  \BibitemOpen
  \bibfield  {author} {\bibinfo {author} {\bibfnamefont {M.~W.}\ \bibnamefont
  {Matth\`es}}, \bibinfo {author} {\bibfnamefont {Y.}~\bibnamefont {Bromberg}},
  \bibinfo {author} {\bibfnamefont {J.}~\bibnamefont {de~Rosny}},\ and\
  \bibinfo {author} {\bibfnamefont {S.~M.}\ \bibnamefont {Popoff}},\ }\bibfield
   {title} {\bibinfo {title} {Learning and avoiding disorder in multimode
  fibers},\ }\href {https://doi.org/10.1103/PhysRevX.11.021060} {\bibfield
  {journal} {\bibinfo  {journal} {Phys. Rev. X}\ }\textbf {\bibinfo {volume}
  {11}},\ \bibinfo {pages} {021060} (\bibinfo {year} {2021})}\BibitemShut
  {NoStop}%
\bibitem [{\citenamefont {H\"upfl}\ \emph
  {et~al.}(2023{\natexlab{a}})\citenamefont {H\"upfl}, \citenamefont
  {Bachelard}, \citenamefont {Kaczvinszki}, \citenamefont {Horodynski},
  \citenamefont {K\"uhmayer},\ and\ \citenamefont {Rotter}}]{HupflPRL}%
  \BibitemOpen
  \bibfield  {author} {\bibinfo {author} {\bibfnamefont {J.}~\bibnamefont
  {H\"upfl}}, \bibinfo {author} {\bibfnamefont {N.}~\bibnamefont {Bachelard}},
  \bibinfo {author} {\bibfnamefont {M.}~\bibnamefont {Kaczvinszki}}, \bibinfo
  {author} {\bibfnamefont {M.}~\bibnamefont {Horodynski}}, \bibinfo {author}
  {\bibfnamefont {M.}~\bibnamefont {K\"uhmayer}},\ and\ \bibinfo {author}
  {\bibfnamefont {S.}~\bibnamefont {Rotter}},\ }\bibfield  {title} {\bibinfo
  {title} {Optimal cooling of multiple levitated particles through far-field
  wavefront shaping},\ }\href {https://doi.org/10.1103/PhysRevLett.130.083203}
  {\bibfield  {journal} {\bibinfo  {journal} {Phys. Rev. Lett.}\ }\textbf
  {\bibinfo {volume} {130}},\ \bibinfo {pages} {083203} (\bibinfo {year}
  {2023}{\natexlab{a}})}\BibitemShut {NoStop}%
\bibitem [{\citenamefont {H\"upfl}\ \emph
  {et~al.}(2023{\natexlab{b}})\citenamefont {H\"upfl}, \citenamefont
  {Bachelard}, \citenamefont {Kaczvinszki}, \citenamefont {Horodynski},
  \citenamefont {K\"uhmayer},\ and\ \citenamefont {Rotter}}]{HupflPRA}%
  \BibitemOpen
  \bibfield  {author} {\bibinfo {author} {\bibfnamefont {J.}~\bibnamefont
  {H\"upfl}}, \bibinfo {author} {\bibfnamefont {N.}~\bibnamefont {Bachelard}},
  \bibinfo {author} {\bibfnamefont {M.}~\bibnamefont {Kaczvinszki}}, \bibinfo
  {author} {\bibfnamefont {M.}~\bibnamefont {Horodynski}}, \bibinfo {author}
  {\bibfnamefont {M.}~\bibnamefont {K\"uhmayer}},\ and\ \bibinfo {author}
  {\bibfnamefont {S.}~\bibnamefont {Rotter}},\ }\bibfield  {title} {\bibinfo
  {title} {Optimal cooling of multiple levitated particles: Theory of far-field
  wavefront shaping},\ }\href {https://doi.org/10.1103/PhysRevA.107.023112}
  {\bibfield  {journal} {\bibinfo  {journal} {Phys. Rev. A}\ }\textbf {\bibinfo
  {volume} {107}},\ \bibinfo {pages} {023112} (\bibinfo {year}
  {2023}{\natexlab{b}})}\BibitemShut {NoStop}%
\bibitem [{\citenamefont {Orazbayev}\ \emph {et~al.}(2024)\citenamefont
  {Orazbayev}, \citenamefont {Mall{\'e}jac}, \citenamefont {Bachelard},
  \citenamefont {Rotter},\ and\ \citenamefont {Fleury}}]{Orazbayev}%
  \BibitemOpen
  \bibfield  {author} {\bibinfo {author} {\bibfnamefont {B.}~\bibnamefont
  {Orazbayev}}, \bibinfo {author} {\bibfnamefont {M.}~\bibnamefont
  {Mall{\'e}jac}}, \bibinfo {author} {\bibfnamefont {N.}~\bibnamefont
  {Bachelard}}, \bibinfo {author} {\bibfnamefont {S.}~\bibnamefont {Rotter}},\
  and\ \bibinfo {author} {\bibfnamefont {R.}~\bibnamefont {Fleury}},\
  }\bibfield  {title} {\bibinfo {title} {Wave-momentum shaping for moving
  objects in heterogeneous and dynamic media},\ }\bibfield  {journal} {\bibinfo
   {journal} {Nature Physics}\ }\href
  {https://doi.org/10.1038/s41567-024-02538-5} {10.1038/s41567-024-02538-5}
  (\bibinfo {year} {2024})\BibitemShut {NoStop}%
\bibitem [{\citenamefont {Mazilu}\ \emph {et~al.}(2011)\citenamefont {Mazilu},
  \citenamefont {Baumgartl}, \citenamefont {Kosmeier},\ and\ \citenamefont
  {Dholakia}}]{Mazilu}%
  \BibitemOpen
  \bibfield  {author} {\bibinfo {author} {\bibfnamefont {M.}~\bibnamefont
  {Mazilu}}, \bibinfo {author} {\bibfnamefont {J.}~\bibnamefont {Baumgartl}},
  \bibinfo {author} {\bibfnamefont {S.}~\bibnamefont {Kosmeier}},\ and\
  \bibinfo {author} {\bibfnamefont {K.}~\bibnamefont {Dholakia}},\ }\bibfield
  {title} {\bibinfo {title} {Optical eigenmodes; exploiting the quadratic
  nature of the energy flux and of scattering interactions},\ }\href
  {https://doi.org/10.1364/OE.19.000933} {\bibfield  {journal} {\bibinfo
  {journal} {Opt. Express}\ }\textbf {\bibinfo {volume} {19}},\ \bibinfo
  {pages} {933} (\bibinfo {year} {2011})}\BibitemShut {NoStop}%
\bibitem [{\citenamefont {McCabe}\ \emph {et~al.}(2011)\citenamefont {McCabe},
  \citenamefont {Tajalli}, \citenamefont {Austin}, \citenamefont {Bondareff},
  \citenamefont {Walmsley}, \citenamefont {Gigan},\ and\ \citenamefont
  {Chatel}}]{McCabe}%
  \BibitemOpen
  \bibfield  {author} {\bibinfo {author} {\bibfnamefont {D.~J.}\ \bibnamefont
  {McCabe}}, \bibinfo {author} {\bibfnamefont {A.}~\bibnamefont {Tajalli}},
  \bibinfo {author} {\bibfnamefont {D.~R.}\ \bibnamefont {Austin}}, \bibinfo
  {author} {\bibfnamefont {P.}~\bibnamefont {Bondareff}}, \bibinfo {author}
  {\bibfnamefont {I.~A.}\ \bibnamefont {Walmsley}}, \bibinfo {author}
  {\bibfnamefont {S.}~\bibnamefont {Gigan}},\ and\ \bibinfo {author}
  {\bibfnamefont {B.}~\bibnamefont {Chatel}},\ }\bibfield  {title} {\bibinfo
  {title} {Spatio-temporal focusing of an ultrafast pulse through a multiply
  scattering medium},\ }\href {https://doi.org/10.1038/ncomms1434} {\bibfield
  {journal} {\bibinfo  {journal} {Nature Communications}\ }\textbf {\bibinfo
  {volume} {2}},\ \bibinfo {pages} {447} (\bibinfo {year} {2011})}\BibitemShut
  {NoStop}%
\bibitem [{\citenamefont {Mounaix}\ \emph {et~al.}(2020)\citenamefont
  {Mounaix}, \citenamefont {Fontaine}, \citenamefont {Neilson}, \citenamefont
  {Ryf}, \citenamefont {Chen}, \citenamefont {{Alvarado-Zacarias}},\ and\
  \citenamefont {Carpenter}}]{mounaix2020time}%
  \BibitemOpen
  \bibfield  {author} {\bibinfo {author} {\bibfnamefont {M.}~\bibnamefont
  {Mounaix}}, \bibinfo {author} {\bibfnamefont {N.~K.}\ \bibnamefont
  {Fontaine}}, \bibinfo {author} {\bibfnamefont {D.~T.}\ \bibnamefont
  {Neilson}}, \bibinfo {author} {\bibfnamefont {R.}~\bibnamefont {Ryf}},
  \bibinfo {author} {\bibfnamefont {H.}~\bibnamefont {Chen}}, \bibinfo {author}
  {\bibfnamefont {J.~C.}\ \bibnamefont {{Alvarado-Zacarias}}},\ and\ \bibinfo
  {author} {\bibfnamefont {J.}~\bibnamefont {Carpenter}},\ }\bibfield  {title}
  {\bibinfo {title} {Time reversed optical waves by arbitrary vector
  spatiotemporal field generation},\ }\href
  {https://doi.org/10.1038/s41467-020-19601-3} {\bibfield  {journal} {\bibinfo
  {journal} {Nature Communications}\ }\textbf {\bibinfo {volume} {11}},\
  \bibinfo {pages} {5813} (\bibinfo {year} {2020})}\BibitemShut {NoStop}%
\bibitem [{\citenamefont {Bouchet}\ and\ \citenamefont
  {Bossy}(2023)}]{Bouchet23}%
  \BibitemOpen
  \bibfield  {author} {\bibinfo {author} {\bibfnamefont {D.}~\bibnamefont
  {Bouchet}}\ and\ \bibinfo {author} {\bibfnamefont {E.}~\bibnamefont
  {Bossy}},\ }\bibfield  {title} {\bibinfo {title} {Temporal shaping of wave
  fields for optimally precise measurements in scattering environments},\
  }\href {https://doi.org/10.1103/PhysRevResearch.5.013144} {\bibfield
  {journal} {\bibinfo  {journal} {Phys. Rev. Res.}\ }\textbf {\bibinfo {volume}
  {5}},\ \bibinfo {pages} {013144} (\bibinfo {year} {2023})}\BibitemShut
  {NoStop}%
\bibitem [{\citenamefont {Ferise}\ \emph {et~al.}(2023)\citenamefont {Ferise},
  \citenamefont {del Hougne},\ and\ \citenamefont {Davy}}]{Ferise}%
  \BibitemOpen
  \bibfield  {author} {\bibinfo {author} {\bibfnamefont {C.}~\bibnamefont
  {Ferise}}, \bibinfo {author} {\bibfnamefont {P.}~\bibnamefont {del Hougne}},\
  and\ \bibinfo {author} {\bibfnamefont {M.}~\bibnamefont {Davy}},\ }\bibfield
  {title} {\bibinfo {title} {Optimal matrix-based spatiotemporal wave control
  for virtual perfect absorption, energy deposition, and scattering-invariant
  modes in disordered systems},\ }\href
  {https://doi.org/10.1103/PhysRevApplied.20.054023} {\bibfield  {journal}
  {\bibinfo  {journal} {Phys. Rev. Appl.}\ }\textbf {\bibinfo {volume} {20}},\
  \bibinfo {pages} {054023} (\bibinfo {year} {2023})}\BibitemShut {NoStop}%
\bibitem [{\citenamefont {Horstmeyer}\ \emph {et~al.}(2015)\citenamefont
  {Horstmeyer}, \citenamefont {Ruan},\ and\ \citenamefont
  {Yang}}]{horstmeyer2015guidestar}%
  \BibitemOpen
  \bibfield  {author} {\bibinfo {author} {\bibfnamefont {R.}~\bibnamefont
  {Horstmeyer}}, \bibinfo {author} {\bibfnamefont {H.}~\bibnamefont {Ruan}},\
  and\ \bibinfo {author} {\bibfnamefont {C.}~\bibnamefont {Yang}},\ }\bibfield
  {title} {\bibinfo {title} {Guidestar-assisted wavefront-shaping methods for
  focusing light into biological tissue},\ }\href
  {https://doi.org/10.1038/nphoton.2015.140} {\bibfield  {journal} {\bibinfo
  {journal} {Nature Photonics}\ }\textbf {\bibinfo {volume} {9}},\ \bibinfo
  {pages} {563} (\bibinfo {year} {2015})}\BibitemShut {NoStop}%
\end{thebibliography}%
\end{document}